\documentclass[twocolumn,numberedappendix,trackchanges]{aastex62}

\usepackage{natbib}
\usepackage{multirow}
\usepackage{gensymb} 
\usepackage[flushleft]{threeparttable} 
\usepackage{amsmath}
\usepackage{url}
\usepackage{breakcites} 
\hypersetup{breaklinks=true} 

\shorttitle{Brightest $z\sim8$ LBGs in COSMOS/UltraVISTA}
\shortauthors{Stefanon et al.}

\begin{document}

\title{The Brightest $z\gtrsim8$ Galaxies over the COSMOS UltraVISTA Field}

\author{Mauro Stefanon}
\affiliation{Leiden Observatory, Leiden University, NL-2300 RA Leiden, Netherlands}

\author{Ivo Labb\'e}
\affiliation{Centre for Astrophysics and SuperComputing, Swinburne, University of Technology, Hawthorn, Victoria, 3122, Australia}

\author{Rychard J. Bouwens}
\affiliation{Leiden Observatory, Leiden University, NL-2300 RA Leiden, Netherlands}

\author{Pascal Oesch}
\affiliation{Observatoire de Gen\`eve, 51 Ch. des Maillettes, 1290 Versoix, Switzerland}

\author{Matthew L. N. Ashby}
\affiliation{Center for Astrophysics | Harvard \& Smithsonian 60 Garden St., Cambridge, MA, 02138, USA}

\author{Karina I. Caputi}
\affiliation{Kapteyn Astronomical Institute, University of Groningen, P.O. Box 800, 9700AV Groningen, The Netherlands}
\affiliation{Cosmic Dawn Center (DAWN)}

\author{Marijn Franx}
\affiliation{Leiden Observatory, Leiden University, NL-2300 RA Leiden, Netherlands}

\author{Johan P. U. Fynbo}
\affiliation{Cosmic Dawn Center (DAWN)}
\affiliation{Niels Bohr Institute, University of Copenhagen, Lyngbyvej 2, 2100 Copenhagen \O, Denmark}

\author{Garth D. Illingworth}
\affiliation{UCO/Lick Observatory, University of California, Santa Cruz, 1156 High St, Santa Cruz, CA 95064, USA}

\author{Olivier Le F\`evre}
\affiliation{Aix-Marseille Universit\'e, CNRS, LAM (Laboratoire d'Astrophysique de Marseille) UMR 7326, 13388 Marseille, France}

\author{Danilo Marchesini}
\affiliation{Department of Physics and Astronomy, Tufts University, Medford, MA 02155, USA}

\author{Henry J. McCracken}
\affiliation{Sorbonne Universit\'e, CNRS, UMR 7095, Institut d'Astrophysique de Paris, 98 bis bd Arago, 75014 Paris, France}

\author{Bo Milvang-Jensen}
\affiliation{Cosmic Dawn Center (DAWN)}
\affiliation{Niels Bohr Institute, University of Copenhagen, Lyngbyvej 2, 2100 Copenhagen \O, Denmark}

\author{Adam Muzzin}
\affiliation{York University, 4700 Keele Street, Toronto, ON, M3J 1P3, Canada}

\author{Pieter van Dokkum}
\affiliation{Astronomy Department, Yale University, 52 Hillhouse Ave, New Haven, CT 06511, USA}

\email{Email: stefanon@strw.leidenuniv.nl}

\begin{abstract}
We present 16 new ultrabright $H_{AB}\lesssim25$ galaxy candidates at $z\sim8$ identified over the COSMOS/UltraVISTA field. The new search takes advantage of the deepest-available ground-based optical and near-infrared observations, including the DR3 release of UltraVISTA and full-depth {\it Spitzer}/IRAC observations from the SMUVS and SPLASH programs.  Candidates are selected using Lyman-break color criteria, combined with strict optical non-detection and SED-fitting criteria, designed to minimize contamination by low-redshift galaxies and low-mass stars. \textit{HST}/WFC3 coverage from the DASH program reveals that one source evident in our ground-based near-IR data has significant substructure and may actually correspond to 3 separate $z\sim8$ objects, resulting in a total sample of 18 galaxies, 10 of which seem to be fairly robust (with a $>97\%$ probability of being at $z>7$). The UV-continuum slope $\beta$ for the bright $z\sim8$ sample is $\beta=-2.2\pm0.6$, bluer but still consistent with that of similarly bright galaxies at $z\sim6$ ($\beta=-1.55\pm0.17$) and $z\sim7$ ($\beta=-1.75\pm0.18$). Their typical stellar masses are 10$^{9.1^{+0.5}_{-0.4}}$ $M_{\odot}$, with the SFRs of  $32^{+44}_{-32}M_{\odot}$/year, specific SFR of $4^{+8}_{-4}$ Gyr$^{-1}$, stellar ages of $\sim22^{+69}_{-22}$\,Myr, and low dust content A$_V=0.15^{+0.30}_{-0.15}$\, mag. Using this sample we constrain the bright end of the $z\sim8$ UV luminosity function (LF). When combined with recent empty field LF estimates at similar redshifts, the resulting $z\sim8$ LF can be equally well represented by either a Schechter or a double power-law (DPL) form. Assuming a Schechter parameterization, the best-fit characteristic magnitude is $M^* = -20.95^{+0.30}_{-0.35}$ mag with a very steep faint end slope $\alpha=-2.15^{+0.20}_{-0.19}$. These new candidates include amongst the brightest yet found at these redshifts, $0.5-1.0$ magnitude brighter than found over CANDELS, providing excellent targets for spectroscopic and longer-wavelength follow-up studies. 
\end{abstract}


\section{Introduction}

The confirmation and characterization of galaxy candidates within the cosmic reionization epoch has been a major challenge for observational extragalactic astronomy for the last few years. The exceptional sensitivity offered by the Wide Field Camera 3 Infrared (WFC3/IR) instrument onboard the \textit{ Hubble Space Telescope (HST)}, combined with efficient photometric selection techniques have enabled the identification of $\gtrsim 700$ faint galaxy candidates at $z=7-11$ (e.g., \citealt{bouwens2011b, bouwens2015, schenker2013, mclure2013, oesch2012, oesch2014, oesch2016, oesch2018, schmidt2014, finkelstein2015a}). These high-redshift galaxy samples have provided a powerful way to investigate the build-up and evolution of galaxies, by imposing new constraints on the evolution of their rest-frame ultra-violet (UV) luminosity functions (LFs) and integrated star formation rate density (SFRD - but see also e.g., \citealt{tanvir2012,  mcguire2016} for a complementary approach using gamma-ray bursts).

The redshift range of $z\sim8-10$ is of particular interest: a number of works suggest a rapid decline of the star-formation rate density (SFRD) from z$\sim$8 to z$\sim$10 (see e.g., \citealt{oesch2012, oesch2014, oesch2015b, oesch2018, ellis2013, bouwens2015} - but see e.g., \citealt{mcleod2015, mcleod2016}). A key question is therefore whether the faint galaxies emit enough ionizing photons to reionize the universe at $z\gtrsim7$ (e.g., \citealt{bolton2007, oesch2009, robertson2010b, shull2012, bouwens2011b, bouwens2015, finkelstein2015a, tanvir2019}). 

Answering the above question requires estimating the faint-end slope of the UV LF during the reionization epoch. For a \citet{schechter1976} parameterization of the LF, because of the correlation between the characteristic luminosity and the faint-end slope, constraining the bright end  of the LF (e.g., through searches in shallow wide-field surveys) will also improve the estimates at the faint end (e.g., \citealt{bouwens2008}).  Furthermore, identifying bright Lyman-break galaxies (LBGs) will help determine whether the LF has an exponential cut-off (with relatively few luminous galaxies, as has been established at  $z < 7$) or is featureless like a power-law (as suggested by a recent works - e.g., \citealt{bowler2015, bowler2017, ono2018}).  Finally, measurements of the bright end encode crucial information about early galaxies, including the effects of dust, star formation feedback, and the duty cycle of galaxies. The evolution of the bright end therefore provides strong tests for models of galaxy evolution at  these redshifts (e.g., \citealt{finlator2011, jaacks2012,  mason2015, trac2015, mashian2016, waters2016}).

Bright $z\gtrsim 8$ candidate LBGs are also important targets for spectroscopic follow-up and in  preparation for  the James  Webb Space Telescope. Spectroscopic confirmation is vital to test the validity of the photometric selection techniques and to identify potential contaminant populations at lower redshift,  given the physical conditions at such early times are potentially very different than at present increasing the uncertainty in photometric redshift determinations. When galaxies are confirmed, spectroscopy enables the study of UV spectral features (e.g., \citealt{stark2015b, stark2015a, stark2017}) and improve estimates of stellar mass and star formation rate. However, spectroscopic confirmation has been very challenging so far, with fewer than expected (e.g., \citealt{stark2011}) normal galaxies with robust redshift measurements at $z > 7$ (e.g., \citealt{vanzella2011, pentericci2011, ono2012, schenker2012, shibuya2012, finkelstein2013, tilvi2014, song2016b, schmidt2016, huang2016, hoag2017, hoag2018, larson2018, pentericci2018}). The likely reason for this is the increased neutral fraction at $z\gtrsim 6$ combined with the faintness of the sources (e.g., \citealt{treu2013, schenker2014, pentericci2014, tilvi2014}). Interestingly, a number of recent works have reported spectroscopic confirmation for bright ($H\sim25$ mag) LBGs at the epoch of the reionization from Ly$\alpha$ detection (\citealt{oesch2015a, roberts-borsani2016, stark2016, zitrin2015}). These observations further suggested that reionization could have happened in a patchy form, rather than homogeneously, and inspired confidence in our ability to reliably select bright sources to the highest possible redshifts.

Perhaps surprisingly, observational progress on the very bright end has been relatively slow. Covering wide areas with HST is very inefficient due to the extremely low surface densities of the brightest $z>8$ galaxies. Some progress has come from pure parallel imaging surveys such as BORG/HIPPIES (\citealt{trenti2011, yan2011}), from targeted follow up over the full CANDELS area (\citealt{oesch2015a, roberts-borsani2016, zitrin2015, stark2016}) and from the RELICS program (\citealt{salmon2017}), which builds on the strong-lensing strategy of the \textit{Hubble} Frontier Field (HFF) and CLASH surveys. Combined together, these wider-area, shallow surveys still only cover $<1300$ arcmin$^2$ and provided only $\lesssim 5$ candidates at $z\gtrsim8$ brighter than  $M_{UV}\lesssim -22.0$ (\citealt{bernard2016, calvi2016, livermore2018, morishita2018}). 

An alternative approach consists in leveraging the on-going wide-field ground-based surveys such as COSMOS/UltraVISTA and UKIDSS/UDS, which benefit from deep ($\sim26$ mag) wide wavelength coverage ($0.3-5\mu$m - e.g., \citealt{bowler2012, bowler2014, bowler2015, bowler2017, stefanon2017c}).

Here we report the full analysis and the results of the search for ultrabright $H \sim 24-26$\,mag galaxy candidates at $z \gtrsim 8$ from the COSMOS/UltraVISTA program. This search takes advantage of the deepest-available ground-based optical+near-infrared observations, in particular the DR3 release of UltraVISTA which provides $\sim1.4$ mag deeper data in $Y,J,H,K_s$ compared to DR1 (\citealt{mccracken2012}). Our study also takes advantage of deep {\it Spitzer}/IRAC  \citep{fazio2004} observations from the {\it Spitzer} Large Area Survey with Hyper-Suprime-Cam (SPLASH, PI: Capak) and the \textit{ Spitzer} Matching survey of the UltraVISTA ultra-deep Stripes (SMUVS, PI: Caputi - \citealt{caputi2017, ashby2018}) programs. The increased depth and the inclusion of {\it Spitzer}/IRAC data, probing the rest-frame optical, now makes it possible to access the galaxy population at $z\gtrsim8$ through reliable sample selections.

In \citet{stefanon2017c} we already presented five candidate bright $z\gtrsim8$ LBGs initially identified in this search. Specifically, in that work we focused on the analysis of those sources with recent \textit{HST}/WFC3 imaging from one of our programs, and showed that the new \textit{HST} observations strengthened the available photometric constraints placing them at $z\sim8$.  The purpose of the present work is to present the parent sample from which those five objects were selected.

This paper is organized as follows. The observations are summarized in Sect. 2, while in Sect. 3 we describe how we performed the photometry. The source selection is detailed in Sect. 4. The sample is presented in Sect. 5 and it is characterized in Sect. 6. We present our conclusions in Sect. 7. Throughout, we adopt $\Omega_M=0.3, \Omega_\Lambda=0.7, H_0=70$ km~s$^{-1}$Mpc$^{-1}$.  Magnitudes are given in the AB system \citet{oke1983} and we adopt a \citet{chabrier2003} initial mass function (IMF).

\section{Observational Data}

 \begin{deluxetable}{rcc}
\tablecaption{Photometric depths of the adopted ground-based and \textit{Spitzer}/IRAC data sets, and corresponding average aperture corrections. \label{tab:depths}}
\tablehead{
\colhead{Filter} & \colhead{Aperture} & \colhead{Depth}  \\
\colhead{name} & \colhead{correction\tablenotemark{a}} & \colhead{$5 \sigma$\tablenotemark{b}} 
}
 \startdata
         CFHTLS $u^*$  & $    2.2 $ & $        26.7$ \\
              SSC $B$  & $    1.7 $ & $        27.4$ \\
              HSC $g$\tablenotemark{c}  & $    2.1 $ & $        26.7$ \\
           CFHTLS $g$  & $    2.1 $ & $        26.8$ \\
              SSC $V$  & $    2.1 $ & $        26.4$ \\
              HSC $r$\tablenotemark{c}  & $    1.7 $ & $        26.8$ \\
           CFHTLS $r$  & $    2.0 $ & $        26.4$ \\
            SSC $r^+$  & $    2.0 $ & $        26.6$ \\
           SSC  $i^+$  & $    1.9 $ & $         26.2$ \\
           CFHTLS $y$  & $    1.9 $ & $        26.1$ \\
           CFHTLS $i$  & $    1.9 $ & $        26.0$ \\
              HSC $i$\tablenotemark{c}  & $    1.8 $ & $        26.3$ \\
           CFHTLS $z$  & $    2.0 $ & $        25.2$ \\
              HSC $z$\tablenotemark{c}  & $    1.7 $ & $        25.9$ \\
           SSC  $z^+$  & $    2.2 $ & $        25.0$ \\
              HSC $y$\tablenotemark{c}  & $    2.1 $ & $        24.9$ \\
           UVISTA $Y$  & $    2.5 $ & $ 25.4 / 24.5 $ \\
          UVISTA $J$   & $    2.3 $ & $ 25.4 / 24.4 $ \\
           UVISTA $H$  & $    2.2 $ & $ 25.1 / 24.1 $ \\
UVISTA $K_\mathrm{S}$  & $    2.1 $ & $ 24.8 / 23.7 $ \\
       IRAC $3.6\mu$m   & $    2.7\tablenotemark{d} $ & $ 25.4/24.9 / 24.5 $ \\
       IRAC $4.5\mu$m   & $    2.7\tablenotemark{d} $ & $ 25.3/24.7 / 24.3 $ \\
       IRAC $5.8\mu$m   & $    3.4\tablenotemark{d} $ & $        20.8$ \\
       IRAC $8.0\mu$m   & $   4.1\tablenotemark{d} $ & $        20.6$ \\
 \enddata
 \tablenotetext{a}{Average multiplicative factors applied to estimate total fluxes.}
\tablenotetext{b}{Average depth over the full field corresponding to $5 \sigma$ flux dispersions in empty apertures of $1\farcs2$ diameter corrected to total using the average aperture correction. The two depths for UltraVISTA correspond to the ultradeep and deep stripes, respectively; the three depths for the \textit{Spitzer/}IRAC $3.6\mu$m and $4.5\mu$m bands correspond to the regions with SMUVS+SCOSMOS+SPLASH coverage (approximately overlapping with the ultradeep stripes) and SPLASH+SCOSMOS only ($\approx$ deep stripes).}
 \tablenotetext{c}{The HyperSuprimeCam data were not available during the initial selection of the sample; we included them in our subsequent analysis applying the same methods adopted for the rest of the ground and \textit{Spitzer/}IRAC mosaics.
 \tablenotetext{d}{Aperture corrections for IRAC bands refer to the $1\farcs8$ diameter.}}
 \end{deluxetable}

\begin{figure}
\includegraphics[width=9.2cm]{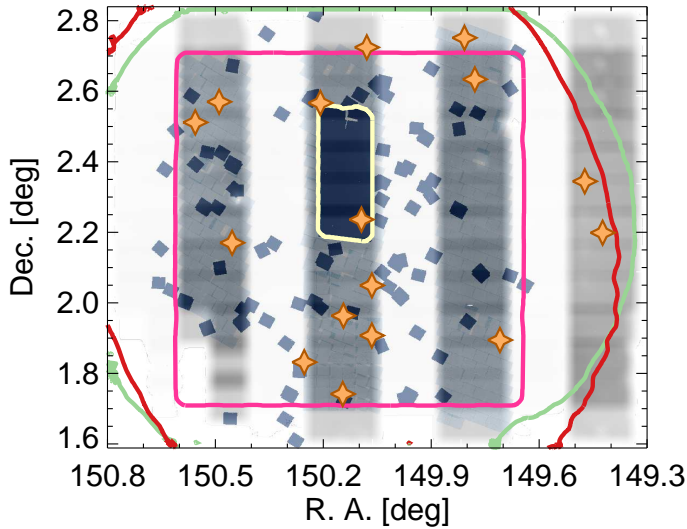}
\caption{Depth and layout of observations relevant to our current search for $z\sim8$ and $z\sim9$ galaxies over the UltraVISTA field.  The gray shaded image represents the UltraVISTA DR3 exposure time map (deeper exposure for darker regions). The colored curves mark the coverage from CFHT Legacy Survey (magenta), ultradeep HSC (green) and the deep {\it Spitzer}/IRAC observations from the SPLASH program (red). Even deeper {\it Spitzer}/IRAC observations are   available over the deep stripes from the SMUVS program. The yellow rectangle in the center demarcates the region with observations from the CANDELS program. The blue-shaded image corresponds to the COSMOS/DASH coverage map (darker regions indicate deeper coverage).  The orange stars mark the position of bright candidate $z\sim8$ galaxies we  have discovered in our search. \label{fig:cosmos}}
\end{figure}

Our analysis is based on ultradeep near-infrared imaging over the COSMOS field~\citep{scoville2007} from the third data release (DR3) of UltraVISTA (McCracken et al., in prep). UltraVISTA provides imaging which covers 1.6 square degrees~\citep{mccracken2012} in the $Y$, $J$, $H$ and $K_s$ filters to $\sim24-25$ mag (AB, $5\sigma$), with DR3 achieving fainter limits over 0.8 square degrees in 4 ultradeep stripes. The DR3 contains all data taken between December 2009 and July 2014 and reaches $Y=25.4, J=25.4, H=25.1, K=24.8$ mag (AB, $5\sigma$ in $1\farcs2$-diameter apertures).  The nominal depth we measure in the $Y$, $J$, $H$, and $K_s$ bands for the UltraVISTA DR3 release is $\sim$0.2 mag, $\sim$0.6 mag, $\sim$0.8 mag, and $\sim$0.2 mag, respectively, deeper than in the UltraVISTA DR2 release.

The optical data consists of CFHT/Megacam in $g$, $r$, $i$, $y$ and $z$ (\citealt{erben2009, hildebrandt2009} from the Canada-France-Hawaii Legacy Survey (CFHTLS), Subaru/Suprime-Cam  $B_j$,$V_j$, $g^+$, $r^+$, $i^+$ and $z^+$-imaging ~\citep{taniguchi2007}, and Subaru HyperSuprimeCam $g$, $r$, $i$, $z$ and $y$ (\citealt{aihara2017a, aihara2017b}).

For this work, we used full-depth \textit{Spitzer}/IRAC $3.6\mu$m and $4.5\mu$m mosaics we built combining observations from all available programs: S-COSMOS (\citealt{sanders2007}), the {\it Spitzer} Extended Deep Survey (\citealt{ashby2013}), the {\it Spitzer}-Cosmic Assembly Near-Infrared Deep Extragalactic Survey (S-CANDELS, \citealt{ashby2015}), the  {\it Spitzer} Large Area Survey with Hyper-Suprime-Cam (SPLASH, PI: Capak), the {\it Spitzer} Matching survey of the UltraVISTA ultra-deep Stripes (SMUVS, \citealt{caputi2017, ashby2018}). Compared to the original S-COSMOS IRAC data, SPLASH provides a large improvement in depth over nearly the whole UltraVISTA area, covering the central 1.2 square degree COSMOS field to 25.5 mag (AB) at 3.6 and $4.5\micron$.  SEDS and S-CANDELS cover smaller areas to even deeper limits, while SMUVS pushes deeper over the ultradeep UltraVISTA stripes.

Finally, we also included measurements in the IRAC $5.8\mu$m and $8.0\mu$m bands from the S-COSMOS program. Even though the coverage in these bands is rather shallow ($\sim20.7$\,mag, $5\sigma$ in $1\farcs8$-diameter aperture), detections in these two bands can be useful to discriminate high-redshift sources from lower-redshift interlopers. We discuss this for our sample at the end of Sect. \ref{sect:sample}.

A summary of all the deep, wide-area data sets along with $5\sigma$ depths is provided in Table \ref{tab:depths}, while in Figure \ref{fig:cosmos} we present the coverage of the different data sets.

\section{Photometry}
\label{sect:photometry}

Source catalogs were constructed using \textsc{SExtractor} v2.19.5 \citep{bertin1996}, run in dual image mode, with source detection performed on the square root of a $\chi^2$ image (\citealt{szalay1999}) built from the combination of the UltraVISTA $J$, $H$ and $K_\mathrm{s}$ images.

The first selection was performed adopting ground-based observations only.  Images were first convolved to the $J$-band point-spread function and carefully registered against the detection image (mean RMS $\sim0\farcs05$). Initial color measurements were made in small \citet{kron1980}-like apertures (SExtractor AUTO and Kron factor 1.2) with typical radius $r_\mathrm{color}\sim0\farcs35-0\farcs50$.

Successively, we refined our selection of $z\sim8$ and $z\sim9$ candidate galaxies using color measurements made in fixed $1.2''$-diameter apertures.  For this step, fluxes from  sources and their nearby neighbors ($12\farcs0\times12\farcs0$ region) are carefully modelled; aperture photometry is then performed after subtracting the neighbours using \textsc{mophongo} (\citealt{labbe2006, labbe2010a, labbe2010b, labbe2013, labbe2015}).  Our careful modeling of the light from neighboring sources improves the overall robustness of our final candidate list to source confusion.  Total magnitudes are derived by correcting the fluxes measured in $1.2''$-diameter apertures for the light lying outside this aperture.  The relevant correction factor is estimated on a source-by-source basis based on the spatial profile of each source and the relevant PSF-correction kernel. Average PSF corrections for each band are listed in Table \ref{tab:depths}.

Photometry on the {\it Spitzer}/IRAC observations is more involved due to the much lower resolution ${\rm FWHM} = 1\farcs7$ compared to the ground-based data (${\rm FWHM} = 0\farcs7$). The lower resolution results in source blending where light from foreground sources contaminates measurements of the sources of interest. Photometry of the IRAC bands was therefore performed with \textsc{mophongo}, adopting $1\farcs8$ apertures. Similarly to the optical bands, IRAC fluxes were corrected to total for missing light outside the aperture using the model profile for the individual sources. The procedure for IRAC photometry employed here is very similar to those of other studies (e.g., \citealt{galametz2013, guo2013, skelton2014,  stefanon2017a, nayyeri2017}). 

Following \citet{stefanon2017c}, the uncertainties associated to the flux densities were estimated from the standard deviation of the flux density measurements in $1\farcs2$-diameter empty apertures, multiplied by the corresponding aperture correction.

\section{Sample selection}
\label{sect:selection}

We require sources to be detected at $>5\sigma$ significance in the $J$, $H$, $K_\mathrm{s}$, $[3.6]$, and $[4.5]$ images after coadding their S/N's in quadrature and in those bands with a positive flux density estimate, and we limit our selection to sources brighter than $H\sim 25.8$ mag.  The combined UltraVISTA and IRAC detection and S/N requirements exclude spurious sources due to noise, detector artifacts, and diffraction features.

We identified candidate $z\sim8$ and $z\sim9$  LBGs using a combination of Lyman-break criteria and photometric redshift selections. While photometric redshifts are a great tool in a number of cases, their quality is a direct consequence of the adopted set of template models. It is not uncommon, for instance, when running photometric redshift codes to obtain solutions at $z\gtrsim6$ represented by red, dusty SEDs. Given our current limited knowledge on the physical properties of high redshift galaxies, the existence of such objects, although unlikely, is still possible. However, their red colors would make the assessment of their nature very difficult with the available data, being unable to effectively exclude (more likely) low redshift solutions. The LBG cuts we applied are strict enough to exclude sources with red, power-law like SEDs, therefore aiming at selecting the most robust sample of star-forming galaxies consistent with at most a small amount of dust attenuation. Furthermore, because the process we applied to measure flux densities heavily  relies on \textsc{mophongo}, it would have required an unfeasible amount of time running it on 24 bands for the full set of sources detected on the $\chi^2$ image ($\sim1$ million sources). For these reasons, we started from a sub-sample selected with Lyman break cuts, and consolidated the selection applying a photometric redshift analysis. The full procedure is detailed below.

We construct a preliminary catalog of candidate $z\sim8$ and $z\sim9$ galaxies using those sources that show an apparent Lyman break due to absorption of UV photons by neutral hydrogen in the IGM blue-ward of the redshifted Ly$\alpha$ line.  At $z > 7.1$, the break results in a significantly lower $Y$-band flux density for candidates, while at $z > 8.7$ it reduces the $J$-band flux densities. Because of this we applied two distinct criteria to select either $z\sim8$ or $z\sim9$ candidte LBGs. Specifically, for the $z\sim8$ sample we applied the following criterion:
\begin{equation}
\label{eq:LBY}
 Y-(J+H)/2 > 0.75
\end{equation}
while for the $z\sim9$ sample we required that:
\begin{equation}
\label{eq:LBJ}
 J-H > 0.8
\end{equation}
In case of a non-detection, the $Y$ or $J$-band flux in these relations was replaced by the equivalent $1\sigma$ upper limit.

\begin{figure}
\hspace{-0.5cm}\includegraphics[width=9.2cm]{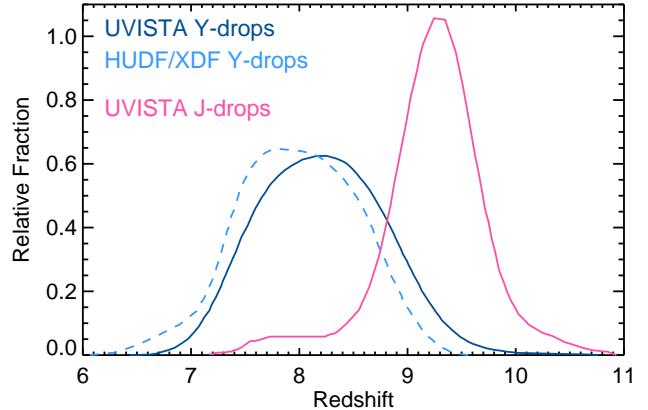}
\caption{Expected redshift distributions (normalized to unit area) for $Y$ and $J$ dropouts recovered from our Monte Carlo simulation (Section \ref{sect:LF}) for the UltraVISTA sample (blue and magenta solid curves, respectively). For comparison, we also show the expected redshift distributions for the $Y$ dropout LBGs samples from \citet{bouwens2015}. The two $z\sim8$ distributions largely overlap, supporting the combination of the UltraVISTA with \citet{bouwens2015} LFs.  \label{fig:zdist}}
\end{figure}

These cuts do not exclusively select $z > 7$ galaxies, but also accept some dust-reddened low redshift galaxies. However, such sources would show a very red continuum and red colors red-ward of the $J-$band or $H-$bands. Therefore, to reject this class of galaxies we also imposed to each one of the sample selected with Equations \ref{eq:LBY} and \ref{eq:LBJ} the requirement of a blue continuuum redward of the break:

\begin{equation}
\label{eq:flat}
 ( H-K < 0.7) ~~ \land ~~  ((K-[3.6] < 1.75) ~ \lor ~ (H-[3.6] < 1.75))
\end{equation}
where $\land$ denotes the logical \textsc{AND} operator, and $\lor$ denotes the logical \textsc{OR} operator. These limits are valuable for excluding a small number of very red sources from our selection.  Nevertheless, it is worth emphasizing that our final sample of $z>7$ bright galaxies shows little dependence on the specific limits chosen here. This initial selection resulted in 2234 candidates (out of $\sim8\times10^5$ detected sources): 2015 $Y$ dropouts, and 183 $J$ dropouts.

We further cleaned our sample from low-redshift sources and Galactic stars by imposing $\chi^2_{opt} < 4$. The $\chi_{opt} ^2$ is defined as $\chi_{opt}^2 = \Sigma_{i} \textrm{SGN}(f_{i}) (f_{i}/\sigma_{i})^2$ \citep{bouwens2011b}, where $f_{i}$ is the flux in any optical band $i$ with uncertainty $\sigma_i$, and SGN($f_{i}$) is $+1$ if $f_{i}>0$ and $-1$ if $f_{i}<0$. The $\chi_{opt}^2$ is calculated in both $1\farcs2$-diameter apertures and in the scaled elliptical apertures.  $\chi^2_{opt}$ is effective in excluding $z=1-3$ low-redshift star-forming galaxies where the Lyman break color selection is satisfied by strong line emission contributing to one of the broad bands (e.g., \citealt{vanderwel2011, atek2011}). We also constructed full depth pseudo $r$-, $i$- and $z$-band mosaics, combining the relevant observations from the CFHTLS, HSC and SSC data sets and excluded sources with a $2\sigma$ detection in either individual ground-based imaging bands or in one of the three full depth optical mosaics, as potentially corresponding to lower-redshift contaminants. This step left 901 candidates LBGs in our sample (791 $Y$ dropouts and 110 $J$ dropouts).

Subsequently, we determined the redshift probability distribution $P(z)$. For this we used the \textsc{EAzY} program (\citealt{brammer2008}), which fits non-negative linear combination of galaxy spectral templates to the observed spectral energy distribution (SED), assuming a flat prior on redshifts. We complemented the standard  \textsc{EAzY\_v1.0} template set with templates extracted from the Binary Population and Spectral Synthesis code (BPASS - \citealt{eldridge2017}) v1.1 for sub-solar metallicity ($Z=0.2Z_\odot$), which include nebular emission from \textsc{cloudy}. Specifically, we adopted templates with equivalent widths EW(H$\alpha$)$\sim1000-5000$\,\AA\ as these extreme EW reproduce the observed $[3.6]-[4.5]$ colors for many spectroscopically confirmed $z\sim7-9$ galaxies (\citealt{ono2012, finkelstein2013, oesch2015a, roberts-borsani2016, zitrin2015, stark2016}). Driven by current observational results  (e.g., \citealt{roberts-borsani2016, oesch2015a, zitrin2015}), we blanketed the Ly$\alpha$ line from those templates with EW(Ly$\alpha$)$\gtrsim40$\AA. Finally, we added templates of 2\,Gyr-old, passively evolving systems from \citet{bruzual2003}, with \citet{calzetti2000} extinction in the range $A_\mathrm{V}=0-8$\, mag to test the robustness of our selected candidates against being lower-redshift interlopers highly attenuated by dust. We imposed an additional constraint, that the integrated probability beyond $z=6$ to be $>50\%$. The use of a redshift likelihood distribution $P(z)$ is very effective in rejecting faint low-redshift galaxies with a strong Balmer/4000\AA\ break and fairly blue colors redward of the break. After this step, the sample resulted composed of 49 candidates (44 $Y$ dropouts and 5 $J$ dropouts).

In Figure \ref{fig:zdist} we present the expected redshift distribution of the $Y-$ (i.e., $z\sim8$) and $J-$ ($z\sim9$) dropout selections obtained from our Monte Carlo simulations described in Section \ref{sect:LF}. The $Y-$dropout selection over UltraVISTA peaks at $z\sim8.2$, but the wings of the $z_\mathrm{phot}$ extend into $z\sim9$. On the other side, the distribution of $z_\mathrm{phot}$ from the $J-$dropout selection presents a wing at lower redshifts, reaching $z\sim7.5$ introduced by the lack of continuity in the coverage of wavelengths between the $Y$ and $J$ bands from the atmospheric absorption.

All the 49 candidates showed compact morphologies. However, the relatively low S/N and coarser spatial resolution of the ground-based data make the distinction between a point source (indicative of a low-mass star nature) and an extended object challenging. Therefore, to further exclude contamination by the coolest low-mass stars we used \textsc{EAzY} to fit all candidates with stellar templates from the SpecX prism library (\citealt{burgasser2014}) and exclude any which are significantly better fit ($\Delta \chi^2 > 1$) by stellar SED models.  The approach we utilized is identical to the SED-fitting approach recently employed by \citet{bouwens2015} for excluding low-mass stars from the CANDELS fields. Through this step we excluded 30 sources as likely brown-dwarf candidates. The sample surviving this selection included 17 candidate $Y$-dropout LBGs and 2 candidates $J$ dropouts.

The IRAC flux densities are particularly crucial for our work, because of the dependence of the $[3.6]-[4.5]$ color on redshift, and because for $z\gtrsim8$ the $3.6\mu$m and $4.5\mu$m bands probe the rest-frame optical red-ward of the Balmer break, thus providing information of the age and stellar mass of the sources. For these reasons, we visually inspected the image stamps containing the original IRAC science frame subtracted of the model sources (hereafter \textit{residual images}). Residual images showed generally clean subtractions, with the exception of two sources (UVISTA-Y7 and UVISTA-Y9). Because the photometric redshifts for these two sources obtained after excluding the IRAC bands still indicated a $z\sim8$ solution, we opted for including the two sources when estimating the luminosity function (see Sect. \ref{sect:LF}), but we excluded them from physical parameter considerations as likely suffering from systematics (Sect. \ref{sect:colors}, \ref{sect:emlines} and \ref{sect:pops}).

Finally, we excluded one $Y$-dropout source which, even though satisfied all the previous criteria, showed a $2.2\sigma$ detection on the image built stacking all the optical data.

When considered together, our selection criteria resulted in very low expected contamination rates.  The nominal contamination rate just summing over the redshift likelihood distribution for the $z\sim8$ sample is $\sim5$\%, based on the assumption our SED templates span the range of colors for the low-z interlopers.  This percentage should just be considered indicative; it does not account for $z<6$ sources scattering into our selection due to the impact of noise.  We will conduct such a quantification in Sect. \ref{sect:contamination}.

In addition to minimizing the impact of contamination in our $z\sim8$ selection, the present selection criteria also likely exclude some bona-fide $z\sim8$ galaxies and thus introduce some incompleteness into our $z\gtrsim8$ samples.  We cope with this incompleteness using selection volume simulations in Sect. \ref{sect:LF}.

\section{Results}

The above selection criteria resulted in a total of 18  LBGs candidates over the UltraVISTA field. Specifically, we identified 16 $Y-$band dropouts (likely $z\sim8$ candidate LBGs) and 2 $J-$band dropouts (likely $z\sim9$ candidate LBGs). These candidates span a range of $H \sim 24.0-26.0$\,mag and constitute the most luminous $z\sim8$ galaxy candidates known to date, $0.5-1.0$ mag brighter than the galaxies recently confirmed through spectroscopy (\citealt{oesch2015a, zitrin2015, roberts-borsani2016}).  

\citet{stefanon2017c} already presented five of them: three $Y$-band dropouts (namely UVISTA-Y1, UVISTA-Y5 and UVISTA-Y6) and the two $J-$band dropouts (UVISTA-J1 and UVISTA-J2), that we had followed-up with \textit{HST}/WFC3 imaging in the F098W, F125W and F160W bands. That analysis further supported the conclusion that the three $Y-$band dropouts are $z\gtrsim8$ LBGs, and showed that the two $J-$band dropout candidates were low-redshift interlopers. In the next sections we present the full sample from which those five sources were extracted. For completeness, we also re-examined the three sources analyzed in \citet{stefanon2017c} (UVISTA-Y1, UVISTA-Y5 and UVISTA-Y6), excluding the flux density measurements in the \textit{HST}/WFC3 bands, and conclude that they are probable $z\gtrsim8$ candidates. We refer the reader to \citet{stefanon2017c} for full details on their analysis including the \textit{HST} flux densities. Nonetheless, high-resolution imaging from \textit{HST} is key in ascertaining the nature of these sources, as we discuss in the next section.

\subsection{High-resolution imaging from HST}
\label{sect:dash}

\begin{figure}
\includegraphics[width=8.6cm]{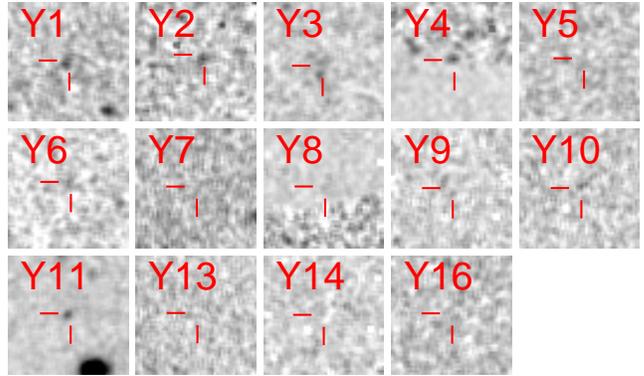}
\caption{Image stamps ($5\farcs0$ side) of those sources with coverage in the WFC3/F160W DASH mosaic (\citealt{momcheva2016, mowla2018}), centered at the nominal location of each object. To improve contrast, each cutout has been smoothed with a $0\farcs2$ Gaussian filter.   \label{fig:dash}}
\end{figure}

\begin{deluxetable}{lclc}
\tablecaption{Candidate $z\sim8$ LBGs with \textit{HST}/WFC3 F160W coverage \label{tab:dash_cov}}
\tablehead{\colhead{ID} & \colhead{PID} & \colhead{PI}  & \colhead{Depth} \\
& &  & \colhead{[mag]}
}
\startdata
UVISTA-Y1  & 14895 & R. Bouwens & $24.7$ \\
UVISTA-Y2  & 14114 & P. van Dokkum & $24.9$ \\
UVISTA-Y3\tablenotemark{a}  & 13868 & D. Kocevski & $26.5$ \\
UVISTA-Y4  & 14114 & P. van Dokkum & $24.9$ \\
UVISTA-Y5  & 14895 & R. Bouwens & $24.9$ \\
UVISTA-Y6  & 14895 & R. Bouwens & $25.0$ \\
UVISTA-Y7  & 14114 & P. van Dokkum & $24.9$ \\
UVISTA-Y8  & 13641 & P. Capak & $25.7$ \\
UVISTA-Y9 & 14114 & P. van Dokkum & $24.8$ \\
UVISTA-Y10 & 14114 & P. van Dokkum & $24.7$ \\
UVISTA-Y11 & 12440 & S. Faber & $26.6$ \\
UVISTA-Y13 & 14114 & P. van Dokkum & $24.9$ \\
UVISTA-Y14 & 14114 & P. van Dokkum & $24.8$\\
UVISTA-Y16 & 14114 & P. van Dokkum & $24.7$ \\
\enddata
\tablecomments{The limiting magnitudes refer to $5\sigma$ fluxes in apertures of $0\farcs6$ diameter corrected to total using the growth curve of point sources.}
\tablenotetext{a}{\textit{HST}/WFC3 imaging suggests this source is potentially multiple. See Sect. \ref{sect:dash} for details.}
\end{deluxetable}

In an effort to further ascertain the nature of the $z\sim8$ LBG sample considered in this work, we also inspected the recent Drift And SHift mosaic (DASH - \citealt{momcheva2016, mowla2018}) at the nominal locations of the selected candidate bright LBGs. This mosaic covers $\sim 0.7$ sq. deg of sky in the WFC3/F160W band  to a depth of $\sim 25.1$\,mag ($0\farcs3$ diameter aperture - \citealt{mowla2018}), and overlaps approximately with three of the four UltraVISTA ultradeep stripes (see Figure \ref{fig:cosmos}). As a bonus, the mosaic also incorporates all the publicly available imaging in the F160W band over the COSMOS/UltraVISTA field. Given the detection of the candidate LBGs was performed on ground-based data (seeing FWHM$\sim0\farcs7$), the finer spacial resolution of \textit{HST}/WFC3 (PSF FWHM$\sim0\farcs2$) is key to test potential multiple components of the candidate bright LBGs, whose blending could artificially increase their measured luminosity (e.g., \citealt{bowler2017, marsan2019}) or systematically affect their redshift estimates.

\begin{figure}
\includegraphics[width=8.7cm]{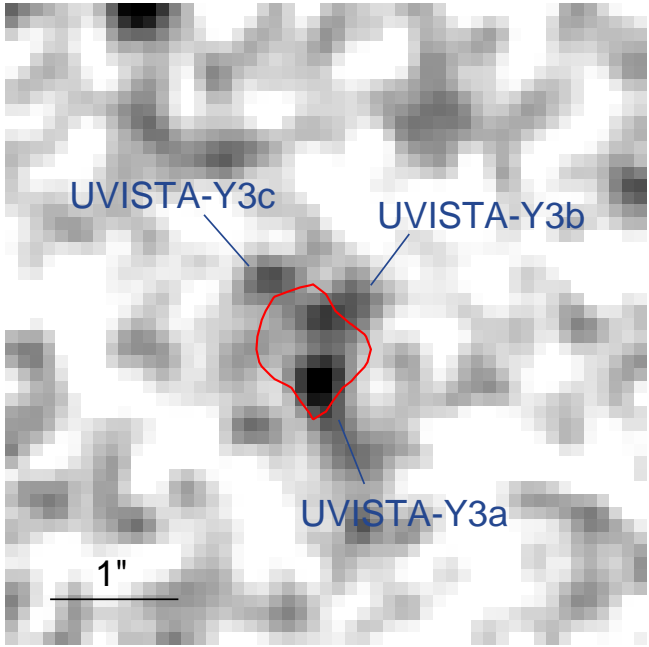}
\caption{Image stamp ($5\farcs0\times5\farcs0$, smoothed with a Gaussian of $0\farcs1$ FWHM) in the WFC3/F160W band extracted from the DASH mosaic (\citealt{momcheva2016, mowla2018}) centered at the position of UVISTA-Y3. Individual components are indicated by the blue labels. The red curve corresponds to the contour of the stacked $J, H$ and $K_\mathrm{s}$ data.  \label{fig:Y3}}
\end{figure}

\begin{figure}
\includegraphics[width=8.7cm]{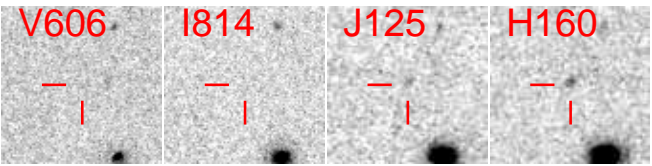}
\caption{Image stamps ($5\farcs0\times5\farcs0$) for UVISTA-Y11 in \textit{HST} bands from the CANDELS program, as labeled at the top-left corner of each panel. No evidence for flux at the nominal location of the source is seen blueward of  the $1.2\mu$m band, consistent with what is seen in the ground-based observations.  \label{fig:Y11}}
\end{figure}

We found that 14 of the 16 candidate LBGs are covered by the DASH mosaic. Their image stamps are presented in Figure \ref{fig:dash}, while in Table \ref{tab:dash_cov} we summarize the coverage details for each source. We note that two sources (UVISTA-Y4 and UVISTA-Y8) fall on or very close to the border between the DASH coverage and deeper WFC3 coverage, resulting in unreliable measurements.

\begin{deluxetable*}{lcccccc}
\tablecaption{Sample of candidate $z\sim8$ LBGs \label{tab:sample}}
\tablehead{\colhead{ID} & \colhead{R.A.} & \colhead{Dec.} & \colhead{$m_\mathrm{H}$\tablenotemark{a}}& \colhead{$Y-J$\tablenotemark{b}}  & \colhead{$[3.6]-[4.5]$\tablenotemark{b}} & \colhead{$z_\mathrm{phot}$\tablenotemark{c}} \\
& \colhead{[J2000]} & \colhead{[J2000]} & \colhead{[mag]} & \colhead{[mag]} & \colhead{[mag]} &
}
\startdata
 UVISTA-Y1\tablenotemark{d,*} & $   09:57:47.900 $  & $   +02:20:43.66 $  & $24.8\pm 0.1 $ & $ > 2.1$ & $ 0.4\pm0.2$ & $8.53^{+0.53}_{-0.62} $ \\
 UVISTA-Y2\tablenotemark{*} & $   10:02:12.558 $  & $   +02:30:45.71 $  & $24.8\pm 0.2 $ & $ > 2.2$ & $ 0.5\pm0.1$ & $8.21^{+0.50}_{-0.49} $ \\
UVISTA-Y3a\tablenotemark{e} & $   10:00:32.324 $  & $   +01:44:30.86 $  & $25.5\pm 0.3 $ & $ > 0.9$ & $ 0.6\pm0.5$ & $8.68^{+0.93}_{-1.21} $ \\
UVISTA-Y3b\tablenotemark{e} & $   10:00:32.317 $  & $   +01:44:31.48 $  & $26.1\pm 0.5 $ & $ > 0.9$ & $ <  0.8$\tablenotemark{f} & $8.90^{+1.24}_{-1.18} $ \\
UVISTA-Y3c\tablenotemark{e} & $   10:00:32.350 $  & $   +01:44:31.73 $  & $26.0\pm 0.5 $ & $ >-0.5$ & $ 0.7\pm0.5$ & $9.29^{+1.58}_{-2.10} $ \\
 UVISTA-Y4\tablenotemark{*} & $   10:00:58.485 $  & $   +01:49:55.96 $  & $24.9\pm 0.2 $  & $ 1.0\pm 0.4 $  & $ 0.1\pm 0.2 $  & $7.42^{+0.19}_{-0.20} $ \\
 UVISTA-Y5\tablenotemark{d,*} & $   10:00:31.886 $  & $   +01:57:50.23 $  & $24.9\pm 0.2 $ & $ > 1.3$ & $ 0.8\pm0.3$ & $8.60^{+0.58}_{-0.65} $ \\
 UVISTA-Y6\tablenotemark{d} & $   10:00:12.506 $  & $   +02:03:00.50 $  & $25.3\pm 0.3 $ & $ > 1.5$ & $ 0.3\pm0.4$ & $8.32^{+0.66}_{-0.92} $ \\
 UVISTA-Y7\tablenotemark{*} & $   09:59:02.566 $  & $   +02:38:06.05 $  & $25.5\pm 0.4 $ & $ > 1.3$ & $ \cdots$\tablenotemark{$\dagger$} & $8.47^{+0.72}_{-0.73} $ \\
 UVISTA-Y8\tablenotemark{*} & $   10:00:47.544 $  & $   +02:34:04.84 $  & $25.4\pm 0.3 $ & $ > 1.4$ & $ 1.0\pm0.8$ & $8.34^{+0.60}_{-0.58} $ \\
UVISTA-Y9 & $   09:59:09.621 $  & $   +02:45:09.68 $  & $25.4\pm 0.3 $ & $ 0.8\pm0.7 $ & $\cdots$\tablenotemark{$\dagger$} & $7.69^{+0.99}_{-0.71} $ \\
UVISTA-Y10\tablenotemark{*} & $   10:01:47.495 $  & $   +02:10:15.37 $  & $25.3\pm 0.3 $ & $ > 1.6$ & $ 0.9\pm0.7$ & $8.25^{+0.61}_{-0.60} $ \\
UVISTA-Y11\tablenotemark{*} & $   10:00:19.607 $  & $   +02:14:13.15 $  & $25.2\pm 0.3 $ & $ > 1.4$ & $ 0.8\pm0.4$ & $8.64^{+0.66}_{-0.72} $ \\
UVISTA-Y12\tablenotemark{*} & $   10:00:15.975 $  & $   +02:43:32.96 $  & $25.6\pm 0.4 $ & $ > 1.2$ & $ 0.2\pm0.8$\tablenotemark{f} & $8.70^{+0.61}_{-0.74} $ \\
UVISTA-Y13 & $   09:58:45.561 $  & $   +01:53:41.79 $  & $25.8\pm 0.4 $ & $ > 1.1$ & $ 0.8\pm0.7$ & $8.54^{+0.79}_{-1.18} $ \\
UVISTA-Y14 & $   10:00:12.568 $  & $   +01:54:28.50 $  & $25.6\pm 0.4 $ & $ > 1.1$ & $ 0.1\pm0.6$ & $7.55^{+1.71}_{-2.68} $ \\
UVISTA-Y15 & $   09:57:35.795 $  & $   +02:11:57.81 $  & $25.6\pm 0.4 $ & $ 1.1\pm0.9 $ & $ < -0.5$\tablenotemark{f,g} & $7.64^{+1.13}_{-1.13} $ \\
UVISTA-Y16\tablenotemark{*} & $   10:01:56.333 $  & $   +02:34:16.25 $  & $25.3\pm 0.3 $  & $ 1.2\pm 0.7 $  & $ 0.6\pm 0.4 $  & $7.90^{+0.74}_{-0.57} $ \\
\enddata
\tablecomments{Measurements for the ground-based bands are $1\farcs2$ aperture flux densities after removing neighbouring sources with \textsc{mophongo} and corrected to total using the PSF and luminosity profile information; measurements for \textit{Spitzer/}IRAC bands are based on $1\farcs8$ aperture flux densities from \textsc{mophongo} corrected to total using the PSF and luminosity profile information. We refer the reader to Tables \ref{tab:flx1}, \ref{tab:flx2} and \ref{tab:flx3} in Appendix \ref{app:flx} for  a complete and more detailed listing of the flux density measurements for all objects in our sample.}
\tablenotetext{a}{$H$-band magnitude and associated $1\sigma$ uncertainty estimated from the UltraVISTA DR3 mosaic.}
\tablenotetext{b}{Upper/lower limits to be intended as $1\sigma$. In computing these colors, we replaced negative fluxes with their corresponding $1\sigma$ uncertainty. See Tables \ref{tab:flx1}, \ref{tab:flx2} and \ref{tab:flx3} for a complete listing of flux densities in all bands.}
\tablenotetext{c}{Photometric redshift and 68\% confidence interval of the best-fitting template from \textsc{EAzY}. }
\tablenotetext{d}{These sources were already presented in \citet{stefanon2017c}. We propose them here again for completeness, noting that their associated parameters in the present work were computed excluding the information from the \textit{HST} bands. We refer the reader to \citet{stefanon2017c} for a more complete analysis.}
\tablenotetext{e}{These candidate LBGs were initially identified as a single source on the UltraVISTA NIR bands. Successive analysis including COSMOS/DASH suggests these are three distinct objects.  The corresponding observables when a single object is assumed are: R.A.= 10:00:32.322; Dec=1:44:31.26, $m_H=25.0\pm0.1$\,mag; $Y-J=1.1\pm0.4$\,mag; $[3.6]-[4.5]=0.3\pm0.1$\,mag and $z_\mathrm{phot}=7.62^{+0.14}_{-0.28}$} 
\tablenotetext{f}{This IRAC color is based on $<2\sigma$ flux density estimate in both bands.}
\tablenotetext{g}{A blue $[3.6]-[4.5]<0$\,mag color might be indicative of a redshift $z\lesssim 7$}
\tablenotetext{\dagger}{After visual inspection, the neighbour-clean image stamps in the IRAC $3.6\mu$m and $4.5\mu$m bands, which constitute the base for our flux density estimates, showed non-negligible residuals that likely systematically affect our estimates. We therefore opted for excluding from our analysis the measurements  involving IRAC for these sources.}
\tablenotetext{*}{These sources have a probability $p(z>7)\ge0.97$, suggesting these may be fairly robust candidates of bright LBGs.}
\end{deluxetable*}

Inspection of the DASH mosaic at the locations of the candidate LBGs discussed in this work resulted in single, isolated sources (for the five sources that are detected at $\gtrsim 4\sigma$) with the important exception of one candidate, UVISTA-Y3. In Figure \ref{fig:Y3} we present an image stamp extracted from DASH with overplotted the contour of the combined $J$, $H$ and $K_\mathrm{s}$ imaging data. A SExtractor run identified three individual objects (with S/N$\sim 4.5, 2.9$ and $2.2$) overlapping with the UltraVISTA footprint of UVISTA-Y3, that we label as UVISTA-Y3a, UVISTA-Y3b and UVISTA-Y3c, for the three components in order of increasing declination, respectively (see Figure \ref{fig:Y3}). The three sources are found to have relative distances of $\sim0\farcs5$. To further ascertain the multiple nature of this source, we run a Monte Carlo simulation, presented in Appendix \ref{app:mc}, consisting in adding to the DASH footprint synthetic sources whose morphologies are similar to those measured for bright $z\gtrsim6$  LBGs. None out of the twenty synthetic sources were split into multiple components by the background noise, increasing our confidence in the multi-component nature of this source. The high resolution provided by the DASH imaging enabled re-running the photometry with \textsc{mophongo} this time adopting the DASH image itself as positional and morphological prior. As we will show in the next section, the single $z\sim8$ source initially identified on the UltraVISTA images resulted in the three objects being at $z\gtrsim8$.

The relatively low S/N significance of the detections of the three components prevents from a comprehensive assessment of their morphology and associated uncertainties.  A  number of works have found that the typical effective radii for LBGs of luminosities similar to those in our sample and at similar redshifts are $r_e\lesssim1$\,kpc (e.g., \citealt{holwerda2015, oesch2016, bowler2017, stefanon2017c, bridge2019}). At $z\sim8$, a separation of $0\farcs5$ correspond to $\sim 2.5$\,kpc, i.e., $\gtrsim2.5\times$ the typical size of bright LBGs at these redshifts. In the spirit of providing further context, we performed an estimate of the sizes for the three sources using the method of \citet{holwerda2015}, and found effective radii of $r_e\sim 0.6, 0.5$ and $0.4$\,kpc, respectively for UVISTA-Y3a, UVISTA-Y3b and UVISTA-Y3c, further supporting our interpretation as three distinct sources. We stress though, that our $r_e$ estimates are only indicative, and should not be considered out of this context.

In our deblending, the flux density of UVISTA-Y3b in the IRAC bands results to be marginal compared to that of the other two components. One possible explanation for this is that while UVISTA-Y3a and UVISTA-Y3c lie at opposite locations with respect to the observed peak of flux density, UVISTA-Y3b is offset from that. In such a configuration, the observed peak of flux density does not coincide with any of the detected sources; instead, it is likely the result of the overlap of the wings of the light profiles of these two components, suggesting the two sources could account for most of the observed flux density. To test this interpretation, we forced the exclusion of either UVISTA-Y3a or UVISTA-Y3c in the deblending process. The result was residual flux at the location of the corresponding component, suggesting these two sources are required to fully account for the observed IRAC flux. However, for a more robust determination of the deblended flux density, higher S/N observations with \textit{HST}/WFC3 and possibly at wavelengths $>3\mu$m (and/or higher spatial resolution) are likely needed. We therefore cannot be sure that our best-fit decomposition is entirely free from systematic errors.

Given that there are 16 $z\sim8$ candidates over the $\sim$0.8 deg$^2$ of the UltraVISTA ultradeep stripes, we would  expect to find only $\sim$1 candidate over the $\sim$190 arcmin$^2$ CANDELS COSMOS field.  Indeed, only one $z\sim8$ candidate from our selection is located over the CANDELS COSMOS field (UVISTA-Y11). In Figure \ref{fig:Y11} we present the image stamps in the $V_{606}$ $I_{814}$, $J_{125}$, $JH_{140}$ and $H_{160}$. The $V_{606}$ mosaic shows a close low-z neighbour just $\sim0\farcs7$ west of UVISTA-Y11, which is not detected in any NIR image (see Figure \ref{fig:Y11} and  Figure \ref{fig:cutouts}). Therefore, we manually included this low-z neighbour when performing the photometry\footnote{Omitting the neighbouring source leads to flux densities systematically over-estimated by $\sim30$\%.}. We do not detect flux at $>1\sigma$ in the $V_{606}$ and $I_{814}$ bands increasing our confidence on its high-z nature.

Finally, we inspected the ACS $I_{814}$-band mosaic of the COSMOS program (\citealt{scoville2007}, $\sim 26.5$\,mag in $0\farcs6$ aperture diameter, $5\sigma$). We found coverage for all sources with the exception of UVISTA-Y1 and UVISTA-Y15. No significant detections exist for any of the sources. We identified a potential low-z galaxy $\sim1\farcs0$ north-west of the nominal location of UVISTA-Y4, which however does not affect our flux density estimates.

The above analysis based on serendipitous deep \textit{HST} coverage for two among the brightest $z\sim8$ LBGs stresses the importance of deep ($\gtrsim1$ orbit) high-resolution multi-band follow-up to further assess the nature of the remarkable LBG candidates identified in the present work.

\subsection{Sample of $z\sim8$ Candidates}
\label{sect:sample}

Figure \ref{fig:cutouts} presents the image stamps of all the candidate $z\sim8$ LBGs. Their positions and main photometry are listed in Table \ref{tab:sample}, while in Appendix \ref{app:flx} we list the flux densities for all objects in all bands. As it is evident from Figure~\ref{fig:cutouts}, the majority of the  sources are clearly detected in the near-infrared, and most of them are also detected in at least one of the {\it Spitzer}/IRAC bands. The brightest source has an $H$-band magnitude of $24.8$\,mag and it is detected at 12$\sigma$, adding in quadrature the detection significance in the $J$, $H$, and $K_\mathrm{s}$ bands.

The observed SEDs of the galaxy candidates are presented in Figure \ref{fig:sed}, along with the \textsc{EAzY} best-fit templates at $z\sim8$ and, to provide contrast, forced fits to model $z<6$ galaxies. The inset in each panel presents the redshift likelihood distribution based on the available optical, infrared and {\it Spitzer}/IRAC photometry. Finally, in Figure \ref{fig:Y3_blended} we show the SED of UVISTA-Y3 when we do not deblend its photometry using the information from the DASH imaging. This SED is best-fitted by a $z\sim8$ solution, consistent with our initial selection.

Four of our 16 $z\sim8$ candidates (or $\sim23$\% of our sample) are located outside the region with the deepest optical observations from the CFHT legacy deep survey.  Because the HSC imaging was not available at the time of the initial sample selection, and given shallower optical observations available in some of the bands to control for contamination (e.g., in the $z$ band), we can ask whether we find an excess of sources over these regions compared to what we would expect from simple Poissonian statistics.  As the outer region contains $\sim$37\% of the area, we find no evidence for a higher surface density of $z\sim8$ candidate galaxies outside those regions providing the best photometric constraints.  This suggests that we can plausibly include the full UltraVISTA search area in quantifying the volume density of bright $z\sim8$ galaxies. Furthermore, the subsequent addition of flux densities from the HSC mosaics did not substantially affect the redshift distributions for these sources, increasing our confidence on their being at $z\gtrsim8$.

Although most of our sample sources are robust $z>8$ candidates, a few have relatively unconstrained redshift probability distributions.  Specifically, 10 sources (when considering UVISTA-Y3 as multiple objects) have a $97\%$ or higher probability of being genuine LBGs at $z>7$, namely UVISTA-Y1, UVISTA-Y2, UVISTA-Y4, UVISTA-Y5, UVISTA-Y7, UVISTA-Y8, UVISTA-Y10, UVISTA-Y11, UVISTA-Y12 and UVISTA-Y16, while the remaining 8 sources, UVISTA-Y3a, UVISTA-Y3b, UVISTA-Y3c, UVISTA-Y6, UVISTA-Y9, UVISTA-Y13, UVISTA-Y14 and UVISTA-Y15, have probabilities $0.6\lesssim p(z>7)\lesssim0.95$.\footnote{When considered as a single object, UVISTA-Y3 has a $p(z>7)=0.99$, suggesting a fairly robust redshift for this source as well.} These tend to have the reddest $J-H$ colors and hence the least certain breaks.  Encouragingly enough, the most uncertain sources are distributed fairly uniformly across the UltraVISTA search area and are not located exclusively over those regions with the poorest observational constraints.

While 14 out of the 16 candidates do not present any significant detection in the $5.8\mu$m and $8.0\mu$m bands, two sources in our $z\sim8$ selection (UVISTA-Y3 and UVISTA-Y13) are formally detected at $>$1$\sigma$ in the combined $5.8\mu$m and $8.0\mu$m observations, with nominal brightnesses of $\sim23-23.5$\, mag at $>5\mu$m.  This could be interpreted as indication of contamination from intrinsically-red $z<3$ galaxies; however,  assuming an intrinsic flux density of $350$\, nJy ($\sim25$\, mag, i.e., an approximately flat $f_\nu$ SED) at $\sim7\mu$m, simple noise statistics  predict 4$\pm$2 sources to be detected at $>1\sigma$.  We therefore conclude that the $>$1$\sigma$ formal detection of two $z\sim8$ candidates in our selection is not a concern.

\begin{figure*}
\includegraphics[width=18cm]{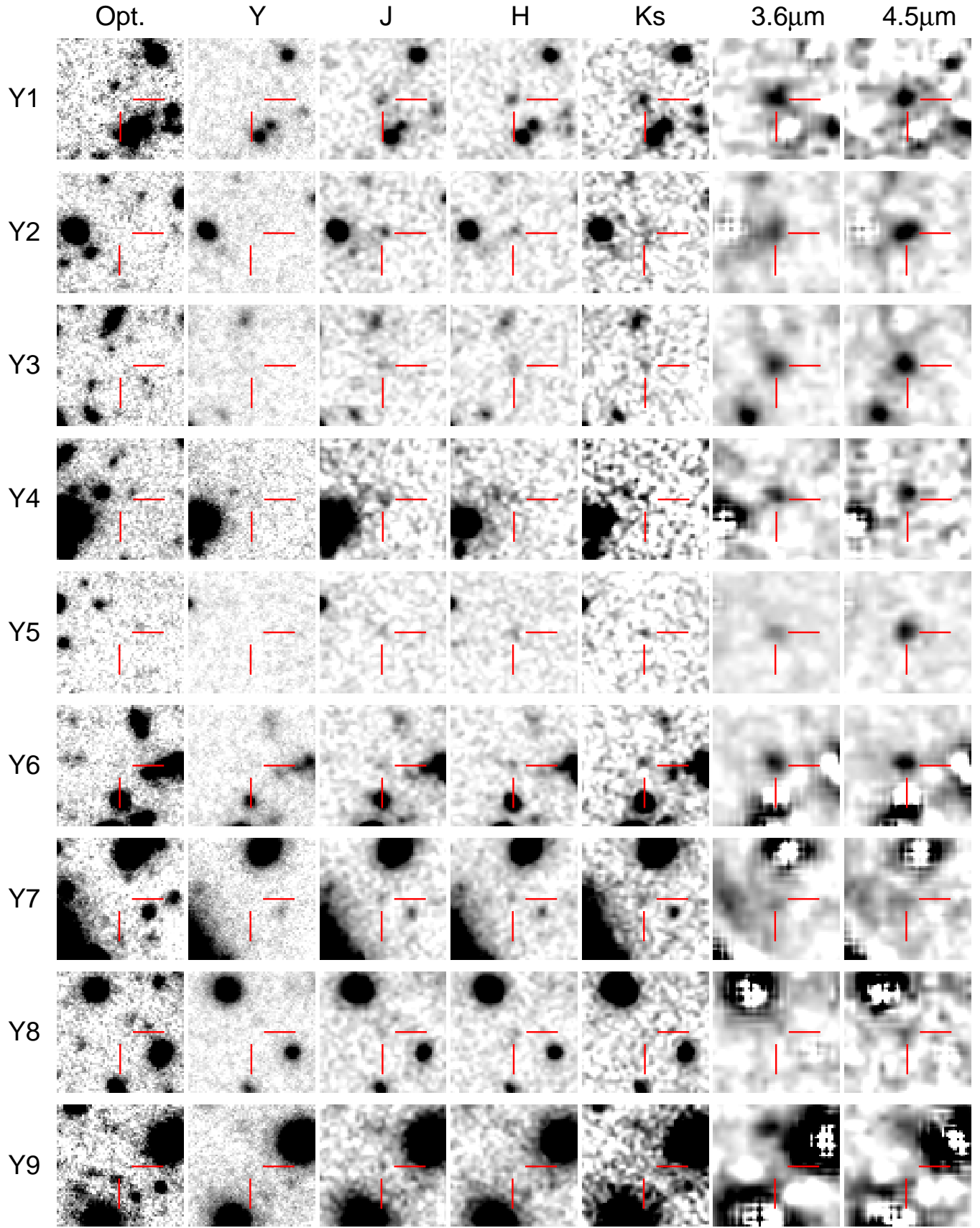}
\caption{Stacked ground-based optical, near-infrared, and \textit{Spitzer}/IRAC image stamps for our bright candidate $z\sim8$ galaxies selected over COSMOS/UltraVISTA. Each image stamp is $10\farcs0\times 10\farcs0$ in size and it is shown in inverted grayscale. Neighbor-subtraction was applied to the IRAC data. \label{fig:cutouts}}
\end{figure*}

\begin{figure*}
\includegraphics[width=18cm]{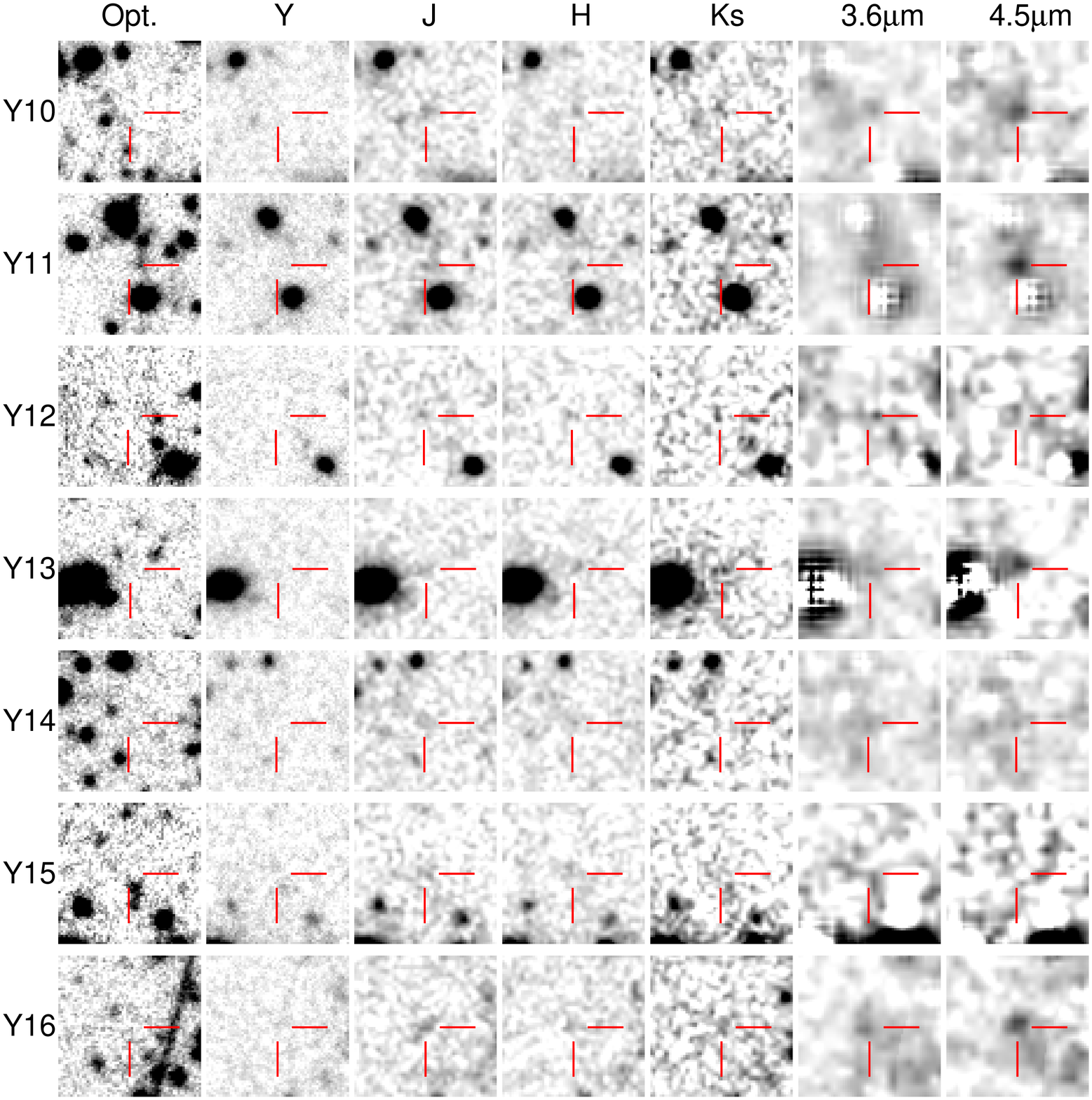}
\figurenum{6}
\caption{-- Continued. \label{fig:cutouts-cont}}
\end{figure*}

\begin{figure*}
\includegraphics[width=18cm]{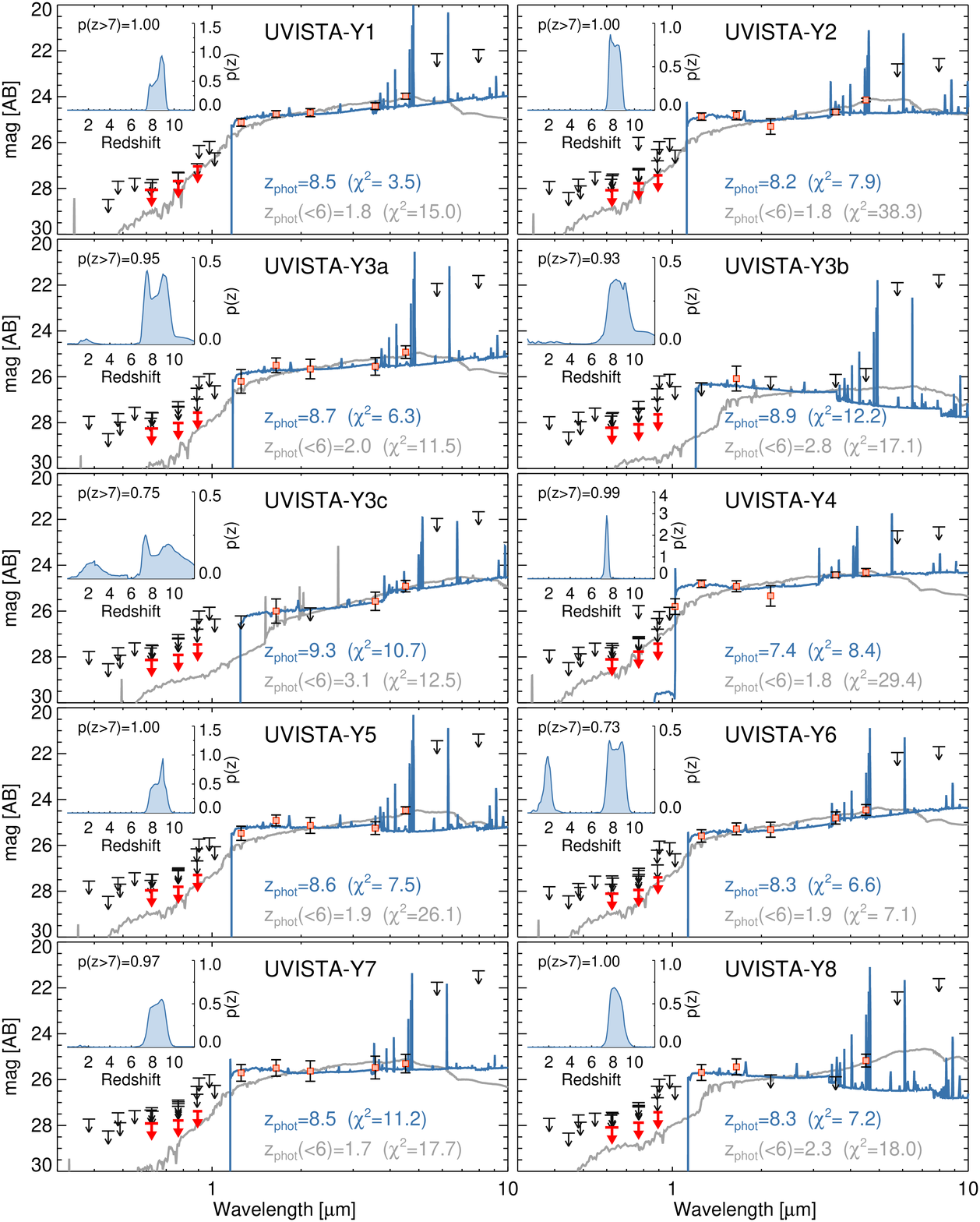}
\caption{Spectral energy distributions from the observed ground-based optical, infrared and \textit{Spitzer}/IRAC photometry (filled red squares with error bars and black $2\sigma$ upperlimits). The red arrows mark $2\sigma$ upper limits in the combined HSC, CFHTLS and SSP $g$, $r$ and $i$ bands. The solid blue curve corresponds to the best-fit SED provided by \textsc{EAzY}, while the grey line shows the best-fit SED when the fit is forced to a $z < 6$ solution. The corresponding redshifts are labeled in matching color, together with the total $\chi^2$. The inset plot on the upper-left corner of each panel presents the redshift probability distributions $P(z)$ for each candidate $z\sim8$ galaxy. \label{fig:sed}}
\end{figure*}

\begin{figure*}
\includegraphics[width=18cm]{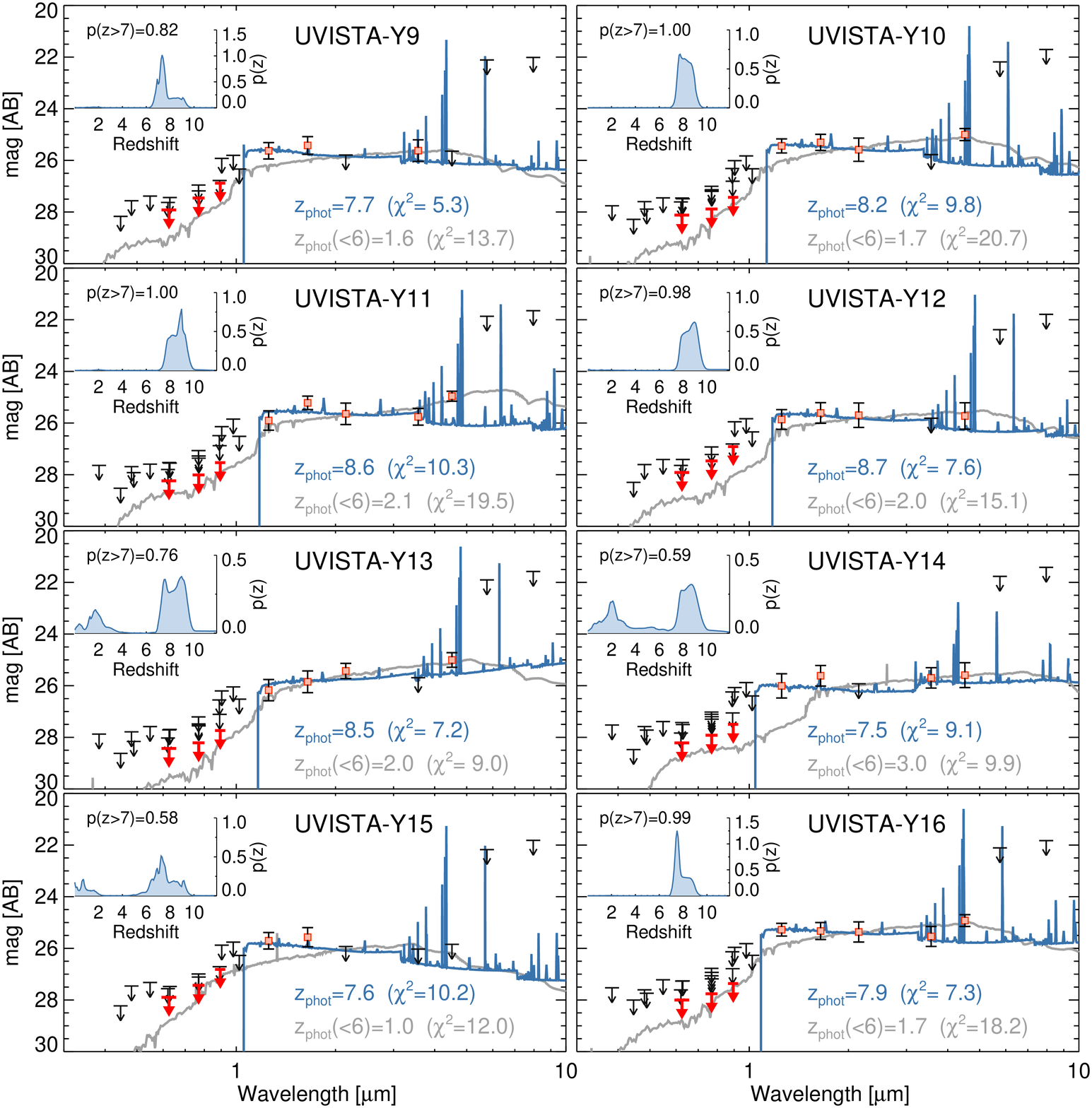}
\figurenum{7}
\caption{-- Continued. \label{fig:sed-cont}}
\end{figure*}

\begin{figure}
\hspace{-0.7cm}\includegraphics[width=9.2cm]{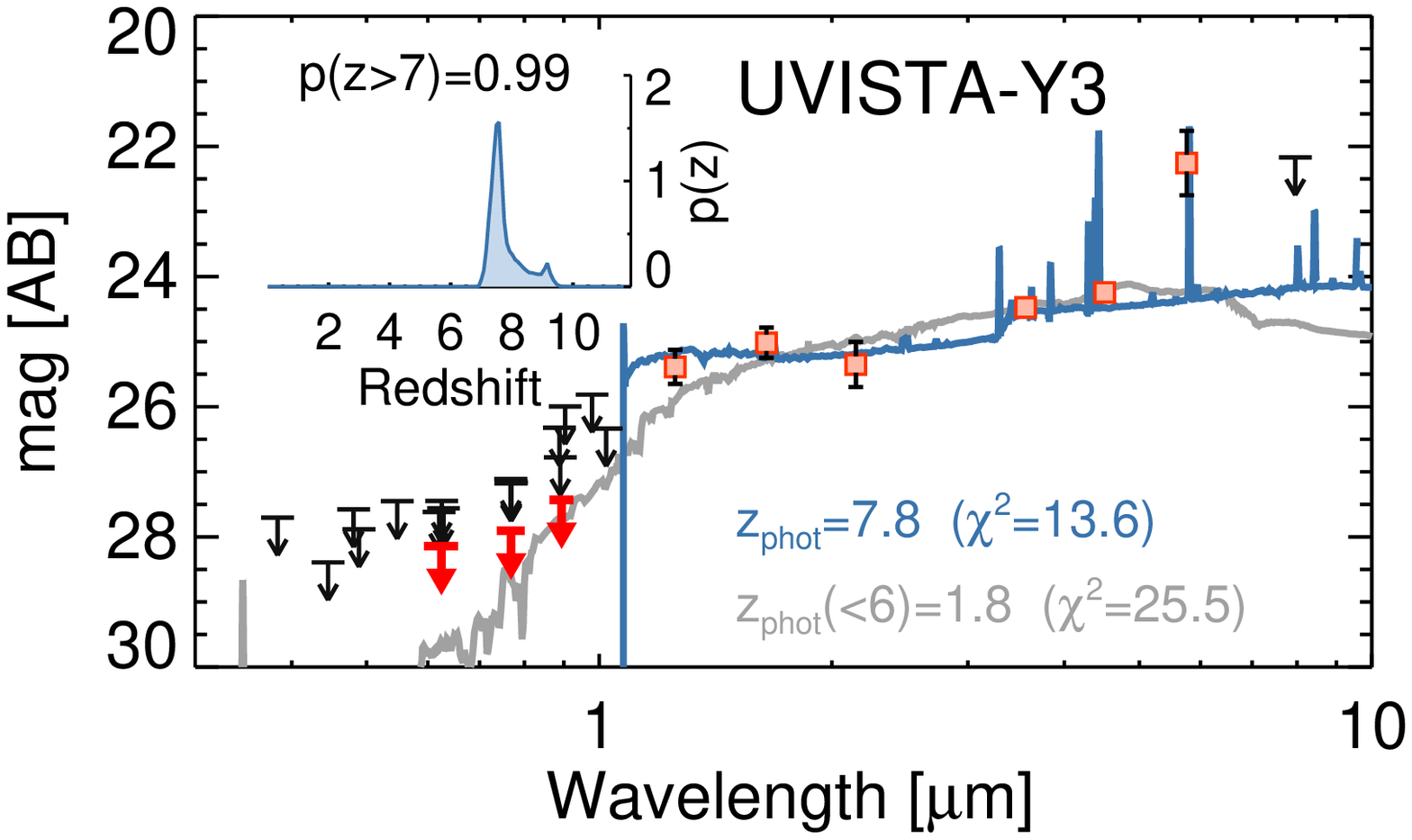}
\caption{Spectral energy distribution of UVISTA-Y3 when we do not deblend its photometry using the higher spatial resolution provided by COSMOS/DASH, but instead consider it as a single source. Same plotting conventions as in Figure \ref{fig:sed}. The solution is still a $z\sim8$ LBG, consistent with our initial selection. \label{fig:Y3_blended}}
\end{figure}

\subsection{Sample of $z\sim9$ Candidates}

The selection criteria expressed by Eq. \ref{eq:LBJ} and Eq. \ref{eq:flat} are designed to select $z\gtrsim9$ LBG candidates. Indeed our initial analysis identified two exceptionally bright ($m_H\sim22.5$\,mag) $J$-dropouts (UVISTA-J1 and UVISTA-J2). However, followup analysis including our \textit{HST}/WFC3 data and presented in \citet{stefanon2017c} revealed that these two sources are likely $z\sim2$ interlopers. For this reason, we omit them from the present sample and refer the reader to \citet{stefanon2017c} for full details.\\

\textit{In summary, to facilitate the comparison of our results to both simulations and observations of LBGs at $z\sim8$, in the rest of this work we consider the 16 $Y$-band dropouts as our fiducial sample of galaxies at $z\sim8$;  specifically, we include in the $z\sim8$ sample those $Y$-dropouts with nominal $z_\mathrm{phot}\sim9$ (see Figure \ref{fig:zdist}). However, in Section \ref{sect:LF} we also consider the contribution of those sources with $z_\mathrm{phot}\sim9$ to the $z\sim9$ LF. We refer the reader to our discussion in Section \ref{sect:LF} for full details.}

\subsection{Expected Contamination in our Bright $z\sim8$ Samples}
\label{sect:contamination}

One potentially important source of contamination for our current $z\sim8$ and $z\sim9$ samples occurs through the impact of noise on the photometry of foreground sources in our search fields.  While noise typically only has a minor impact on the apparent redshift of various foreground sources, the rarity of bright $z\sim 8-10$ galaxies makes it possible for the noise to cause some lower-redshift galaxies to resemble high-redshift galaxies similar to those we are trying to select.  This issue tends to be most important for very wide-area surveys where there exist large numbers of sources which could scatter into our input catalog.  

To determine the impact that noise can have on our samples, we started with an input catalog of $z\leq6$ sources (13000 in total) extracted from the CANDELS/3D-HST catalogs (\citealt{skelton2014, momcheva2016}) over the deep regions in the GOODS North and GOODS South fields, and with apparent magnitudes ranging from $H_{160} = 23$ to $26$\,mag. The procedure was replicated 25 times randomly varying the flux densities according to the measured uncertainties to increase the statistical confidence and to simulate the expected number of sources in the 3000 arcmin$^2$ of the UltraVISTA field.

Fitting the photometry of each source to a redshift and the SED template set described in Sect. \ref{sect:selection}, we derived an SED model for each source in the catalog based on the available photometry and the \textsc{EAzY} SED templates.  We then used that to estimate the equivalent flux for each source in the ground-based imaging bands available over UltraVISTA and perturbed those model fluxes according to the measured noise over the shallow and deep regions over UltraVISTA and according to the depth available over SPLASH, SEDS, and SMUVS.  Finally, we reselected sources using the same selection criteria as we applied to the actual observations.  In perturbing the fluxes of individual sources, we considered both Gaussian and non-Gaussian noise (the latter of which we implemented by increasing the size of noise perturbations by a factor of $\sim$1.3).

Our simulations suggested a very low contamination fraction for our $z\sim8$ samples.  Over the ultradeep stripes where 95\% of the sources in our $z\sim8$ sample were found, these simulations predicted just one $z<6$ contaminant for the entire $\sim$0.8 sq. deg. area, equivalent to a contamination fraction of 5\% for our $z\sim8$ samples.  The typical $H$-band magnitude of the expected contaminants ranged from H$\sim$25 to $25.5$\,mag.

\subsection{Possible Lensing Magnification}
\label{sect:lensing}

A number of recent works has shown that gravitational lensing from foreground galaxies could have a particularly significant effect in enhancing the surface density of bright $z\geq 6$ galaxies (e.g., \citealt{wyithe2011, barone-nugent2015, mason2015, fialkov2015}).  This is especially true for the brightest sources due to the intrinsic rarity and the large path length available for lensing by foreground sources. It has thus become increasingly common to look for possible evidence of lensing amplification in samples of $z\sim6-10$ LBGs (e.g., \citealt{oesch2014, bowler2014, bowler2015, zitrin2015, bouwens2016, roberts-borsani2016, bernard2016, ono2018, morishita2018}).

Even though the fraction of lensed sources among bright samples does not seem to be particularly high (\citealt{bowler2014, bowler2015}), we explicitly considered whether individual sources in our bright $z\sim8$ galaxy compilation showed evidence for being gravitational lensed.  For convenience, we used the \citet{muzzin2013a} catalogs providing stellar mass estimates for all sources over the UltraVISTA area we have searched. These catalogs use the diverse multi-wavelength data over Ultra-VISTA, including {\it GALEX} near and far ultraviolet, {\it HST} optical, near-infrared, {\it Spitzer}/IRAC, and ground-based observations, to provide flux measurements of a wide wavelength range and then use these flux measurements to estimate the redshifts and stellar masses. We also verified that the values obtained did not differ substantially ($\lesssim15\%$) from those obtained adopting the stellar mass estimates of \citet{laigle2016}.

As in \citet{roberts-borsani2016}, we model the foreground objects as singular isothermal spheres (SIS) to assess their influence on the $z\sim8$ galaxy luminosities, and we use the measured half-light radius (\citealt{leauthaud2007}) and inferred stellar mass to derive a velocity dispersion estimates for individual galaxies in these samples.  For cases where size measurements were not available from HST $I_{814}$-band imaging over the COSMOS field, we estimated the half-light radius relying on the mean relation derived by \citet{vanderwel2014}.  Of the 16 $z\sim8$  in our primary sample, only four appear likely to have their flux boosted ($>$0.1 mag) by lensing amplification. 

One of the main advantages of the SIS model is the availability of analytic expressions for the main observables (e.g., magnification, shear, convergence) at the expense of a simplified (spherically symmetric) gravitational potential. For all of our candidate LBGs with the exception of Y6, the lenses have compact, quasi-spheroidal morphology (minor-to-major axis ratio $b/a\gtrsim0.9$) supporting the adoption of a SIS model. For Y6 instead, of three lensing sources, only one has a spheroidal morphology, while the remaining two have elongated shapes ($b/a\sim0.5$), with a position angle of the LBG relative to the main axes of the two ellipses of $\sim29.6$\,degrees and $\sim 4.5$\,degrees, respectively. 

More realistic magnification factors could be obtained for Y6 assuming a singular isothermal ellipsoid model (SIE - e.g., \citealt{kormann1994,kochanek2004}) for the two elongated lensing galaxies. In particular, if the major axis of the ellipsoid is oriented towards the high redshift source, the magnification from a SIE model could be sensibly higher than the magnification from a SIS model. For the two elliptical lenses, the magnifications from the SIE model are $4.5\%$ and $8.4\%$ higher than the corresponding estimates from the SIS model, corresponding to $\sim 0.08$ and $\sim0.04$\,mag difference. Given the small contribution to the magnification estimates, and because  the increase in magnification relative to SIS  are just a fraction of the systematic uncertainties from the stellar mass estimates of the lensing sources ($\sim15\%$), in this work we adopt magnification factors from the SIS model for all lenses.

In the following, we present in more detail our estimates of lensing magnification for the four sources:\\

\begin{figure}
\hspace{-0.5cm}\includegraphics[width=9.2cm]{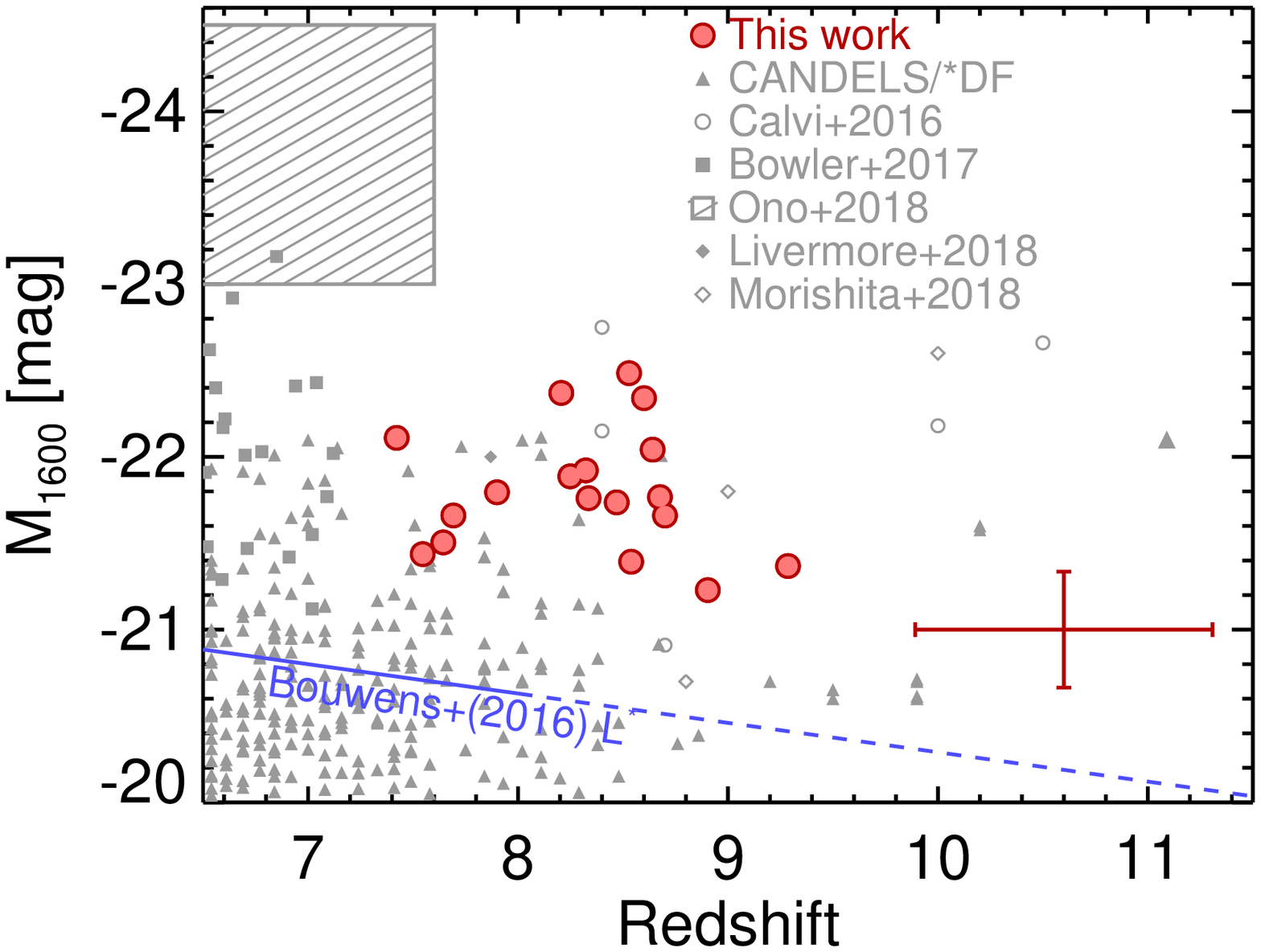}
\caption{Inferred UV luminosities and redshifts for the present sample of bright $z\sim8$ LBGs (filled red circles). The red cross at the bottom-right corner corresponds to the median uncertainties for our sample. For comparison, we also present galaxies from the high-redshift samples compiled by \citet[CANDELS/*DF]{bouwens2015, bouwens2016}, \citet{calvi2016}, \citet{bowler2017}, \citet{ono2018}, \citet{livermore2018} and \citet{morishita2018}. The blue line marks the evolution of the characteristic magnitude of the UV LF of \citet{bouwens2015} up to $z = 8$ (solid blue line) and its extrapolation to $z\sim11$ (dashed blue line). Our candidate LBGs lie at the high-luminosity end of all candidate $z \sim 8$ star-forming galaxies discovered to date, being at least $\sim0.5-1.0$\,mag brighter than the typical bright star-forming galaxy identified over the CANDELS fields.\label{fig:zMUV}}
\end{figure}

\noindent UVISTA-Y6:  This source is estimated to be amplified by $\sim$1.4$\times$, $\sim$1.16$\times$ and $\sim$1.14$\times$ from a $10^{10.7}$ $M_\odot$, $z=1.76$ galaxy (10:00:12.51, 02:02:57.3), $10^{10.6}$ $M_{\odot}$, $z=1.6$ galaxy (10:00:12.15, 02:02:59.6) and a $10^{10.3}$ $M_{\odot}$, $z=1.65$ galaxy (10:00:12.18, 02:03:00.7), respectively, that lie within $4\farcs9$, $3\farcs2$ and $5\farcs4$ of this source.  Their velocity dispersions are estimated to be 259 km/s, 225 km/s, and 206 km/s, respectively.\\\vspace{-0.2cm}

\noindent UVISTA-Y8: This source is estimated to be amplified by 1.39$\times$ from a $10^{10.8}$ $M_{\odot}$ (264 km/s), $z=1.33$ galaxy (10:00:47.68, 02:34:08.4) that lies within $4\farcs1$ of this source.\\\vspace{-0.2cm}

\noindent UVISTA-Y9: This source is estimated to be amplified by 1.37$\times$ and 1.43$\times$ by a $10^{11.0}$ $M_\odot$ (265 km/s), $z=0.91$ galaxy (09:59:09.35, 02:45:11.8) and $10^{11.0}$ $M_\odot$ (268 km/s), $z=0.93$ galaxy, respectively, that lie within $5\farcs0$ and $4\farcs6$ of the source.\\\vspace{-0.2cm}

\noindent UVISTA-Y13: This source is estimated to be amplified by 1.6$\times$ by a $10^{11.15}$ $M_\odot$ (330 km/s), $z=1.63$ galaxy (09:58:45.83,01:53:40.6) that lies within $4\farcs2$ of the source.\\\vspace{-0.2cm}

We discuss the potential impact of lensing on our inferred value for the characteristic magnitude of the UV luminosity function, $M^*$, at the end of Sect. \ref{sect:LF_shape}.

\begin{figure}
\hspace{-0.5cm}\includegraphics[width=9.2cm]{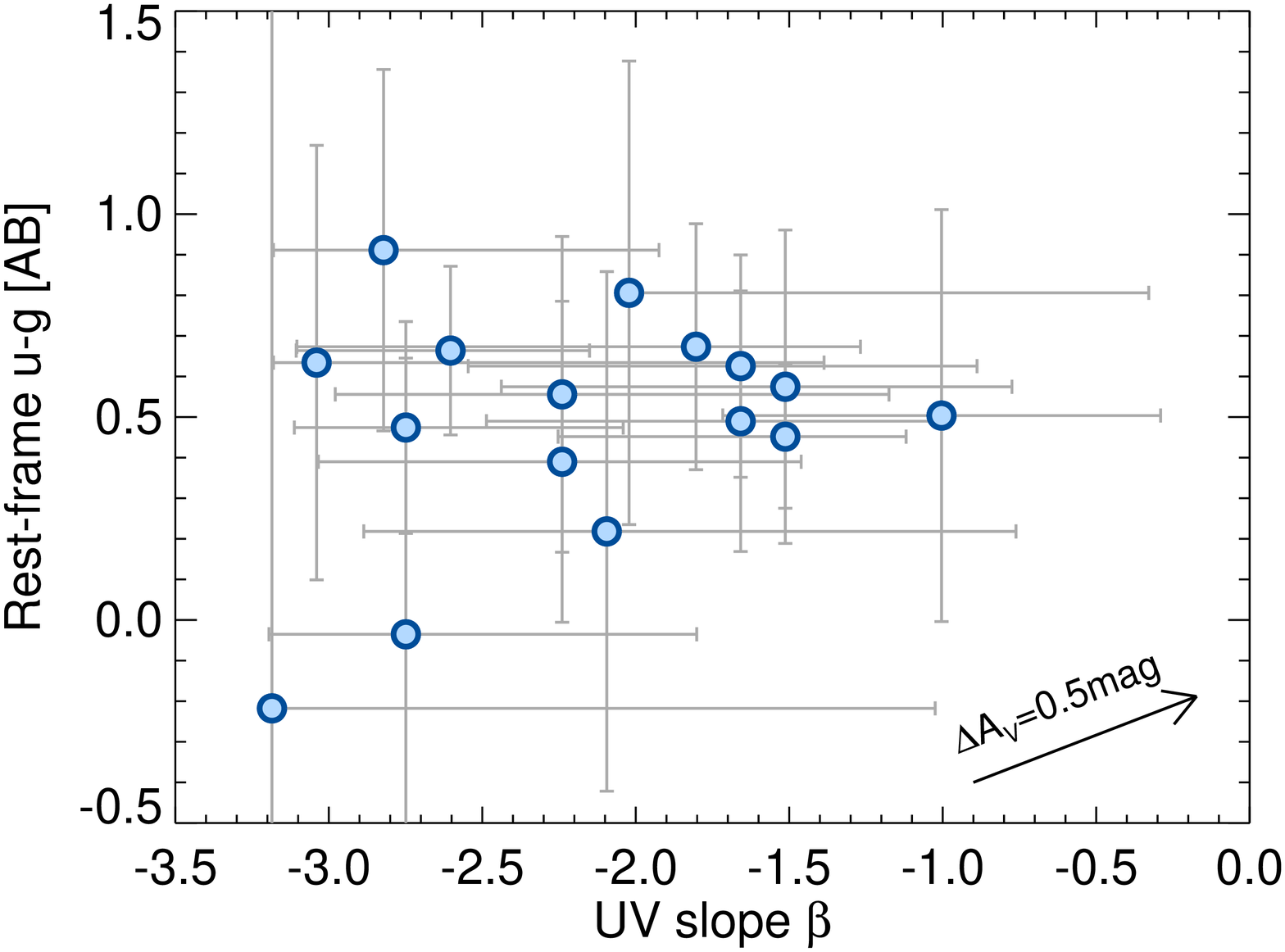}
\caption{Distribution of UV-continuum slopes   and rest-frame $u-g$ colors for the bright $z\sim8$ sample. The vector at the bottom-right corner shows the impact of adding a \citet{calzetti2000} extinction of $A_V=0.5$\,mag. The scatter of points likely reflects a mixture of intrinsic variation and measurement uncertainties. There is no apparent correlation between $\beta$  and rest-frame $u-g$ as might be expected if dust were primarily responsible for the variation in both colors. \label{fig:rest_colors}}
\end{figure}

\section{Discussion}
\subsection{Bright candidate LBGs at $z\sim8$}
In Figure \ref{fig:zMUV} we present our sample of candidate $z\sim8$ LBGs in the redshift-$M_\mathrm{UV}$ plane. For context, we also show recent samples of bright LBGs at similar redshifts from \citet{bouwens2015, bouwens2016}, \citet{calvi2016}, \citet{bowler2017}, \citet{ono2018}, \citet{livermore2018} and \citet{morishita2018}. Our sample of luminous galaxies is among the most luminous galaxies identified at these redshifts, and $\sim0.5-1$\,mag brighter than typical samples selected from CANDELS.

\subsection{Rest-frame Colors of Bright $z\sim8$ Galaxies}
\label{sect:colors}

In this section we present our measurements of two among the most fundamental observables that the deep near-IR and IRAC observations allow us to investigate, i.e.  the spectral slope of the $UV$-continuum light and the rest-frame $u-g$ color.

The spectral slope of the $UV$-continuum light is typically parameterized using the so-called $UV$-continuum slope $\beta$ (where $\beta$ is defined such that $f_{\lambda}\propto \lambda^{\beta}$, \citealt{meurer1999}).  A common way of deriving the $UV$-continuum slope is by considering power-law fits to all photometric constraints in the $UV$ continuum (\citealt{bouwens2012, castellano2012}). Here we take a slightly different approach. First we derive $\beta$'s for a grid of redshifted \citet[hereafter BC03]{bruzual2003} stellar population models with an age of 10 Myr and a range of visual attenuation $A_V=0-2$\, mag. Then for each individual galaxy we fit the predicted $J$, $H$ and $K_\mathrm{s}-$band fluxes to the observations. Uncertainties are derived by randomly scattering the observed fluxes and photometric redshifts by their errors and refitting. This procedure allows us to make full use of the near-IR data and to naturally take into account redshift uncertainties and the Lyman-break entering the $J-$band at $z>8.5$. We caution that, for a small fraction of sources with $z>8.5$,  $\beta$'s derived in this way could still be affected by the Ly$\alpha$ emission line shifting into the $J$-band. We note, however, that observed Ly$\alpha$ equivalent widths of bright $z\sim8-9$ galaxies are modest, $10-30$\,\AA\  (\citealt{roberts-borsani2016, oesch2015a, zitrin2015}). As an exercise, we also computed  the UV slopes by directly fitting the power law to the flux densities in those bands whose effective wavelength was redder than the redshifted 1300\AA\  of  each object (typically  $J, H$ and $K_\mathrm{s}$). These new estimates ($\beta_\mathrm{phot}$) resulted in values essentially equal to those from the method we initially applied (median $\beta_\mathrm{phot}-\beta_{BC03}\sim0.1$), although with large scatter for $\sim30\%$ of the sources ($\Delta\beta\gtrsim 1$).  Nonetheless, the large associated uncertainties make the two measurements consistent with each other. However, we believe that the UV slope measurements recovered  with the initial method are more robust as they better model the effects of redshift on the observed flux density of each source.

\begin{figure*}
\includegraphics[width=18cm]{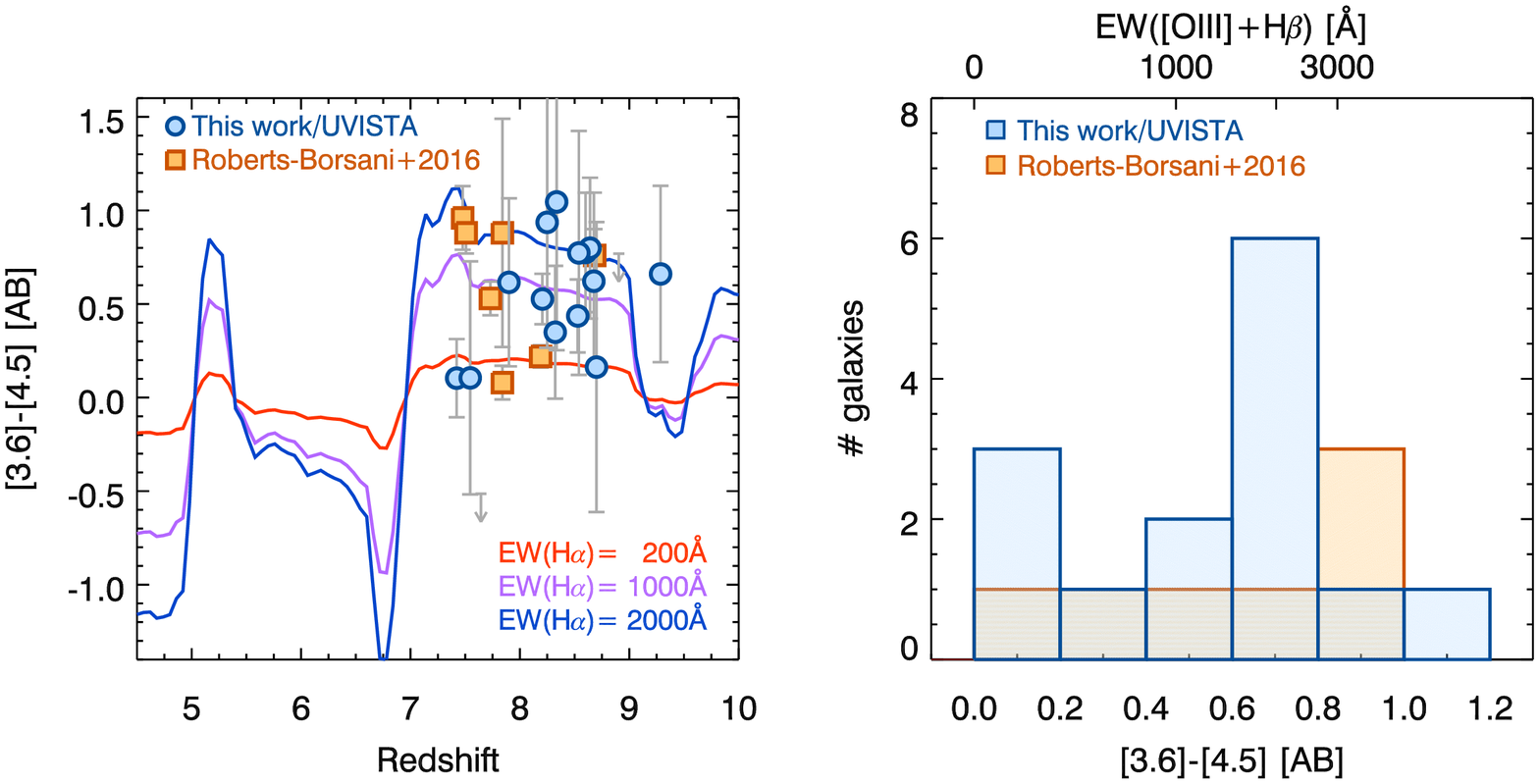}
\caption{\textit{(left)} Observed $[3.6]-[4.5]$ colors vs. photometric redshift for our $z\sim8$ sample (blue circles) and those from  \citet[yellow squares]{roberts-borsani2016}. The predicted dependence of the $[3.6]-[4.5]$ color on redshift is also shown for H$\alpha$ EWs of $200$\,\AA\ (red), 1000\,\AA\ (purple), and 2000\,\AA\ (blue). \textit{(right)} Number of sources in our $z\sim8$ sample (blue histogram) and that of \citet[yellow histogram]{roberts-borsani2016} with a given $[3.6]-[4.5]$ color. The median $[3.6]-[4.5]$ color is $0.62$\, mag. On the upper horizontal axis, we present the EW([\ion{O}{3}+H$\beta$]) corresponding to a given $[3.6]-[4.5]$ color, assuming an intrisic stellar continuum color of 0\,mag. \label{fig:EW}}
\end{figure*}

Figure~\ref{fig:rest_colors} shows the distribution of $UV$ slopes $\beta$ and rest-frame $u-g$ colors for the bright $z\sim8$ sample.  The $z\sim8$ galaxies span a substantial range in $UV$ spectral slope and color. The large uncertainties however, suggest that the observed scatter is likely the combination of intrinsic variation and measurement uncertainties. The average slope of the UV continuum is $\beta=-2.2\pm0.6$, is bluer but still consistent with  the $UV-$continuum slopes found for bright $-22 < M_\mathrm{UV} < 21$ galaxies at $z=6$ ($\beta=-1.55\pm0.17$) and $z\sim7$ ($\beta=-1.75\pm0.18$) by \citet{bouwens2014} and suggests a continuing trend towards bluer $\beta$'s at higher redshifts.

Recently, \citet{oesch2013} analyzed the rest-frame UV and optical properties of a sample of $z\sim4$ LBGs selected from the GOODS-N/S and HUDF fields and spanning a wide range of UV luminosities, $M_\mathrm{UV}\sim -18$ to $\sim -22$ AB. Their $J_{125}-[4.5]$ color (corresponding to approximately rest-frame $u-z$ at $z\sim4$) shows a correlation with the UV slope $\beta$ (see e.g., their Figure 4), likely driven by dust extinction. The uniform scatter observed at $z\sim8$ then may suggest rapidly evolving physical mechanisms responsible for the production of dust during the $\sim 800$\,Myr between the two epochs.

\subsection{Constraints on the EWs of the [OIII]+H$\beta$ lines}
\label{sect:emlines}

Recent observational studies have found that the $[3.6]-[4.5]$ color of galaxies depends dramatically on the redshift of the source (\citealt{shim2011, stark2013, labbe2013, smit2014, smit2015, bowler2014, faisst2016, harikane2018}),  with some sources showing extreme colors (\citealt{ono2012, finkelstein2013, laporte2014, laporte2015, roberts-borsani2016, faisst2016}). A number of works have suggested that these extreme colors are likely due to very strong line emission (\citealt{labbe2013, smit2014}) whereas the intrinsic color of the stellar continua in the absence of emission lines is $[3.6]-[4.5]\sim0$\,mag (\citealt{labbe2013, smit2014, rasappu2016}).

\begin{deluxetable*}{lcccccccc}
\tablecaption{Main physical parameters for the sample of candidate $z\sim8$ LBGs \label{tab:params}}
\tablehead{\colhead{ID} & \colhead{$M_\mathrm{UV}$} & \colhead{UV slope $\beta$} & \colhead{$u-g$} & \colhead{$\log(M_\star)$}& \colhead{$\log(\mathrm{SFR})$}  & \colhead{$\log(\mathrm{sSFR})$} & \colhead{$\log(\mathrm{age})$} & \colhead{$A_\mathrm{V}$} \\
& \colhead{[mag]} &   & \colhead{[mag]} & \colhead{[$M_\odot$]} & \colhead{[$M_\odot \mathrm{yr}^{-1}$]} & \colhead{[yr$^{-1}$]} & \colhead{[yr]} &\colhead{[mag]}
}
\startdata
   UVISTA-Y1 & $-22.48\pm0.15$ & $-1.5^{+0.4}_{-0.7} $ & $ 0.45\pm0.18$ & $10.0^{+0.9}_{-0.4} $ & $ 1.59^{+ 1.02}_{ -9.55} $ & $ -8.4^{+1.8}_{ -9.8} $ & $ 7.30^{+1.42}_{-0.61} $ & $  0.9^{+0.0}_{-0.9} $ \\
   UVISTA-Y2 & $-22.37\pm0.20$ & $-2.6^{+0.5}_{-0.5} $ & $ 0.66\pm0.21$ & $ 9.0^{+0.3}_{-1.2} $ & $ 1.98^{+ 0.65}_{ -7.57} $ & $ -7.0^{+0.8}_{ -8.6} $ & $ 7.00^{+1.80}_{-0.50} $ & $  0.4^{+0.3}_{-0.4} $ \\
  UVISTA-Y3a\tablenotemark{a} & $-21.77\pm0.32$ & $-1.5^{+0.7}_{-0.9} $ & $ 0.57\pm0.39$ & $ 9.8^{+1.3}_{-0.3} $ & $-1.34^{+ 4.06}_{ -7.01} $ & $-11.1^{+4.9}_{ -7.0} $ & $ 8.00^{+0.80}_{-1.50} $ & $  0.0^{+1.1}_{-0.0} $ \\
  UVISTA-Y3b\tablenotemark{a} & $-21.23\pm0.54$ & $-3.2^{+2.2}_{-0.0} $ & $-0.22\pm1.82$ & $ 8.7^{+0.1}_{-0.0} $ & $-0.28^{+ 0.00}_{ -0.07} $ & $ -9.0^{+0.0}_{ -0.0} $ & $ 7.40^{+0.00}_{-0.02} $ & $  0.0^{+0.0}_{-0.0} $ \\
  UVISTA-Y3c\tablenotemark{a} & $-21.37\pm0.53$ & $-2.0^{+1.7}_{-0.0} $ & $ 0.81\pm0.57$ & $10.2^{+1.6}_{-0.5} $ & $ 1.69^{+ 1.86}_{-28.23} $ & $ -8.5^{+2.3}_{-28.1} $ & $ 8.70^{+0.10}_{-2.20} $ & $  0.8^{+1.1}_{-0.8} $ \\
   UVISTA-Y4 & $-22.11\pm0.24$ & $-2.7^{+0.7}_{-0.4} $ & $ 0.47\pm0.26$ & $ 9.9^{+0.5}_{-0.2} $ & $ 1.23^{+ 0.61}_{ -6.97} $ & $ -8.6^{+0.9}_{ -7.0} $ & $ 8.50^{+0.30}_{-1.31} $ & $  0.0^{+0.7}_{-0.0} $ \\
   UVISTA-Y5 & $-22.34\pm0.24$ & $-1.7^{+0.8}_{-0.9} $ & $ 0.63\pm0.27$ & $ 9.0^{+0.4}_{-1.1} $ & $ 1.99^{+ 0.55}_{ -7.68} $ & $ -7.0^{+0.8}_{ -8.6} $ & $ 7.30^{+1.40}_{-0.80} $ & $  0.4^{+0.2}_{-0.4} $ \\
   UVISTA-Y6 & $-21.92\pm0.26$ & $-1.7^{+0.7}_{-0.8} $ & $ 0.49\pm0.32$ & $ 9.7^{+1.1}_{-0.5} $ & $ 1.36^{+ 1.33}_{-12.70} $ & $ -8.4^{+2.2}_{-12.9} $ & $ 7.30^{+1.50}_{-0.80} $ & $  0.9^{+0.3}_{-0.9} $ \\
   UVISTA-Y7 & $-21.74\pm0.36$ & $-2.0^{+0.7}_{-0.5} $ & $ \cdots$\tablenotemark{$\dagger$} & $ \cdots$\tablenotemark{$\dagger$} & $ \cdots$\tablenotemark{$\dagger$} & $ \cdots$\tablenotemark{$\dagger$} & $ \cdots$\tablenotemark{$\dagger$} & $ \cdots$\tablenotemark{$\dagger$} \\
   UVISTA-Y8 & $-21.76\pm0.35$ & $-2.8^{+0.9}_{-0.4} $ & $ 0.91\pm0.45$ & $ 8.3^{+0.1}_{-1.4} $ & $ 1.90^{+ 0.35}_{ -1.41} $ & $ -6.4^{+0.2}_{ -2.6} $ & $ 6.50^{+2.29}_{-0.00} $ & $  0.0^{+0.5}_{-0.0} $ \\
  UVISTA-Y9 & $-21.66\pm0.34$ & $-2.6^{+0.9}_{-0.6} $ & $ \cdots$\tablenotemark{$\dagger$} & $ \cdots$\tablenotemark{$\dagger$} & $ \cdots$\tablenotemark{$\dagger$} & $ \cdots$\tablenotemark{$\dagger$} & $ \cdots$\tablenotemark{$\dagger$} & $ \cdots$\tablenotemark{$\dagger$} \\
  UVISTA-Y10 & $-21.89\pm0.31$ & $-2.2^{+1.1}_{-0.7} $ & $ 0.56\pm0.39$ & $ 8.3^{+0.0}_{-1.4} $ & $ 1.80^{+ 0.47}_{ -4.34} $ & $ -6.5^{+0.3}_{ -5.6} $ & $ 6.70^{+2.10}_{-0.20} $ & $  0.0^{+0.5}_{-0.0} $ \\
  UVISTA-Y11 & $-22.04\pm0.26$ & $-1.8^{+0.5}_{-1.3} $ & $ 0.67\pm0.30$ & $ 8.7^{+0.4}_{-1.2} $ & $ 1.76^{+ 0.60}_{ -7.62} $ & $ -7.0^{+0.8}_{ -8.6} $ & $ 7.30^{+1.44}_{-0.80} $ & $  0.3^{+0.2}_{-0.3} $ \\
  UVISTA-Y12 & $-21.66\pm0.40$ & $-2.1^{+1.3}_{-0.8} $ & $ 0.22\pm0.64$ & $ 9.1^{+0.9}_{-0.4} $ & $ 0.17^{+ 2.22}_{ -2.88} $ & $ -9.0^{+2.8}_{ -3.1} $ & $ 7.40^{+1.30}_{-0.90} $ & $  0.2^{+0.3}_{-0.2} $ \\
  UVISTA-Y13 & $-21.39\pm0.42$ & $-1.0^{+0.7}_{-0.7} $ & $ 0.50\pm0.51$ & $ 9.8^{+1.3}_{-0.3} $ & $ 0.70^{+ 1.82}_{ -9.08} $ & $ -9.1^{+2.8}_{ -9.1} $ & $ 7.50^{+1.28}_{-0.96} $ & $  0.8^{+0.3}_{-0.8} $ \\
  UVISTA-Y14 & $-21.44\pm0.40$ & $-3.0^{+1.7}_{-0.1} $ & $ 0.63\pm0.54$ & $ 9.3^{+1.2}_{-0.4} $ & $ 0.52^{+ 2.03}_{ -9.14} $ & $ -8.8^{+2.6}_{ -9.3} $ & $ 8.20^{+0.63}_{-1.70} $ & $  0.0^{+0.8}_{-0.0} $ \\
  UVISTA-Y15 & $-21.50\pm0.37$ & $-2.7^{+0.9}_{-0.4} $ & $-0.04\pm0.68$ & $ 8.8^{+0.2}_{-0.0} $ & $-0.16^{+ 0.41}_{ -0.02} $ & $ -9.0^{+0.6}_{ -0.0} $ & $ 7.40^{+0.01}_{-0.10} $ & $  0.0^{+0.0}_{-0.0} $ \\
  UVISTA-Y16 & $-21.80\pm0.33$ & $-2.2^{+0.8}_{-0.8} $ & $ 0.39\pm0.40$ & $ 8.6^{+0.3}_{-0.9} $ & $ 1.62^{+ 0.56}_{ -0.98} $ & $ -7.0^{+0.8}_{ -1.8} $ & $ 7.30^{+1.50}_{-0.80} $ & $  0.1^{+0.4}_{-0.1} $ \\
\enddata
\tablenotetext{a}{These three candidate LBGs were originally identified as a single source, successively de-blended using data from the COSMOS/DASH program (see Sect. \ref{sect:dash} and Figure \ref{fig:Y3_blended}). When we do not deblend the source, we obtain $M_\mathrm{UV}=-22.00\pm0.16$\,mag, $\beta=-1.8\pm 0.7$, $u-g=0.58\pm0.16$\,mag, $\log(M_\star/M_\odot)=9.9^{+0.6}_{-0.3}$, $\log(\mathrm{SFR/M_\odot/\mathrm{yr}^{-1}})=1.63^{+ 0.38}_{ -3.77}$, $\log(\mathrm{sSFR/yr}^{-1})=-8.2^{+0.9}_{ -3.8}$, $\log(\mathrm{age/yr})=8.20^{+0.60}_{-1.16}$ and $A_V=0.5^{+0.5}_{-0.5}$\,mag.}
\tablenotetext{\dagger}{After visual inspection, the neighbour-cleaned image stamps in the IRAC $3.6\mu$m and $4.5\mu$m bands showed non-negligible residuals that likely systematically affected our estimates. Photometric redshifts resulted to be robust against the exclusion of the flux densities in these two bands, but stellar population parameters heavily rely on the IRAC colors. Because of the unreliability of the IRAC flux density estimates for these objects, we discard their physical parameters.}
\end{deluxetable*}

\begin{figure}
\hspace{-1cm}\includegraphics[width=9.2cm]{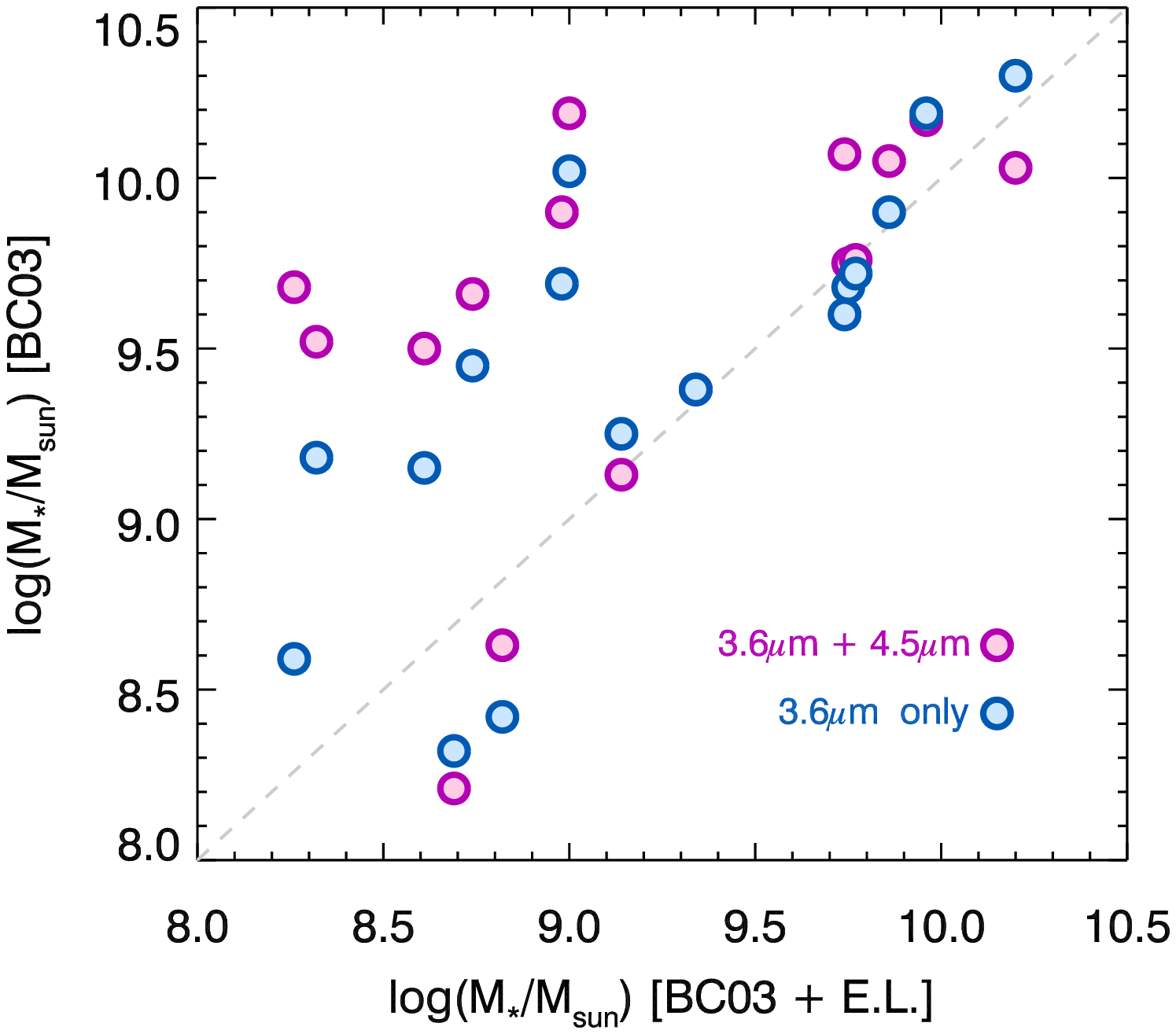}
\caption{The best-fit stellar masses with emission lines included compared to those derived with the standard BC03 models without emission lines. The latter masses (where the models ignore line emisison) are higher by $\sim0.43$ dex on average, consistent with the results of \citet{labbe2013}, with individual galaxies differing by up to 1 dex. One might expect more accurate masses from standard BC03 models if one excludes the $4.5\mu$m band (contaminated by [\ion{O}{3}]+H$\beta$ emission) when performing the fitting, but the estimated stellar masses are still found to be $0.23$~dex higher on average.  This mismatch between the BC03 model fit results (without the emission lines) and the fit results with emission lines included may be due to the contribution of the [\ion{O}{2}] line to the $3.6\mu$m band flux measurements. From the present exercise, we can see how important it is to fully consider nebular emission when estimating stellar population parameters.  \label{fig:mstar}}
\end{figure}

At redshift $z=7.0-9.1$, the  [\ion{O}{3}]+H$\beta$ line emission contributes to the {\it Spitzer}/IRAC 4.5$\mu$m band in galaxies, producing red $[3.6]-[4.5]$ colors. Figure \ref{fig:EW} shows examples of model colors as a function of redshift for lines with very high equivalent width. Using a small sample of $z\sim8$ galaxies selected from the CANDELS survey, \citet{roberts-borsani2016} reported a very red median $[3.6]-[4.5]\sim0.8$\,mag color at bright $H<26$ magnitudes. Using a simple spectral model,  consisting of a flat rest-frame $0.3-0.6\mu$m continuum in  f$_{\nu}$ (i.e., a continuum $[3.6]-[4.5]=0$\,mag or f$_{\lambda}\propto\lambda^{-2}$), with the strongest emission lines ([\ion{O}{2}]$_{3727}$, H$\beta$, [\ion{O}{3}]$_{4959,5007}$, H$\alpha$, [\ion{N}{2}]$_{6548,6583}$, [\ion{S}{2}]$_{6716,6730}$), empirical emission lines ratios from \citet{anders2003} for 0.2 Z$_{\odot}$ metallicity, they inferred a median [\ion{O}{3}]+H$\beta$ EW of $\sim2000$\,\AA. However, the sample of \citet{roberts-borsani2016} was very small, and possibly biased as it was compiled from IRAC-selected $[3.6]-[4.5]>0.5$ galaxies and galaxies with confirmed Ly$\alpha$ emission. So it is unclear if those results were representative of the general bright $z\sim8$ population.

With the UltraVISTA sample and the deep IRAC observations from SPLASH, SEDS, and SMUVS, we have an opportunity to revisit the analysis of \citet{roberts-borsani2016}  with a larger sample. In Figure~\ref{fig:EW}, we present the $[3.6]-[4.5]$ color distribution for bright $z\sim8$ galaxies from both our study and that of \citet{roberts-borsani2016}.  The $[3.6]-[4.5]$ color distribution spans a range of more $1$\, mag, with the UltraVISTA sample showing a median $[3.6]-[4.5]=0.62$\, mag; this color remains unchanged when also combining it with the CANDELS sample.

Adopting the same model of \citet{roberts-borsani2016} (see also \citealt{smit2014}) and supposing that the $3.6\mu$m band receives only a negligible contribution from line emission, a $[3.6]-[4.5]$ color of $\sim0.6$~mag corresponds to an  [\ion{O}{3}]+H$\beta$ EW of $\sim1500$\,\AA. Such a result is consistent with \citet{labbe2013} and \citet{smit2014, smit2015}, and with the recent estimates of \citet{stefanon2019} and \citet{debarros2018} based on samples of $z\sim8$ $L<L^*$ LBGs selected over the GOODS-N/S fields, which benefit from among the deepest IRAC $3.6\mu$m and $4.5\mu$m observations of the GREATS program (PI: I. Labb\'e; \citealt{labbe2018}). 

Under the assumption that the extreme IRAC colors are due to nebular emission, our results combined with those from the literature indicate that strong emission lines might be ubiquitous at these redshifts in galaxies spanning $\sim 3$\,mag range in luminosity. Nevertheless, significant systematic uncertainties remain depending on the assumed continuum shape and line flux ratios. For example, including the full line list of \citet{anders2003}, contribution from the higher order Balmer lines, and assuming a more realistic spectral continuum (e.g., BC03 and scaling emission lines by the flux in hydrogen ionising photons N$_\mathrm{LyC}$), and allowing for \citet{calzetti2000} dust, produces a different $[3.6]-[4.5]$ color versus redshift relation by up to $0.2-0.4$ mag. Also, emission line ratios, in particular [\ion{O}{3}]$_{5007}$, depend strongly on metallicity (e.g., \citealt{inoue2011}). Considering this, we estimate that simple approximations are probably uncertain by factors of $2-3$.

\subsection{Stellar Populations of Bright $z\sim8$ Galaxies}
\label{sect:pops}

\begin{deluxetable*}{lcccccc}
\tablecaption{Observed and rest-frame properties for candidate $z\sim 8$ galaxies identified in the UltraVISTA DR3 observations \label{tab:samplestat}}
\tablehead{\colhead{Quantity} & \colhead{25\%} & \colhead{Median} & \colhead{75\%}  & \colhead{25\% uncertainties} & \colhead{Median uncertainties} & \colhead{75\% uncertainties} 
}
\startdata
   $z_\mathrm{phot}$ & $ 8.05 $  & $ 8.40 $  & $ 8.62 $  & $+0.60/{-0.61} $  & $+0.69/{-0.73} $  & $+0.96/{-1.15} $  \\
        $M_\mathrm{UV}$ [mag]  & $ -22.0 $  & $ -21.8 $  & $ -21.6 $  & $\pm0.3 $  & $\pm0.3 $  & $\pm0.4 $  \\
         UV $\beta$  & $ -2.68 $  & $ -2.17 $  & $ -1.73 $  & $+0.70/{-0.40} $  & $+0.77/{-0.65} $  & $+1.01/{-0.79} $  \\
  $(u-g)_\mathrm{rest}$  [mag] & $  0.42 $  & $  0.53 $  & $  0.65 $  & $\pm0.29 $  & $\pm0.39 $  & $\pm0.55 $  \\
      $\log(M_\star/M_\odot)$  & $   8.71 $  & $   9.07 $  & $   9.76 $  & $+0.32/{-0.24} $  & $+0.46/{-0.44} $  & $+1.14/{-1.15} $  \\
 $M_\star/L_\mathrm{UV} [M_\odot/L_\odot]$  & $   0.005 $  & $   0.010 $  & $   0.044 $  & $+  0.008/{-  0.004} $  & $+  0.015/{-  0.014} $  & $+  0.034/{-  0.098} $  \\
  $M_\star/L_\mathrm{u} [M_\odot/L_\odot]$  & $   0.013 $  & $   0.048 $  & $   0.101 $  & $+  0.026/{-  0.010} $  & $+  0.038/{-  0.056} $  & $+  0.098/{-  0.244} $  \\
  $M_\star/L_\mathrm{g} [M_\odot/L_\odot]$  & $   0.017 $  & $   0.064 $  & $   0.133 $  & $+  0.026/{-  0.007} $  & $+  0.043/{-  0.089} $  & $+  0.078/{-  0.188} $  \\
    $\log(\mathrm{SFR}/M_\star/\mathrm{yr}^{-1})$  & $    0.3 $  & $    1.5 $  & $    1.8 $  & $+0.5/{-2.1} $  & $+0.6/{-7.3} $  & $+1.8/{-9.1} $  \\
$\log(\mathrm{sSFR/yr}^{-1})$  & $   -9.0 $  & $   -8.4 $  & $   -7.0 $  & $+0.7/{-2.9} $  & $+0.9/{-7.8} $  & $+2.5/{-9.2} $  \\
      $\log(\mathrm{age/yr})$  & $   7.30 $  & $   7.35 $  & $   7.75 $  & $+0.47/{-0.35} $  & $+1.35/{-0.80} $  & $+1.50/{-1.13} $  \\
          $A_\mathrm{V}$ [mag] & $   0.00 $  & $   0.15 $  & $   0.60 $  & $+0.22/{-0.00} $  & $+0.32/{-0.15} $  & $+0.56/{-0.60} $  \\
\enddata
\tablecomments{Estimates of $z_\mathrm{phot}$, $M_\mathrm{UV}$ and $L_X$ were obtained from \textsc{EAzY} (see Sect. \ref{sect:selection}); $M_\star$, SFR, sSFR, age and $A_\mathrm{V}$ were measured with \textsc{FAST} (see Sect. \ref{sect:pops}); the UV continuum slope $\beta$ were measured following the procedure described in Sect. \ref{sect:colors}. The last two columns present the first and third quartiles of uncertainties, respectively.}
\end{deluxetable*}

\begin{figure*}
\includegraphics[width=18cm]{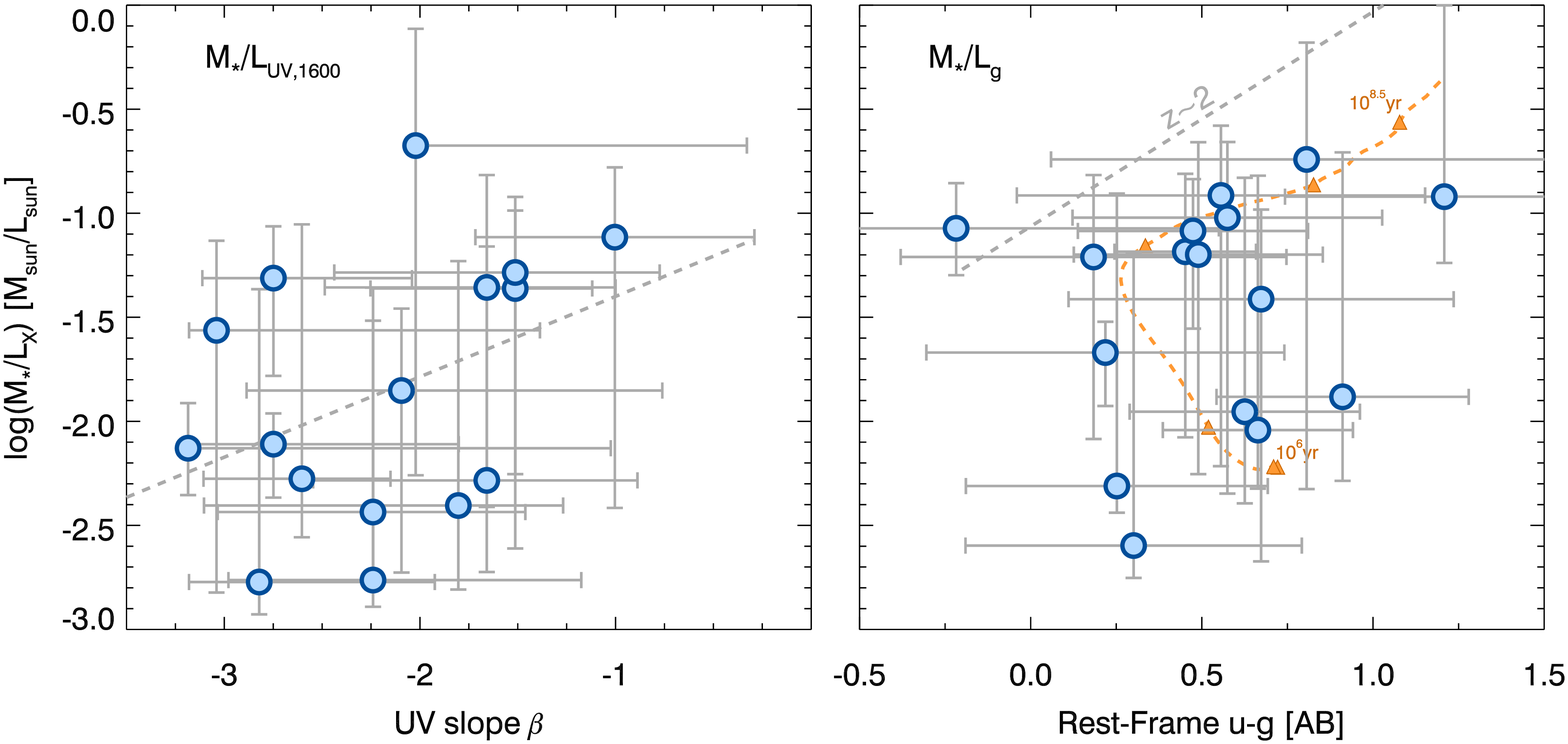}
\caption{\textit{Left panel:} The rest-frame $UV$ continuum slope $\beta$ versus  the best-fit stellar mass-to-UV light ratio. The dashed line marks a tentative linear correlation. \textit{Right panel:} The rest-frame $u-g$ color versus the best-fit stellar mass-to-optical light ratio.  At low to intermediate redshift $z\sim2$ a tight relation exists between rest-frame $u-g$ colors and $M_\star/L_g$ ratios with unity slope, such that redder galaxies exhibit higher $M_\star/L$. The grey dashed line shows the relation derived by \citet{szomoru2013} at $z\sim2$. The orange dashed curve shows the relation for our BC03 models including emission lines. The orange triangles mark the age of the stellar population, starting from $\log(\mathrm{age/yr})=6$ to $\log(\mathrm{age/yr})=8.5$, in steps of $0.5$\,dex. While we find a marginal positive correlation with $UV$ slope for our $z\sim8$ sample, there is no clear relation between $u-g$ colors and $M_\star/L$. Instead, the optically reddest galaxies tend to have the lowest $M_\star/L$ ratios.  This likely reflects the impact of strong emission lines on the $g-$band fluxes.    \label{fig:ML}}
\end{figure*}

In this section we present our estimates of stellar population parameters for the bright $z\sim8$ galaxies. Measurements were performed with the \textsc{FAST} code \citep{kriek2009}, adopting \citet{bruzual2003} models for sub-solar 0.2$Z_{\odot}$ metallicity, a \citet{chabrier2003} IMF, constant star formation, and the \citet{calzetti2000} dust law. As discussed above, gaseous emission lines contribute significantly to the integrated broadband fluxes. Given standard BC03 models do not include nebular emission, line and continuum nebular emission were added following the procedure of  \citet{salmon2015} and assuming line flux ratios relative to $H\beta$ from the models calculated by \citet{inoue2011}. The luminosity in $H\beta$ is taken to be proportional to the luminosity in hydrogen ionising photons N$_\mathrm{LyC}$, assuming ionization-recombination equilibrium (case B). The emission line ratios of \citet{inoue2011} agree well with the empirical compilations of \citet{anders2003}, with observations of the local galaxy I Zw 18 (\citealt{izotov1999}), and the $z=2.3$ galaxy from \citet{erb2010}, in particular for the strongest metal line [\ion{O}{3}]$_{5007}$. In Table~\ref{tab:params} we present the results of our stellar population modeling, specifically the stellar mass, star formation rate, specific star formation rate, age and extinction together with the $UV_\mathrm{1600}$ absolute magnitude, the $UV$ continuum slope $\beta$ and the rest-frame $u-g$ color for each individual candidate bright $z\sim8$ LBG. A summary of the physical properties is presented in Table \ref{tab:samplestat}.

As we already introduced in Sect. \ref{sect:selection}, the neighbour-cleaned IRAC $3.6\mu$m- and $4.5\mu$m-band image sections for two sources (UVISTA-Y7 and UVISTA-Y9) presented residuals that might be systematically affecting our estimates of stellar population parameters (see Figure \ref{fig:cutouts}). We therefore recomputed the redshift likelihood distributions for these two sources after excluding the IRAC flux densities. The photometric redshifts we derived were consistent with the estimates obtained adopting the full set of measurements. However, the stellar population parameters heavily rely on the IRAC colors because at $z\sim8$ these probe the rest-frame optical red-ward of the Balmer break and the emission line properties, both affecting their age and the stellar mass measurements. As a result, the physical parameters for the two sources have not been included in Table \ref{tab:params} or Figures presenting these parameters (i.e., Figures \ref{fig:rest_colors}, \ref{fig:EW}, \ref{fig:mstar} and \ref{fig:ML})

In Sect. \ref{sect:emlines} we showed that our sample is characterized by extreme $[3.6]-[4.5]\sim0.6$\,mag colors, likely the result of strong  [\ion{O}{3}]+H$\beta$ emission entering the $4.5\mu$m band. A number of studies have shown that nebular emission can systematically bias stellar mass estimates (e.g., \citealt{stark2013}). Figure~\ref{fig:mstar} compares the best-fit stellar masses to those derived with the standard BC03 models without emission lines for our sample. Those masses are higher by $\sim0.4$\,dex on average (scatter $\sim 0.6$\,dex), with individual galaxies differing by up to 1 dex. This is consistent with \citet{labbe2013}, who estimate that $z \sim 7-8$ galaxies' average stellar masses decrease by $\sim 0.5$\,dex if the contributions of emission lines to their broadband fluxes are accounted for. However, the discrepancy appears to be related not only to the strong contribution of [\ion{O}{3}]$_{5007}$ to the $4.5\mu$m band.   Indeed, if we refit the galaxies with the standard BC03 models (without emission lines) while omitting the flux in the $4.5\mu$m band, the offset is marginally reduced to $0.23$\,dex (scatter $0.43$\,dex) compared to the BC03 and emission lines fit to all bands.  This residual offset is likely due to the effect of nebular emission (mainly [\ion{O}{2}]$_{3727}$) characteristic of young stellar populations which still substantially contaminates the $3.6\mu$m band. This result stresses once more the importance of accounting for nebular emission in estimating the physical parameters of $z\gtrsim8$ galaxies.

The typical estimated stellar masses for bright sources in our $z\sim8$ selection (see Table \ref{tab:samplestat}) are 10$^{9.1^{+0.5}_{-0.4}}$ $M_{\odot}$, with the SFRs of  $32^{+44}_{-32}M_{\odot}$/year, specific SFR of $4^{+8}_{-4}$ Gyr$^{-1}$, stellar ages of $\sim22^{+69}_{-22}$\,Myr, and low dust content A$_V=0.15^{+0.30}_{-0.15}$\, mag. As evident from Table \ref{tab:samplestat},  individual galaxies shows a broad range in each of these properties, with interquartile masses, ages, and specific star formation rates spanning $\sim 1$ dex.

In Figure~\ref{fig:ML} we compare the rest-frame properties with the best-fit stellar mass-to-light ratios for luminosities in the rest-frame UV$_{1600}$ and rest-frame $g$ band. These quantities are not completely independent, as both are derived from the same photometry, but provide useful insights in how color relates to stellar mass. Overall, the mass-to-light ratios are very low, as expected for very young stellar ages ($<100$\,Myr), but span quite a wide range, between $0.1$ and $0.01 M_{\odot}$/L$_{\odot}$.

We find a positive  although marginal correlation of the $M_\star/L_\mathrm{UV,1600}$ with the $UV$ slope for our $z\sim8$ sample as it could be expected from older and/or dustier stellar populations characterized by redder UV slope (e.g., \citealt{bouwens2014}).

 A number of works have shown that at low redshift there exists a tight relation between rest-frame optical colors and $M_\star/L$ ratios, such that redder galaxies exhibit  higher $M_\star/L$, and that this empirical relation is not sensitive to details of the stellar population modeling (e.g., \citealt{bell2001}). This relation appears to hold even at intermediate redshifts $z\sim2$ (e.g., \citealt{szomoru2013}). Remarkably, in contrast to the situation at low-redshift, redder rest-frame $u-g$ colors of the $z\sim8$ sample do not correspond to higher $M_\star/L$. Instead, the optically reddest galaxies tend to have the lowest $M_\star/L$.  This likely reflects the effect of strong emission lines in the $g-$band. The fact that age and dust have very different effects on the colors of the high redshift galaxies studied here probably also explains the lack of correlation between $\beta$ and $u-g$ in Figure~\ref{fig:rest_colors}.

\subsection{Volume Density of Bright $z\sim8$ and $z\sim9$ Galaxies}
\label{sect:LF}

In this section we present our measurements of the UV LF based on the sample presented in this work. Our main result is the UV LF at $z\sim8$ based on the sample of $Y-$band dropouts (i.e., considering UVISTA-Y3 as three independent sources) presented in Sect. \ref{sect:sample}. However, because some objects have a nominal photometric redshift $z_\mathrm{phot}\sim9$, we also explored the contribution to the  UV LF at $z\sim 9$ from the five sources with $z_\mathrm{phot}\ge8.6$ (namely UVISTA-Y3a, UVISTA-Y3b, UVISTA-Y3c, UVISTA-Y11 and UVISTA-Y12). Because the nominal photometric redshift of UVISTA-Y5 is $z_\mathrm{phot}=8.596$, this object was initially excluded by our redshift selection criterion. Considering the very marginal difference of its photo-z from the selection threshold, we also forced its inclusion into the sample adopted for the estimate of the $z\sim9$ LF, bringing to six the total number of sources used for the $z\sim9$ LF.

\begin{figure}
\hspace{-1cm}\includegraphics[width=9.2cm]{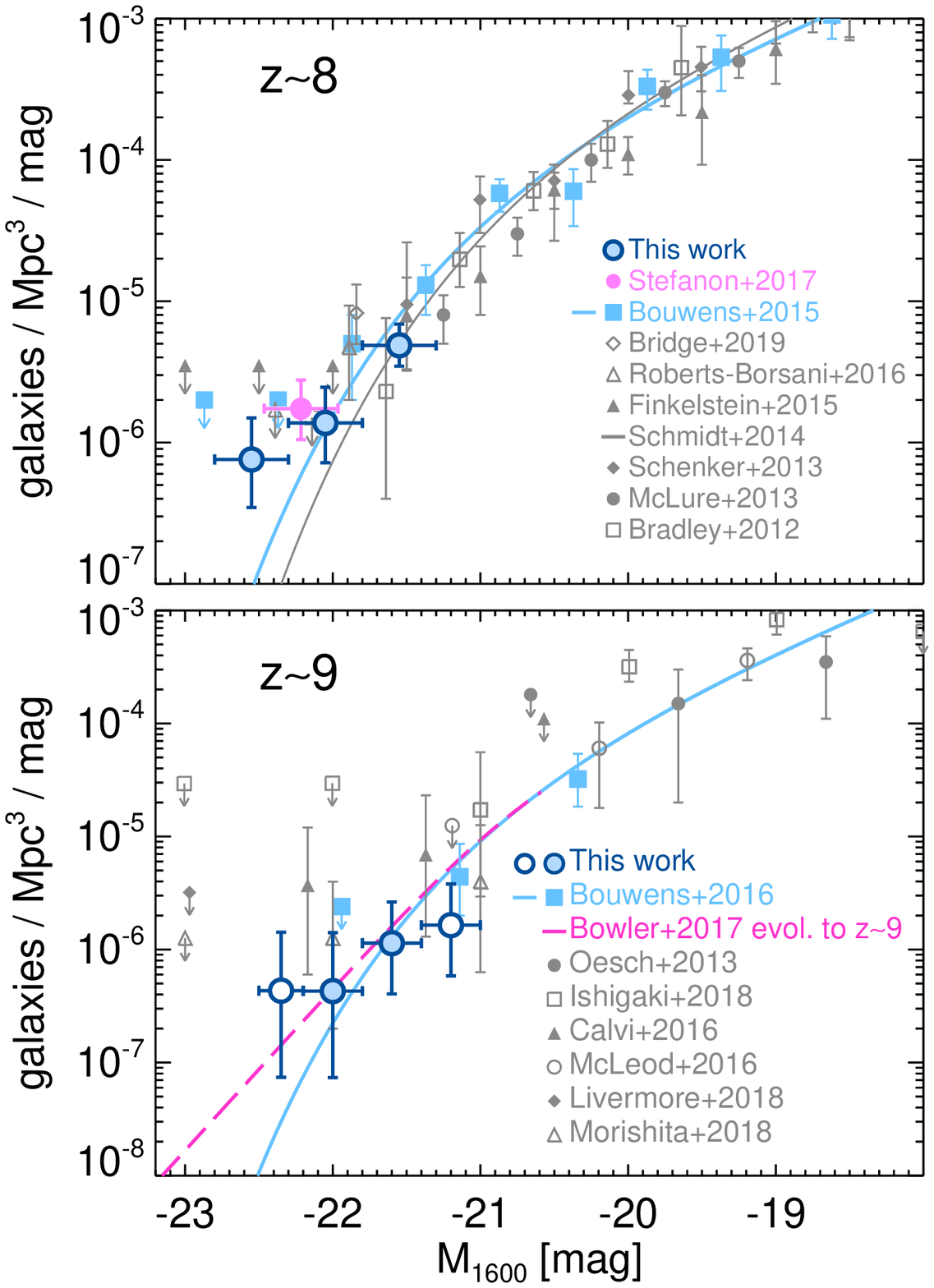}
\caption{\textit{Top panel:} The blue points with errorbars mark our estimates of volume density associated to the sample of candidate luminous $z\sim8$ galaxies considered in this work. For comparison, we also present recent UV LF determinations at $z\sim8$ from empty field studies, as indicated by the legend (we arbitrarily shifted the measurement of \citealt{bridge2019} - by $+0.05$\,mag to improve readability). \textit{Bottom panel:} Here we compare our $z\sim9$ volume density estimate from the five sources with $z_\mathrm{phot}\ge 8.6$ (blue points) to measurements of the UV LF at $z\sim9$. The blue open circle at the bright end corresponds to the measurement obtained forcing UVISTA-Y5 into the $z\sim9$ sample (see Section \ref{sect:LF} for details), while the blue open circle at $M_\mathrm{UV}=-21.2$\,mag marks the volume density measurement for the faintest luminosity bin, where our sample is likely incomplete. The magenta curve presents the bright end of the dual power law from \citet{bowler2017} evolved to $z\sim9$ following \citet{bouwens2016} and whose characteristic density has been adjusted to match that of the Schechter function at the characteristic luminosity. \label{fig:LF}}
\end{figure}

\begin{deluxetable}{cc}
\tablecaption{V$_\mathrm{max}$ determinations of the UV LF \label{tab:LF}}
\tablehead{\colhead{$M_\mathrm{UV}$} & \colhead{$\phi$}  \\
\colhead{[mag]} & \colhead{[$ \times 10^{-6} \mathrm{mag}^{-1} \mathrm{Mpc}^{-3} $]} 
}
\startdata
\multicolumn{2}{c}{$z\sim8$} \\
 $ -22.55$ & $ 0.76^{+0.74}_{-0.41} $ \\
 $ -22.05$ & $ 1.38^{+1.09}_{-0.66} $ \\
 $ -21.55$\tablenotemark{a} & $ 4.87^{+2.01}_{-1.41} $ \\
& \\
\multicolumn{2}{c}{$z\sim9$} \\
 $ -22.35$\tablenotemark{b} & $ 0.43^{+0.99}_{-0.36} $ \\
 $ -22.00$ & $ 0.43^{+0.98}_{-0.36} $ \\
 $ -21.60$ & $ 1.14^{+1.50}_{-0.73} $ \\
 $ -21.20$\tablenotemark{a} & $ 1.64^{+2.16}_{-1.06} $ \\
\enddata
\tablenotetext{a}{This luminosity bin includes sources from the deblending of UVISTA-Y3, which fall below our nominal detection threshold. The sample in this luminosity bin is therefore likely incomplete.}
\tablenotetext{b}{The volume density in this luminosity bin was obtained forcing UVISTA-Y5 into the sample of galaxies at $z_\mathrm{phot}\ge8.6$ (i.e., our $z\sim9$ LBG sample). Its nominal $z_\mathrm{phot}=8.596$ would exclude it from the sample of $z\sim9$ sources when the redshift selection criteria is strictly enforced; however, considering the very small difference with the $z_\mathrm{phot}$ threshold, here we include it for completeness.}
\end{deluxetable}

To infer the volume densities of the galaxies we first estimate the detection completeness and selection function through simulations. Following \citet{bouwens2015}, we generated catalogs of mock sources with realistic sizes and morphologies by randomly selecting images of $z\sim4$ galaxies from the Hubble Ultra Deep Field (\citealt{beckwith2006, illingworth2013}) as templates. The images were scaled to account for the change in angular diameter distance with redshift and for evolution of galaxy sizes at fixed luminosity $\propto(1+z)^{-1}$ (e.g., \citealt{oesch2010, ono2013, holwerda2015, shibuya2015}). The template images are then inserted into the observed images, assigning colors expected for star forming galaxies in the range $6 < z < 11$. The colors were based on a $UV$ continuum slope distribution of $\beta = -1.8 \pm 0.3$ to match the measurements for luminous $6<z<8$ galaxies (\citealt{bouwens2012, bouwens2014, finkelstein2012, rogers2014}). The simulations include the full suite of {\it HST}, ground-based, and {\it Spitzer}/IRAC images. For the ground-based and {\it Spitzer}/IRAC data the mock sources were convolved with appropriate kernels to match the lower resolution PSF. To simulate IRAC colors we assume a continuum flat in $f_{\nu}$ and strong emission lines with fixed rest-frame EW(H$\alpha$+[\ion{N}{2}]+[\ion{S}{2}]) = 300\AA\ and rest-frame EW([\ion{O}{3}]+H$\beta$) = 500\AA\ consistent with the results of \citet{labbe2013, stark2013, smit2014, smit2015} and \citet{rasappu2016}. We included the effect of other nebular lines following the recipe of \citet{anders2003} for sub-solar metallicity.

The same detection and selection criteria as described in Sect. \ref{sect:selection} were then applied to the simulated images to calculate the completeness as a function of recovered magnitude and the selection as a function of magnitude and redshift  (see Figure 8 of \citealt{stefanon2017c} for the selection functions over the UltraVISTA deep and ultradeep stripes).

The total selection volume over our UltraVISTA area for galaxies with $H\sim24.0-24.5$\, mag and $24.5-25.0$ is $5.3\times10^{6}$ Mpc$^{3}$ and $2.6\times10^{6}$ Mpc$^{3}$, respectively.\\

We estimate constraints on the bright end of the UV LF adopting the  $V_\mathrm{max}$ formalism of \citet{avni1980} in 0.5\,mag bins, optimizing the range in UV luminosities of the sample. Following \citet{moster2011} we increase by $24\%$ the Poisson uncertainties to account for cosmic variance. The resulting $z\sim8$ LF is shown in the top panel of Figure \ref{fig:LF} and the corresponding number densities are listed in Tab. \ref{tab:LF}. In our measuring, we only included sources more luminous than $M_\mathrm{UV}\le -21.3$\,mag, for a total of 17 sources, excluding UVISTA-Y3b due to its extremely low luminosity which makes the estimate of the completeness at that luminosity uncertain. Nonetheless, we stress that the volume density we derive in the faintest luminosity bin is likely a lower limit, as the actual incompleteness may be larger than what we estimate. Considering that six sources in our sample are characterized by redshifts $z\ge 8.6$ (after forcing the inclusion of UVISTA-Y5), we considered these galaxies to belong to the $z\sim9$ redshift bin and  computed the associated number densities accordingly. The resulting $z\sim9$ LF is presented in the bottom panel of Figure \ref{fig:LF} and in Table \ref{tab:LF}.

In Figure \ref{fig:LF} we also compare our LF estimates with other recent estimates of the bright end of the LF from empty field searches at $z\sim8$ (\citealt{bradley2012, mclure2013, schenker2013, schmidt2014, bouwens2015, finkelstein2015a, roberts-borsani2016, stefanon2017c, bridge2019}) and $z\sim9$ (\citealt{oesch2013, bouwens2016, calvi2016, mcleod2016, ishigaki2018, livermore2018, morishita2018}). The volume density of $z\sim8$ LBGs probed here corresponds to a luminosity range which exhibits only a modest overlap with earlier LF studies (i.e. \citealt{bouwens2010, bouwens2011b, schenker2013, mclure2013, schmidt2014, finkelstein2015a}), where essentially all $z\sim8$ candidates have apparent magnitudes fainter than  $H\sim 25.5$\, mag. Nonetheless, our luminosity regime overlaps with the widest-area searches available to date from the CANDELS fields (\citealt{bouwens2015} and \citealt{roberts-borsani2016} which includes the spectroscopically confirmed $z\sim8$ LBGs of \citealt{oesch2015a} and \citealt{zitrin2015})  and from the BoRG program (\citealt{trenti2011, calvi2016, bridge2019, livermore2018, morishita2018}).

Perhaps quite unsurprisingly, the new estimate of the $z\sim8$ LF is consistent with the previous measurement at $M_\mathrm{UV} \lesssim -22$\, mag of \citet{stefanon2017c} based on a partly different analysis of the six among the brightest sources presented in this work (UVISTA-Y1 through UVISTA-Y6), and where we also considered  \textit{HST}/WFC3 imaging  for three of them from one of our \textit{HST} programs. The availability of \textit{HST}/WFC3 DASH data allowed us to ascertain that UVISTA-Y3 is likely a triple system of fainter ($\sim L^*$) LBGs. However, the revised analysis performed for the current work showed that one of the sources previously considered to be at $z_\mathrm{phot}\sim7.5$ is actually at $z_\mathrm{phot}\gtrsim8$, thus increasing its luminosity and balancing the final volume density.

For $M_\mathrm{UV} \lesssim-22$\, mag sources, our new results are also consistent with the upper limits of \citet{bradley2012}, \citet{bouwens2015}, \citet{finkelstein2015a} and of \citet{roberts-borsani2016}; our measurements are in excess of what is expected extrapolating the \citet{bouwens2015} results to brighter magnitudes by a factor of $8$, but are nevertheless consistent within $2\sigma$.  

At $M_\mathrm{UV} \gtrsim -22$\, mag our new estimates are consistent with the volume densities of bright LBGs over the CANDELS fields reported by \citet{mclure2013}, \citet{bouwens2015}, \citet{finkelstein2015a} and by \citet{roberts-borsani2016} and with the measurements of \citet{bradley2012}, \citet{schenker2013} and \citet{schmidt2014} from the BoRG program (\citealt{trenti2011, yan2011}).  Recently, \citet{bridge2019} presented the $z\sim8$ LF from eight $M_\mathrm{UV}\gtrsim -22$\,mag sources identified over BoRG fields for which \textit{Spitzer}/IRAC data were collected in the $3.6\mu$m  and $4.5\mu$m bands. The associated volume density is $\sim 5\times$ higher than what we estimate for our sample and their measurements are only consistent at $\sim 3\sigma$. However, the steepness of the LF at the bright end significantly increases the challenges in comparing volume density estimates due to the sensitive dependence on the precise luminosity range probed in different studies. Furthermore, this discrepancy could in part be explained by the different median cosmic times probed by the two samples, considering that the median redshift of the \citet{bridge2019} sample, $z_\mathrm{phot,med}=7.76$ is lower than the median redshift of our sample ($ z_\mathrm{phot}\sim8.4$).

In the lower panel of Figure \ref{fig:LF} we present our estimates of the $z\sim9$ LF. Here we mark with an open symbol the point corresponding to the faintest bin of luminosity because our selection in that luminosity range is likely very incomplete.

At  $M_\mathrm{UV} \lesssim -22$\, mag our new bright $z\sim9$ results are consistent with the upper limits of \citet{bouwens2016} from CANDELS and of \citet{ishigaki2018} from the Hubble Frontier Field initiative (\citealt{lotz2017}). Our measurement at  $M_\mathrm{UV} \sim -22$\, mag  is consistent at $1\sigma$ with the measurement of \citet{morishita2018} based on BoRG observations partly supported by \textit{Spitzer}/IRAC observations, while it is consistent with that of \citet{calvi2016} at $\sim2\sigma$, our density being $\sim9\times$ lower than the corresponding measurement of \citet{calvi2016}. One possible explanation for differences between our results and those of \citet{calvi2016} would be if the \citet{calvi2016} samples suffer from significant contamination. This is especially a concern since few of candidate $z\sim9$ sources have available \textit{Spitzer}/IRAC or deep $Y_{098}$ imaging to aid in source selection.  In fact, \citet{livermore2018} find that one especially bright candidate reported by \citet{calvi2016} appeared to be clearly a low-redshift candidate after further examination.

In the same panel we also plot a double power law that we evolved to $z\sim9$ applying the relations of \citet{bouwens2016} to the double power law found at $z\sim7$ by \citet[see also \citealt{stefanon2017c}]{bowler2017}. Indeed, the excess in number density we observe for $M_\mathrm{UV}<-22$\,mag at $z\sim9$ (introduced by forcing UVISTA-Y5 into the $z\sim9$ sample) seems to be better described by the double power-law. We remark, however, that the still large uncertainties do not allow us to fully remove the degeneracy on the shape of the LF at $z\sim9$, which instead needs larger samples. We will discuss the shape of the $z\sim8$ LF in more detail in Sect.~\ref{sect:LF_shape}.

\subsection{Combination of Present Constraints with Faint $z\sim8$ LF Results}

\begin{figure}
\includegraphics[width=9.2cm]{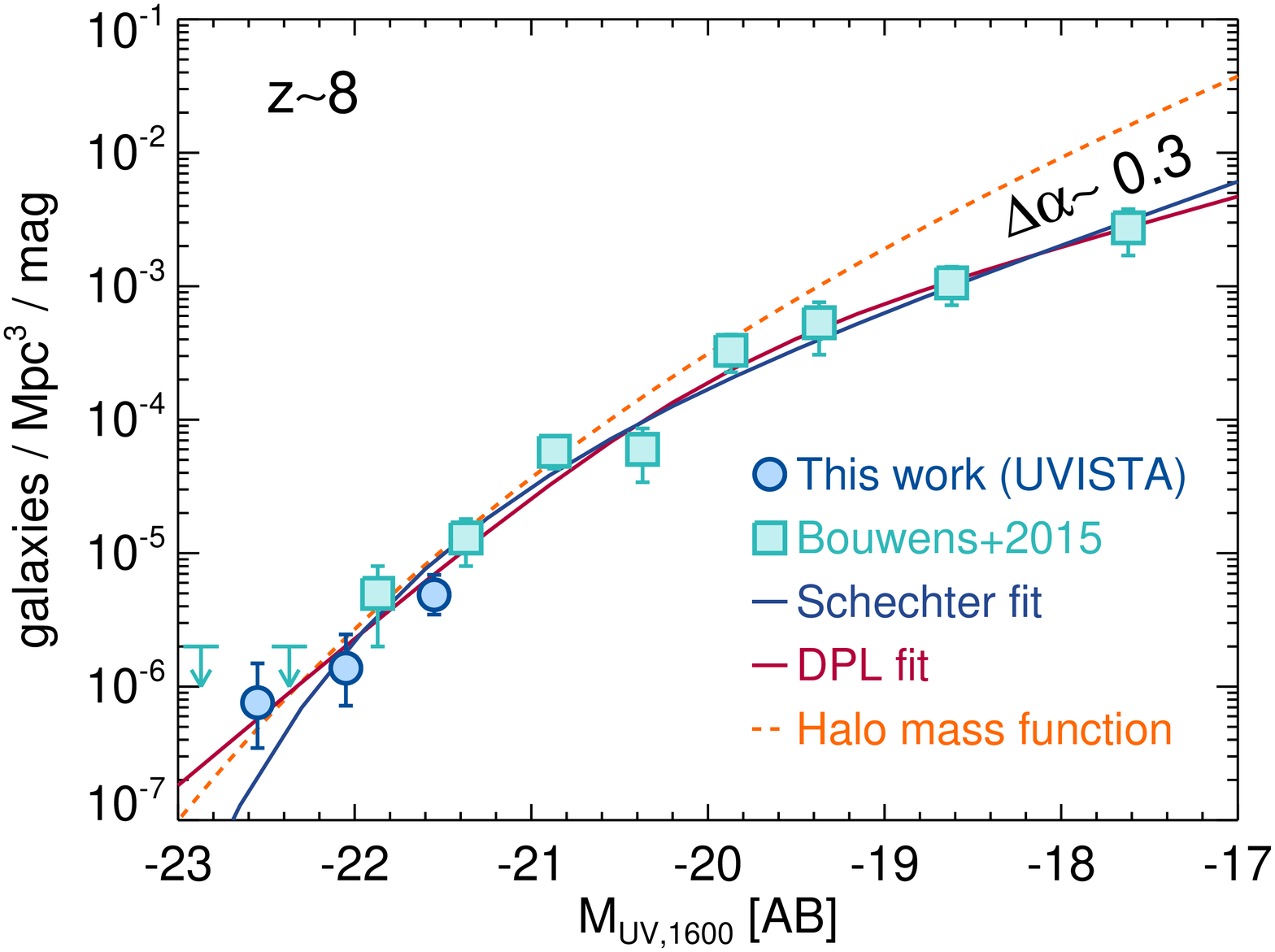}
\caption{Step-wise LF at $z\sim8$ obtained combining the $V_\mathrm{max}$ estimates at the bright end combined to those of \citet{bouwens2015} at $M_\mathrm{UV}>-22$\,mag. The discrete LF measurements can be represented by either a \citet[blue solid curve]{schechter1976} or double power law (red solid curve) form, with a marginal preference for this latter model. We also show the halo mass function at $z\sim8$ (orange dashed line) scaled by a fixed $M_\mathrm{halo}/L_\mathrm{UV}$ to match the knee of our derived $UV$ LF at $z\sim8$.  The high-mass-end slope of the halo mass function is similar to the effective slope of the UV LF at the bright end.  The difference between the low-mass-end slope of the halo mass function and the  faint-end slope of the LF is $\Delta\alpha\sim0.3$. \label{fig:LF_Schechter}}
\end{figure}

\begin{figure*}
\includegraphics[width=18cm]{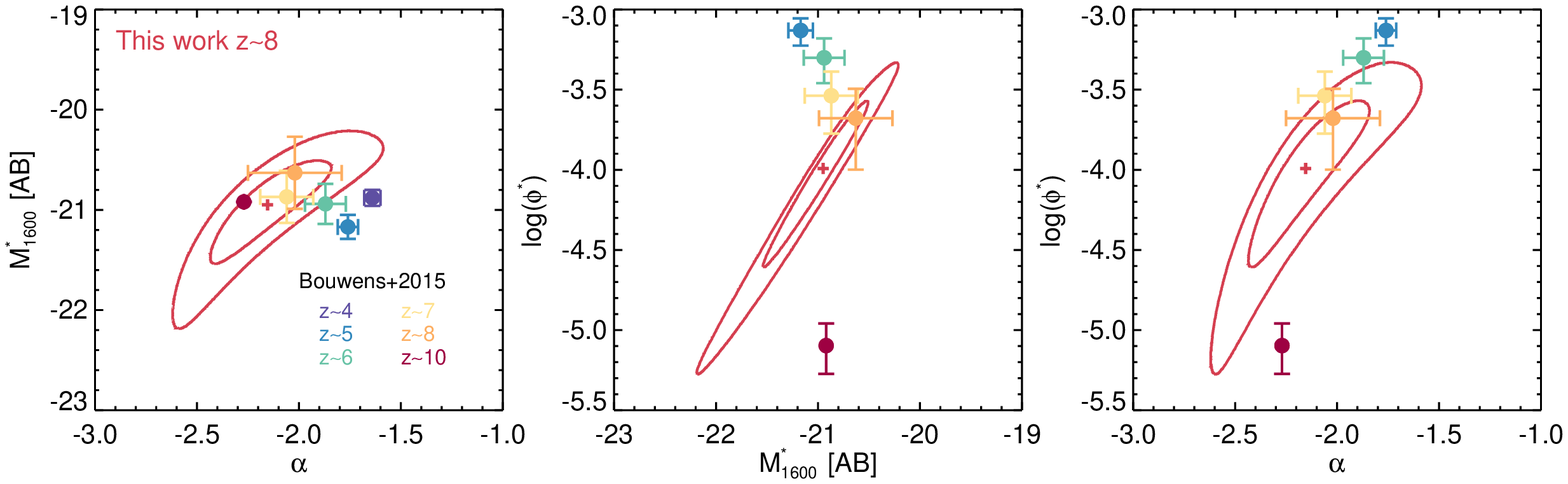}
\caption{68\% and 95\% likelihood contours on the Schechter parameters at $z\sim8$ (red) derived in the present work shown relative to the Schechter parameter estimates found for the LF results at $z\sim4$-to-$10$ by \citet{bouwens2015}, as specified by the legend.  The observations seem to point towards a clear increase in $\phi^*$ and flattening of $\alpha$ with cosmic time. \label{fig:LF_chi2}}
\end{figure*}

The bright candidates found over UltraVISTA alone are not sufficient to constrain the overall shape of the UV LF due to lack of dynamic range.  In the case of a \citet{schechter1976} function where the shape is determined by the faint-end slope $\alpha$ and turn over magnitude $M^*$, both bright and faint objects are needed to constrain $\alpha$ and $M^*$.  The similar redshift distributions expected for bright galaxies selected by our $z\sim8$ criteria and those selected in the fainter  \citet{bouwens2015} samples (see Figure \ref{fig:zdist}) make it possible to combine our $z\sim8$ LF with the corresponding estimates of \citet{bouwens2015}, based on the CANDELS, HUDF09, HUDF12, ERS, and BoRG/HIPPIES programs.

The combined step-wise determination of the $UV$ LF at $z\sim8$ is presented in Figure \ref{fig:LF_Schechter}. We determined the Schechter function parameters $M^*$, $\alpha$, and $\phi^*$  minimizing the $\chi^2$, and obtaining $\log(\phi^*)=-3.99^{+0.29}_{-0.37}$, $M^*_{1600} = -20.95^{+0.30}_{-0.35}$ mag, and $\alpha = -2.15^{+0.20}_{-0.19}$. The 68\% and 95\% confidence level contours are presented in Figure \ref{fig:LF_chi2}. 

Our sample of bright $z\sim8$ LBGs make the characteristic luminosity $M^*$ is brighter by $\sim 0.5$\, mag compared to the most recent estimates of \citet{bouwens2015}, even though this result is significant only at $\sim1.2\sigma$, while the faint-end slope $\alpha$ is consistent at $1\sigma$.

In Figure \ref{fig:LF_chi2}  we also compare our estimated Schechter parameters to their evolution over a wide range of redshift, $4\lesssim z \lesssim 10$ from \citet{bouwens2015}. Our result confirm the picture of marginal evolution of $M^*$ for $z\gtrsim 3-4$, but significant evolution of $\alpha$ and $\phi^*$. This conclusion was first drawn by \citet{bouwens2015} using LF results from $z\sim7$ to $z\sim4$ (see also \citealt{finkelstein2015a}), although they are in modest tension with the results of \citet{bowler2015} who suggest an evolution of $dM^* / dz \sim 0.2$ from $z\sim7$ to $z\sim5$.

\subsection{The shape of the LF at $z\sim8$}
\label{sect:LF_shape}

One significant area of exploration over the last few years has regarded the form on the $UV$ LF at the bright end.  In particular, there has been interest in determining whether the $UV$ LF shows more of an exponential cut-off at the bright end or a power-law-like cut-off.  The higher number densities implied by a power-law-like form might indicate that the impact of either feedback or dust is less important at high redshifts than it is at later cosmic times.  Successfully distinguishing a power-law-like form for the bright end of the LF from a sharper exponential-like cut-off is challenging, as it requires very tight constraints on the bright end of the LF and hence substantial volumes for progress.

The simplest functional form to use in fitting the $UV$ LF is a power law and can be useful when very wide-area constraints are not available for fitting the bright end.  One of the earliest considerations of a power-law form in fitting the $UV$ LF at $z>6$ was by \citet{bouwens2011b}, and it was shown that such a functional form satisfactorily fit all constraints on the $z\sim8$ LF from {\it HST} available at the time (Figure 9 from that work).  

Here we consider three functional forms that can potentially be adopted to describe the number density of galaxies at $z\sim8$: a single power law, a double power law and the \citet{schechter1976} form. The parameterization for a double power-law is as follows (see also \citealt{bowler2012, ono2018}):
\begin{displaymath}
\phi(M) = \frac{\phi^*}{10^{0.4(\alpha+1)(M-M^*)} + 10^{0.4(\beta+1)(M-M^*)}}
\end{displaymath}
where $\alpha$ and $\beta$ are the faint-end and bright-end slopes, respectively, $M^*$ is the transition luminosity between the two power-law regimes, and $\phi^*$ is the normalization.

A quick inspection of Figure~\ref{fig:LF_Schechter} suggests already that the $UV$ LF at $z\sim8$ cannot be well represented by a power-law form. Indeed, a $\chi^2$ test,  as previously adopted by e.g., \citet{bowler2012, bowler2015, bowler2017} at $z\sim7$, results in reduced $\chi^2$, $\chi^2_\nu= 3.5, 1.04$ and $1.05$ for the single power law, double power law and Schechter functional form, respectively. The double power-law parameters are $\alpha=-1.92 \pm 0.50$, $\beta=-3.78 \pm 0.48$, $M^*=-20.04 \pm 1.00$\,mag and $\phi^*=3.88^{+5.80}_{-3.88} \times 10^{-4}$\,Mpc$^{-3}$ mag$^{-1}$.

The above results suggest that we can not yet properly distinguish between a Schechter and a double power-law form, a result which might be driven by the higher volume density we measured in the brightest absolute magnitude bin. Nevertheless, this result is in line with recent UV LF estimates at $z\lesssim 7$ from large area surveys (UltraVISTA DR2 - \citealt{bowler2014, bowler2015}, HSC Survey - \citealt{ono2018}), who found an excess in the volume densities of $z\sim 4-7$ galaxies for $L>L^*$ compared to the Schechter exponential decline.

Even though our favoured interpretation consists in considering UVISTA-Y3 composed by three independent sources, in Appendix \ref{app:lf_bdb} for completeness we also present the $V_\mathrm{max}$ estimates obtained from the blended UVISTA-Y3.  We note however that these new estimates do not change significantly from those presented in this section.

In Sect. \ref{sect:lensing} we identified four sources whose flux was likely amplified by massive foreground galaxies.  Indeed, recent studies  have found that gravitational lensing magnification could explain, at least in part, the excess in number density observed at the bright end of $z\sim4-7$ UV LF (see e.g., \citealt{ono2018}). Correcting the apparent magnitude of the four impacted sources for the estimated magnitude and re-deriving the Schechter parameters from the $z\sim8$ constraints, we find $M^*=-20.81$\,mag, $\sim 0.14$ mag fainter than with no correction (see Appendix \ref{app:lens} for details), possibly indicating that lensing is likely playing a modest role in shaping the bright end ($M_\mathrm{UV}\lesssim -22.5$\,mag) of the $z\sim8$ LF.   As the impact of lensing amplification is uncertain and also model dependent, we follow \citet{bowler2015} in ignoring the impact for our fiducial determinations.

An alternative way of making sense of the overall shape of the $UV$ LF is to compare it to the halo mass function.  To this aim, we scaled the halo mass function by a fixed $M_\mathrm{halo}/L_\mathrm{UV}$ ratio to match the UV LF.  We present the result in Figure~\ref{fig:LF_Schechter}, adopting the \citet{sheth2001} halo mass function generated by \textsc{HMFcalc}\footnote{http://hmf.icrar.org/hmf\_finder/form/create/} \citep{murray2013}, assuming a \citet{planck2016_cosmology} cosmology, with $\Omega_m  = 0.2678$, $\Omega_b = 0.049$, $H_0=67.04$, $n_s=0.962$, and $\sigma_8 = 0.8347$.

The scaled halo-mass function looks similar to the $UV$ LF at $z\sim8$.  We observe only a small $\Delta\alpha\sim0.4$ difference in the faint-end slope.  We also observe a slight difference in the effective slope at the bright end when a Schechter form is considered ($\Delta\beta\sim 0.2-0.3$). Interestingly, the bright end of the double power-law overlaps with the HMF for $M_\mathrm{UV}\lesssim -22$\,mag, and might suggest different feedback efficiencies (see also \citealt{bowler2014, ono2018}). However, as we concluded earlier in this section, our results do not allow us to ascertain whether the cut-off we observe is exponential in form or has a more power-law-like form.

We will not conduct a similar quantitative assessment of the shape of the $UV$ LF at $z\sim9$, due to the challenges in determining the total number of bright $z\sim9$ galaxies over UltraVISTA.  Clearly, if any significant number of the candidates do prove to be bona-fide $z\sim9$ galaxies, they would favour more of a power-law form to the bright end of the $z\sim9$ LF.

\subsection{Evolution of the UV Luminosity Density for Luminous Galaxies from $z\sim10$ to $z\sim4$}
\label{sect:LD}

\begin{figure}[!t]
\includegraphics[width=9.2cm]{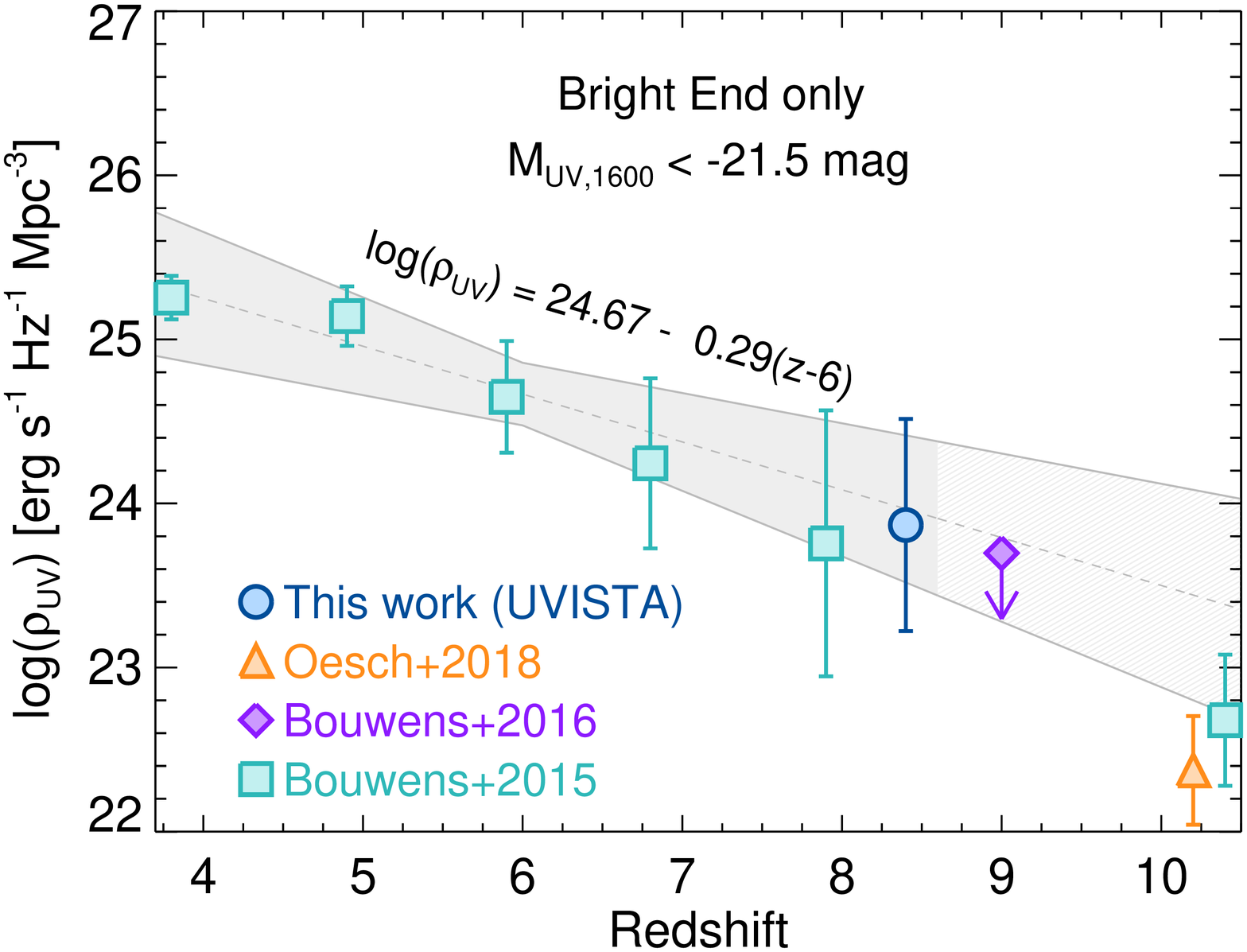}
\caption{The evolution of the UV luminosity density of galaxies at $4 < z < 9$ (filled blue circle) bright-ward of $-21.5$ mag using the present search over COSMOS/UltraVISTA. The green squares mark the luminosity density of galaxies obtained from the Schechter parameterization by \citet{bouwens2015} at $z\sim4-10$. Also shown are the recent estimates at $z\sim9$ of \citet{bouwens2016} and at $z\sim10$ of \citet{oesch2018}. The shaded region shows a linear fit to the evolutionary trend to $z\sim10$ in the UV luminosity density preferred at 68\% confidence. \label{fig:LD}}
\end{figure}

The evolution of the UV luminosity density with cosmic time provides an insight into the rate at which galaxies are building up and how this rate might depend on cosmic time or galaxy/halo mass.

In Figure~\ref{fig:LD}, we present the $UV$ luminosity densities we infer to the effective faint-end limit of our present search (i.e., $\sim-21.5$\, mag) at $z\sim8$ obtained assuming a Schechter form, given a power-law at the bright end would imply an infinite luminosity density. We note however, that we obtain identical results if we compute the luminosity density over $M_\mathrm{UV}=[-23, -21.5]$\,mag with both the Schechter and double power law parameterization.  The luminosity density constraint we find at $z\sim8$ is $10^{23.87^{+0.58}_{-0.68}}$ ergs/s/Hz/Mpc$^{3}$ ($68\%$ c.l.).

In Figure~\ref{fig:LD} we also present the luminosity density determinations and uncertainties that we derive integrating to $M_\mathrm{UV}=-21.5$\,mag the recent LF results of \citet{bouwens2015} at $z\sim4$, $z\sim5$, $z\sim6$, $z\sim7$ and $z\sim8$, the results of \citet{bouwens2016} at $z\sim9$,  and at $z\sim10$ from \citet{oesch2018}, together with our best-fit constraints and $1\sigma$ uncertainties on the evolution of the luminosity density with redshift, assuming that the logarithm of the luminosity density decreases linearly with increasing the redshift.

Our estimate is consistent with that derived from the $z \sim8$ LF of \citet{bouwens2015} and with the upper limit at $z\sim9$ based on \citet{bouwens2016} LF. However, the luminosity densities recovered from the LF of \citet{oesch2018} and \citet{bouwens2015} at $z\sim10$ are  $\sim1.7\sigma$ and $\sim1.1\sigma$ respectively lower than the extrapolation of the linear relation discussed above, hinting at a potentially even more rapid buildup of the brightest galaxies in the first $\sim500$\,Myr of cosmic history (see e.g., the extensive discussion in \citealt{oesch2018}).

\section{Conclusions}

Using deep infrared data from the COSMOS/ UltraVISTA program, we have identified 16 new ultrabright $H\sim24.8-25.6$\,mag galaxy candidates at $z\sim8$.  The new candidates are amongst the brightest yet found at these redshifts, $\gtrsim0.5$ magnitude brighter than found over CANDELS, providing improved constraints at the bright end of the UV luminosity function, and providing excellent targets for follow up at longer wavelengths and with spectroscopy.

The spectral slope of the $UV$-continuum $\beta$, parameterized as $f_{\lambda}\propto \lambda^{\beta}$; \citealt{meurer1999}) for the bright $z\sim8$ sample is $\beta=-2.2\pm0.6$, which is bluer but still consistent with the $UV-$continuum slopes found for bright $-22 < M_\mathrm{UV} < -21$ galaxies at $z=6$ ($\beta=-1.55\pm0.17$) and $z\sim7$ ($\beta=-1.75\pm0.18$) by \citet{bouwens2014} and suggests a continuing trend towards bluer $\beta$'s at higher redshifts. The typical estimated stellar masses for bright sources in our $z\sim8$ selection are 10$^{9.1^{+0.5}_{-0.4}}$ $M_{\odot}$, with the SFRs of  $32^{+44}_{-32}M_{\odot}$/year, specific SFR of $4^{+8}_{-4}$ Gyr$^{-1}$, stellar ages of $\sim22^{+69}_{-22}$\,Myr, and low dust content A$_V=0.15^{+0.30}_{-0.15}$\, mag, with the properties of individual galaxies spanning a large range of values. 

Using public catalogs we checked the lensing magnification from close, lower redshift sources. We find that four sources are likely subject to magnifications of approximately $1.5\times$. Nevertheless, the effect on the UV LF is marginal.

We use the candidate galaxies to constrain the bright end of the $z\sim8$   UV luminosity function.  Combining our ultrabright sample with candidates found over CANDELS, HUDF and HFF field data allows us to constrain on the $z\sim8$  LF. Assuming a Schechter function, the best-fit characteristic magnitude is $M^*(z=8) = -20.95^{+0.30}_{-0.35}$\,mag with a very steep faint end slope $\alpha = -2.15^{+0.20}_{-0.19}$.  Our $z\sim8$ LF results can be equally well represented adopting a functional form where the effective slope is steeper at the bright end of the LF than at the faint end, such as for a double power law.  Our results rule out the use of a single power-law in representing the $z\sim8$ LF.

We note that, despite much recent progress, the lack of spectra and deep high-resolution imaging still limit us in establishing the reliability of high redshift galaxy selections, in particular for rare luminous galaxies that constrain the bright end of the UV luminosity and mass functions where any contamination has a very large impact. While care is taken in estimating the completeness and contamination rates, these still rely on assumed spectral energy distributions. Ultimately, spectroscopy is needed to validate these assumptions. While recent results suggest ALMA as a potentially efficient machine for the study of emission lines (e.g., \citealt{smit2018}), currently this is still hard in the rest-frame UV and optical, due to long integration times, low multiplexing, and the reduced observable emission of Ly$\alpha$ likely caused by the increasing neutral hydrogen fraction $z > 6$ (e.g., \citealt{schenker2014}),  but will be possible in the future with JWST and next generation of extremely large ground-based telescopes.

\acknowledgments
The authors would like to thank the referee for their careful reading and for the many helpful, positive and very constructive comments that helped improving the quality of the paper. The authors wish to acknowledge the UltraVista team who conducted and followed the observations and image processing. The authors are thankful to the COSMOS collaboration for their continued efforts in making available part of the data used in this work. KIC acknowledges funding from the European Research Council through the award of the Consolidator Grant ID 681627-BUILDUP. Based on data products from observations made with ESO Telescopes at the La Silla Paranal Observatory under ESO programme ID 179.A-2005 and on data products produced by TERAPIX and the Cambridge Astronomy Survey Unit on behalf of the UltraVISTA consortium.   The Cosmic Dawn Center is funded by the Danish National Research Foundation. This work is based in part on observations made with the Spitzer Space Telescope, which is operated by the Jet Propulsion Laboratory, California Institute of Technology under a contract with NASA. Based in part on data collected at the Subaru Telescope and retrieved from the HSC data archive system, which is operated by Subaru Telescope and Astronomy Data Center at National Astronomical Observatory of Japan.The Hyper Suprime-Cam (HSC) collaboration includes the astronomical communities of Japan and Taiwan, and Princeton University. The HSC instrumentation and software were developed by the National Astronomical Observatory of Japan (NAOJ), the Kavli Institute for the Physics and Mathematics of the Universe (Kavli IPMU), the University of Tokyo, the High Energy Accelerator Research Organization (KEK), the Academia Sinica Institute for Astronomy and Astrophysics in Taiwan (ASIAA), and Princeton University. Funding was contributed by the FIRST program from Japanese Cabinet Office, the Ministry of Education, Culture, Sports, Science and Technology (MEXT), the Japan Society for the Promotion of Science (JSPS), Japan Science and Technology Agency (JST), the Toray Science Foundation, NAOJ, Kavli IPMU, KEK, ASIAA, and Princeton University.

\appendix

\section{Monte Carlo assessment of the multi-component nature of UVISTA-Y3}
\label{app:mc}

Because the three components of UVISTA-Y3 are characterized by low S/N  on the DASH footprint adopted for their identification (S/N$\sim 4.5, 2.9$ and $\sim2.2$), one might wonder whether the detected splitting into three components is instead the result of background noise acting on a single, extended source.

To test this hypothesis we implemented the following Monte Carlo procedure. We generated a table of twenty random positions on the footprint of the DASH mosaic with similar background noise properties. For each one of these positions we created an elongated disk, with minor-to-major axis ratio $b/a$ drawn from a pool of random values $0.05<b/a<0.25$ with an exponential luminosity profile with effective radius $r_\mathrm{e}=1$\,kpc, consistent with recent rest-frame UV size estimates for bright LBGs at $z>6$ (e.g., \citealt{oesch2016, bowler2017, stefanon2017c}).  The choice of low values for $b/a$ was guided by the relatively large separation between the three components, which is difficult to obtain when more compact morphologies are considered. The total flux density of each exponential disk was set to be equal to the $H$ band flux of UVISTA-Y3 when considered as single source. Each exponential disk was then convolved with the WFC3 $H_{160}$-band PSF, randomly rotated and added to the original DASH image.

\begin{figure*}
\includegraphics[width=18cm]{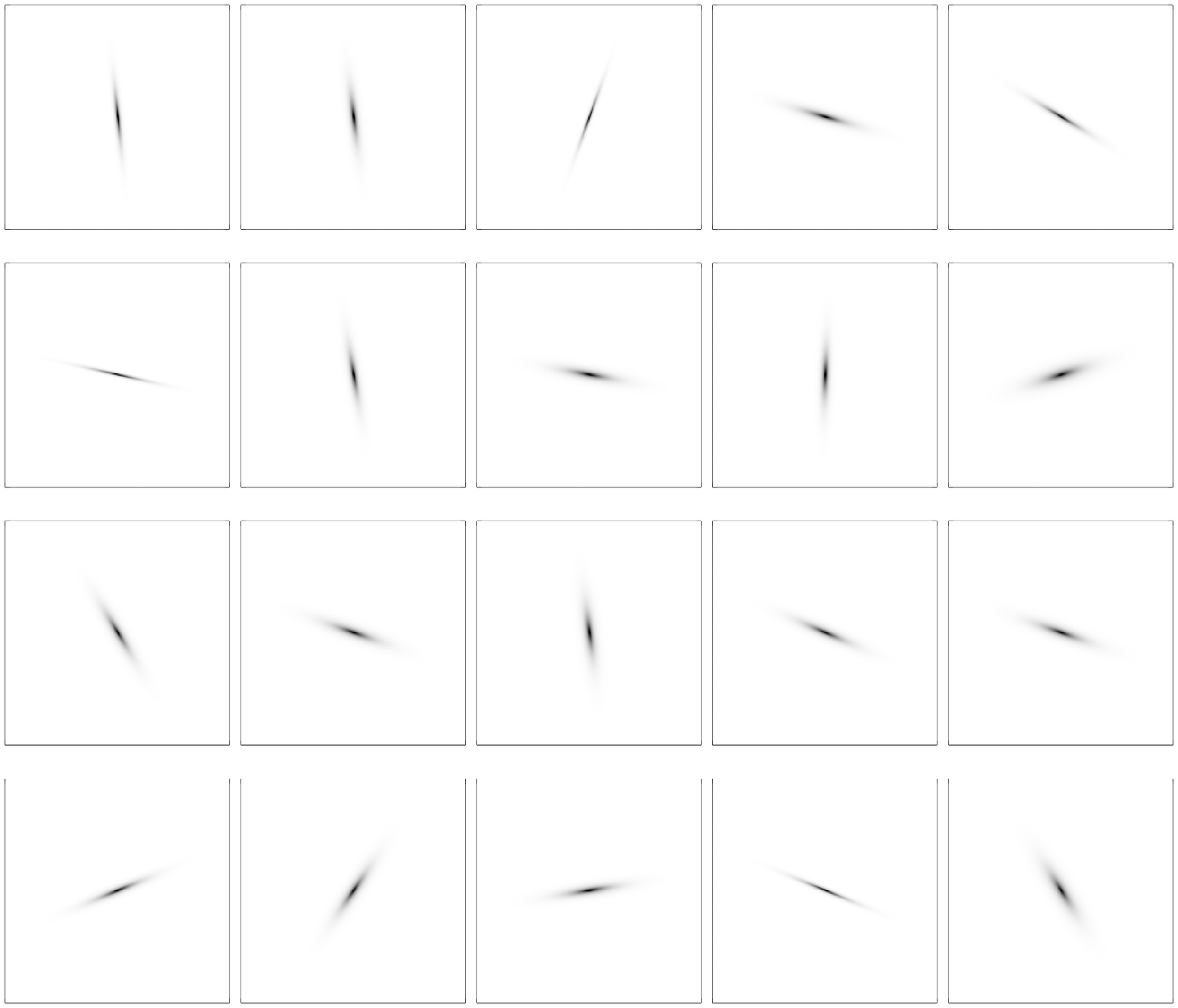}
\caption{Each panel presents one of the random realizations of the intrinsic (i.e., before being convolved with the WFC3 $H_{160}$-band PSF and added to the DASH mosaic) exponential disk created to test the multi-component nature of UVISTA-Y3. The side of each stamp is $2\farcs0$.  \label{fig:mc_synth}}
\end{figure*}

\begin{figure*}
\includegraphics[width=18cm]{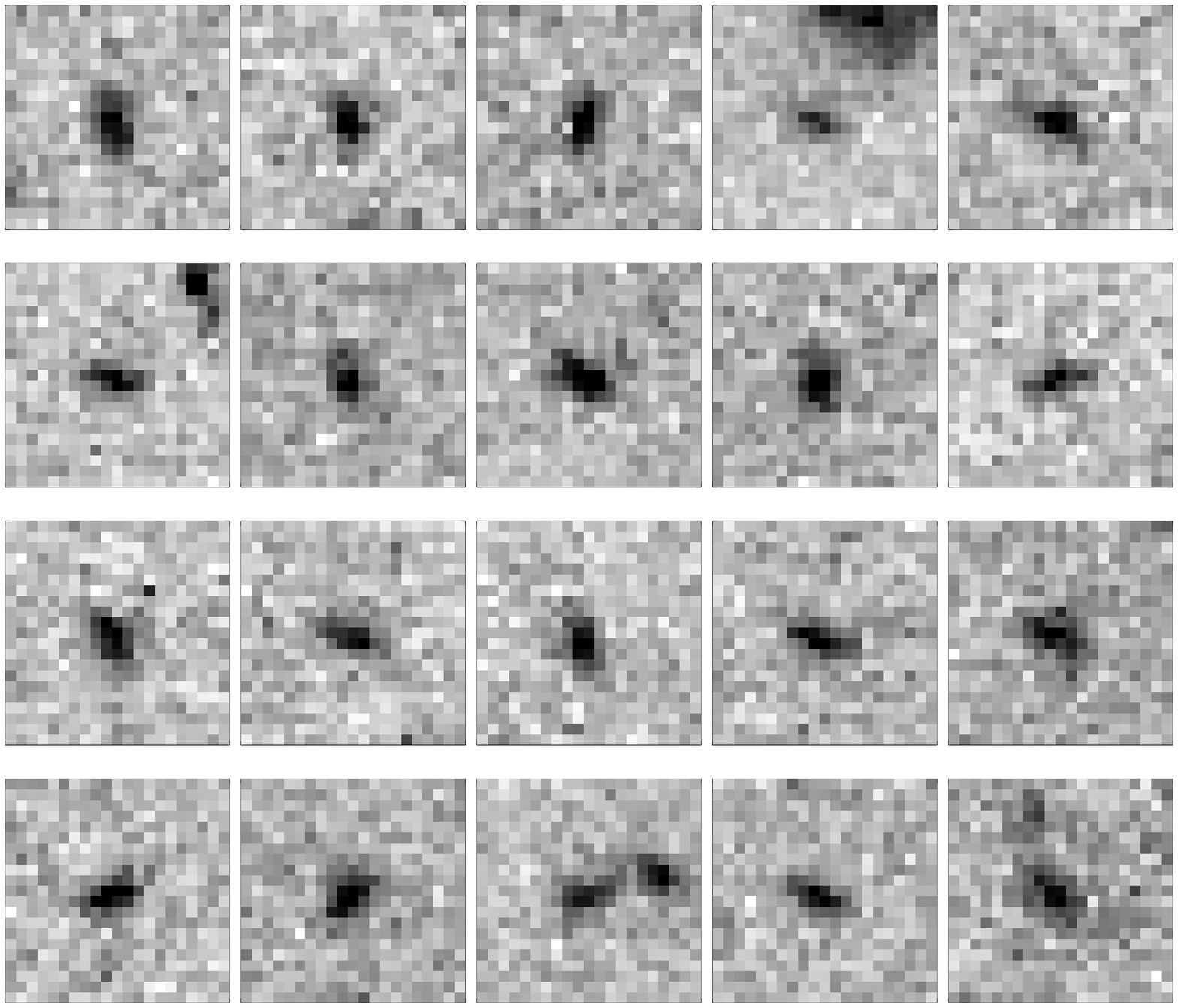}
\caption{Each image stamp ($2\farcs0$ side) is centered on the corresponding synthetic exponential disk of Figure \ref{fig:mc_synth}, after being convolved with the WFC3 PSF and added to the DASH mosaic at locations with noise properties similar to those where UVISTA-Y3 lies. In the central panel of the bottom row, the object on the right is a real source present in the observations (and therefore bears no relation to the Monte-Carlo simulations we perform). None of the simulated sources show apparent multi-component structure (as UVISTA-Y3 seems to show) as a result of noise in the background.  \label{fig:mc_onImage}}
\end{figure*}

In Figure \ref{fig:mc_synth} we present the twenty random realizations of the exponential disk, before convolution with the WFC3 PSF, while in Figure \ref{fig:mc_onImage} we present the image stamps of the DASH mosaics after the synthetic exponential disks have been added. Based on simple visual inspection, we see no indication in these simulated images for a multiple component structure. Finally we run SExtractor using the same set of parameters adopted for the original deblending and found that none of the synthetic sources were split into two or more components. This test therefore increased our confidence on the multiple nature of UVISTA-Y3. It is worth remarking that as a result of the low S/N of the deblended photometry for each component there are substantially larger uncertainties in the derived physical parameters for each component.

\begin{figure*}
\includegraphics[width=18cm]{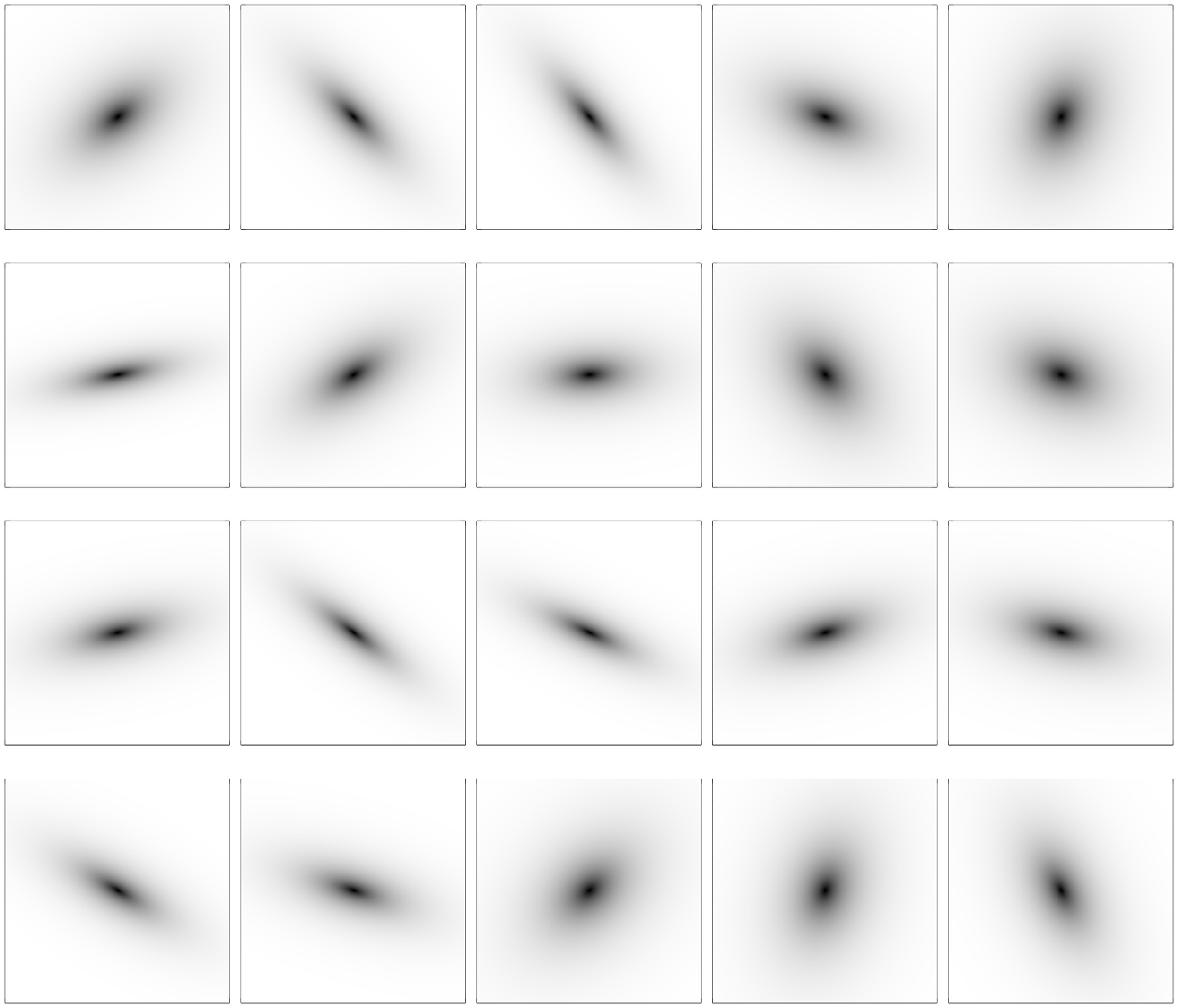}
\caption{Similar to Figure \ref{fig:mc_synth} but with $r_\mathrm{e}=3$\,kpc.  \label{fig:mc_synth3}}
\end{figure*}

\begin{figure*}
\includegraphics[width=18cm]{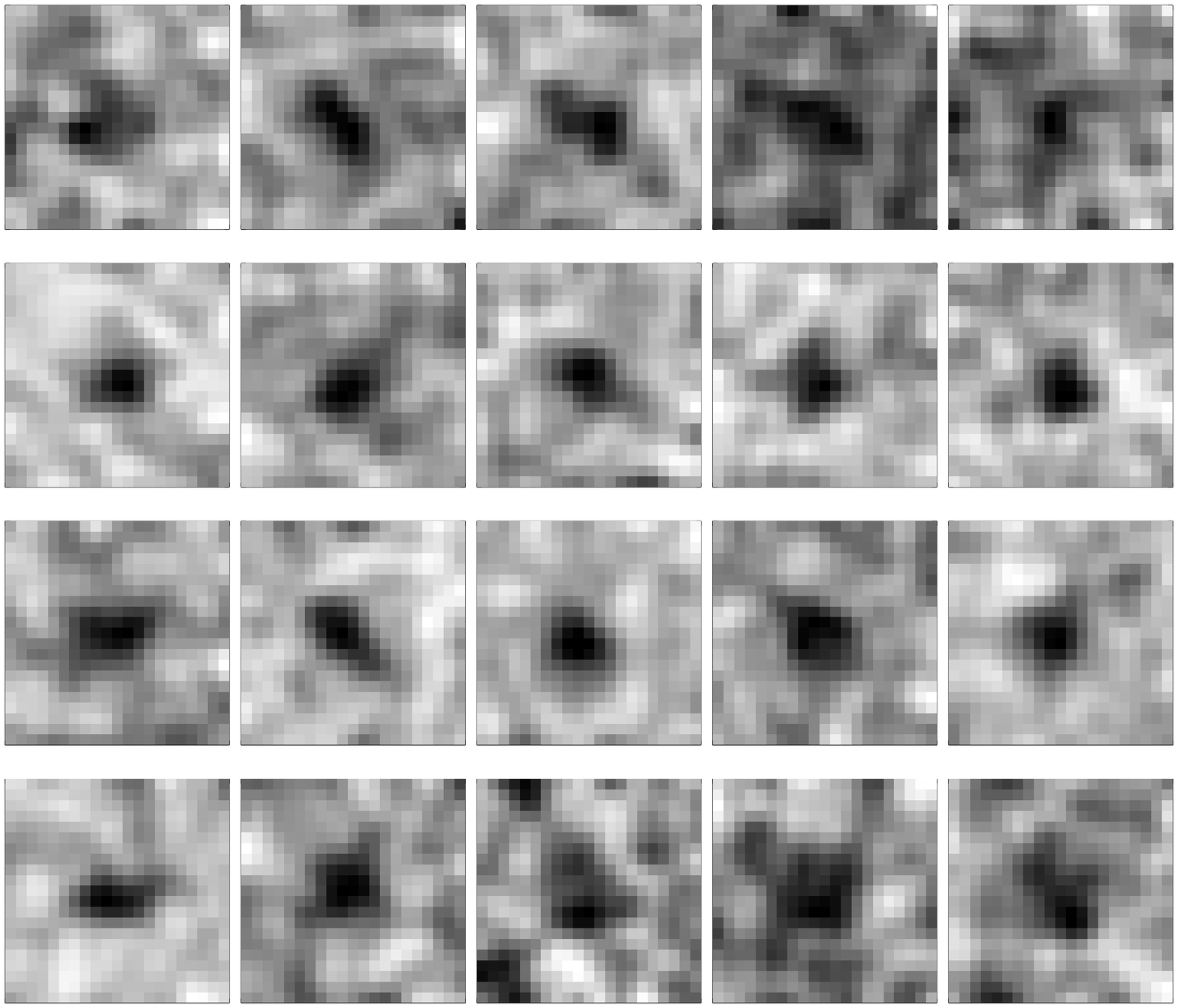}
\caption{Similar to Figure \ref{fig:mc_onImage} but with $r_\mathrm{e}=3$\,kpc. The image stamps have been smoothed with a gaussian of $0\farcs1$ FWHM (equal to the value used for Figure \ref{fig:Y3}) to improve the contrast.  \label{fig:mc_onImage3}}
\end{figure*}

We complemented this first assessment with a second Monte Carlo simulation in which we adopt an effective radius $r_\mathrm{e}=3$\,kpc, similar to the sizes of  luminous high redshift LBGs when potentially multiple sources are considered as a single object (e.g., \citealt{bowler2017}), and  $0.1<b/a<0.7$. We present the result of this simulation in Figures \ref{fig:mc_synth3} and \ref{fig:mc_onImage3}. Using a procedure similar to that applied for the $r_\mathrm{e}=1$\,kpc case, we do not find evidence for multiple components even when large $r_\mathrm{e}$ are considered, increasing the confidence on our interpretation of the three components in UVISTA-Y3.

\section{Flux density estimates}
\label{app:flx}

In Table \ref{tab:flx1}, \ref{tab:flx2} and \ref{tab:flx3} we list the flux density estimates and associated $1\sigma$ uncertainties for the full sample of  $z\sim8$ candidate galaxies presented in this work.

\begin{deluxetable*}{lr@{ $\pm$}rr@{ $\pm$}rr@{ $\pm$}rr@{ $\pm$}rr@{ $\pm$}rr@{ $\pm$}rr@{ $\pm$}r}
\tablecaption{Flux density measurements for UVISTA-Y1 to UVISTA-Y5  \label{tab:flx1}}
\tablehead{\colhead{Filter} & \twocolhead{UVISTA-Y1} & \twocolhead{UVISTA-Y2} & \twocolhead{UVISTA-Y3a} & \twocolhead{UVISTA-Y3b} & \twocolhead{UVISTA-Y3c} & \twocolhead{UVISTA-Y4} & \twocolhead{UVISTA-Y5} \\
\colhead{name} & \twocolhead{[nJy]} & \twocolhead{[nJy]} & \twocolhead{[nJy]} & \twocolhead{[nJy]} & \twocolhead{[nJy]} & \twocolhead{[nJy]} & \twocolhead{[nJy]}
}
\startdata
            CFHTLS $u^*$  & \multicolumn2c{$\cdots$}  & $      -6 $  & $      15 $  & $       7 $  & $      15 $  & $      -7 $  & $      15 $  & $      -5 $  & $      14 $  & $     -10 $  & $      15 $  & $     -13 $  & $      17 $  \\
                 SSC $B$  & $      -2 $  & $       7 $  & $      -1 $  & $      10 $  & $       1 $  & $       7 $  & $      -6 $  & $       8 $  & $     -10 $  & $       9 $  & $       5 $  & $       9 $  & $       4 $  & $       9 $  \\
                 HSC $g$  & $       8 $  & $      15 $  & $       2 $  & $      17 $  & $     -22 $  & $      16 $  & $       5 $  & $      17 $  & $       2 $  & $      17 $  & $      -8 $  & $      17 $  & $     -10 $  & $      20 $  \\
              CFHTLS $g$  & \multicolumn2c{$\cdots$}  & $      -1 $  & $      13 $  & $     -10 $  & $      12 $  & $     -11 $  & $      12 $  & $      -6 $  & $      12 $  & $       2 $  & $      13 $  & $       3 $  & $      15 $  \\
                 SSC $V$  & $     -12 $  & $      17 $  & $     -14 $  & $      22 $  & $      -5 $  & $      21 $  & $       0 $  & $      21 $  & $      -7 $  & $      20 $  & $      29 $  & $      20 $  & $      -1 $  & $      24 $  \\
                 HSC $r$  & $      -7 $  & $      14 $  & $      -4 $  & $      17 $  & $      -5 $  & $      14 $  & $      -5 $  & $      15 $  & $       1 $  & $      16 $  & $       7 $  & $      17 $  & $      -3 $  & $      18 $  \\
              CFHTLS $r$  & \multicolumn2c{$\cdots$}  & $     -12 $  & $      19 $  & $      -1 $  & $      17 $  & $     -14 $  & $      17 $  & $      14 $  & $      19 $  & $       7 $  & $      20 $  & $     -15 $  & $      23 $  \\
               SSC $r^+$  & $       7 $  & $      16 $  & $     -15 $  & $      20 $  & $       1 $  & $      17 $  & $       5 $  & $      17 $  & $      10 $  & $      18 $  & $      10 $  & $      18 $  & $     -29 $  & $      23 $  \\
              SSC  $i^+$  & $      -9 $  & $      22 $  & $    -139 $  & $      89 $  & $     -17 $  & $      27 $  & $     -27 $  & $      27 $  & $     -21 $  & $      28 $  & $    -103 $  & $      90 $  & $     -36 $  & $      26 $  \\
              CFHTLS $y$  & \multicolumn2c{$\cdots$}  & $     -18 $  & $      25 $  & $     -13 $  & $      22 $  & $      17 $  & $      20 $  & $       2 $  & $      24 $  & $     -14 $  & $      25 $  & $     -11 $  & $      28 $  \\
               CFHTLS $i$ & \multicolumn2c{$\cdots$}  & $     -28 $  & $      26 $  & $     -20 $  & $      22 $  & $     -19 $  & $      21 $  & $     -11 $  & $      24 $  & $       3 $  & $      26 $  & $     -11 $  & $      29 $  \\
                 HSC $i$  & $      21 $  & $      21 $  & $     -44 $  & $      24 $  & $     -12 $  & $      21 $  & $     -21 $  & $      20 $  & $      -2 $  & $      24 $  & $     -10 $  & $      24 $  & $     -20 $  & $      27 $  \\
              CFHTLS $z$  & \multicolumn2c{$\cdots$}  & $     -13 $  & $      55 $  & $     -22 $  & $      46 $  & $       0 $  & $      45 $  & $     -15 $  & $      52 $  & $       7 $  & $      54 $  & $       6 $  & $      63 $  \\
                 HSC $z$  & $       9 $  & $      31 $  & $      -1 $  & $      35 $  & $       1 $  & $      30 $  & $     -30 $  & $      29 $  & $     -51 $  & $      35 $  & $     -50 $  & $      35 $  & $     -27 $  & $      39 $  \\
              SSC  $z^+$  & $      51 $  & $      64 $  & $    -102 $  & $      75 $  & $      19 $  & $      72 $  & $     -52 $  & $      75 $  & $      79 $  & $      72 $  & $      43 $  & $      71 $  & $     -51 $  & $      93 $  \\
                 HSC $y$  & $     -31 $  & $      76 $  & $      31 $  & $      85 $  & $      -5 $  & $      81 $  & $     -86 $  & $      80 $  & $     -83 $  & $      84 $  & $      33 $  & $      84 $  & $     -91 $  & $      98 $  \\
              UVISTA $Y$  & $      18 $  & $      48 $  & $      29 $  & $      53 $  & $      55 $  & $      50 $  & $     -12 $  & $      49 $  & $      80 $  & $      53 $  & $     173 $  & $      54 $  & $     -42 $  & $      68 $  \\
             UVISTA $J$   & $     324 $  & $      50 $  & $     410 $  & $      61 $  & $     120 $  & $      57 $  & $     111 $  & $      57 $  & $      33 $  & $      59 $  & $     432 $  & $      65 $  & $     235 $  & $      66 $  \\
              UVISTA $H$  & $     455 $  & $      61 $  & $     432 $  & $      78 $  & $     229 $  & $      68 $  & $     135 $  & $      67 $  & $     145 $  & $      70 $  & $     392 $  & $      86 $  & $     393 $  & $      86 $  \\
   UVISTA $K_\mathrm{S}$  & $     480 $  & $      77 $  & $     275 $  & $      86 $  & $     197 $  & $      77 $  & $      68 $  & $      72 $  & $     110 $  & $      82 $  & $     266 $  & $     110 $  & $     321 $  & $     102 $  \\
          IRAC $3.6\mu$m  & $     623 $  & $      85 $  & $     492 $  & $      50 $  & $     220 $  & $      78 $  & $      50 $  & $      80 $  & $     214 $  & $      79 $  & $     620 $  & $      68 $  & $     289 $  & $      74 $  \\
          IRAC $4.5\mu$m  & $     931 $  & $     109 $  & $     799 $  & $      57 $  & $     389 $  & $      99 $  & $     -15 $  & $     101 $  & $     393 $  & $      90 $  & $     682 $  & $     108 $  & $     589 $  & $      86 $  \\
          IRAC $5.8\mu$m  & $   -2893 $  & $    2568 $  & $     688 $  & $    1702 $  & $   -1028 $  & $    3081 $  & $    4174 $  & $    3169 $  & $    2700 $  & $    2979 $  & $   -1686 $  & $    1819 $  & $   -1978 $  & $    4831 $  \\
          IRAC $8.0\mu$m  & $    1423 $  & $    3021 $  & $    1384 $  & $    2105 $  & $   -3418 $  & $    4231 $  & $    -753 $  & $    4341 $  & $   -1036 $  & $    3907 $  & $    -795 $  & $    2123 $  & $     499 $  & $    6310 $  \\
\enddata
\end{deluxetable*}

\begin{deluxetable*}{lr@{ $\pm$}rr@{ $\pm$}rr@{ $\pm$}rr@{ $\pm$}rr@{ $\pm$}rr@{ $\pm$}rr@{ $\pm$}r}
\tablecaption{Flux density measurements for UVISTA-Y6 to UVISTA-Y12  \label{tab:flx2}}
\tablehead{\colhead{Filter} & \twocolhead{UVISTA-Y6} & \twocolhead{UVISTA-Y7} & \twocolhead{UVISTA-Y8} & \twocolhead{UVISTA-Y9} & \twocolhead{UVISTA-Y10} & \twocolhead{UVISTA-Y11} & \twocolhead{UVISTA-Y12} \\
\colhead{name} & \twocolhead{[nJy]} & \twocolhead{[nJy]} & \twocolhead{[nJy]} & \twocolhead{[nJy]} & \twocolhead{[nJy]} & \twocolhead{[nJy]} & \twocolhead{[nJy]}
}
\startdata
            CFHTLS $u^*$  & $       8 $  & $      14 $  & $     -10 $  & $      15 $  & $     -15 $  & $      15 $  & \multicolumn2c{$\cdots$}  & $     -10 $  & $      14 $  & $      -3 $  & $      16 $  & \multicolumn2c{$\cdots$}  \\
                 SSC $B$  & $     -11 $  & $       9 $  & $     -12 $  & $       9 $  & $       0 $  & $       8 $  & $       4 $  & $      10 $  & $       0 $  & $       9 $  & $       3 $  & $       7 $  & $     -12 $  & $       9 $  \\
                 HSC $g$  & $       4 $  & $      16 $  & $     -13 $  & $      19 $  & $       0 $  & $      17 $  & $      -2 $  & $      17 $  & $      -3 $  & $      17 $  & $     -19 $  & $      15 $  & $       1 $  & $      16 $  \\
              CFHTLS $g$  & $       1 $  & $      13 $  & $     -17 $  & $      13 $  & $      -5 $  & $      13 $  & \multicolumn2c{$\cdots$}  & $     -21 $  & $      13 $  & $       6 $  & $      12 $  & \multicolumn2c{$\cdots$}  \\
                 SSC $V$  & $       0 $  & $      21 $  & $      11 $  & $      21 $  & $       2 $  & $      19 $  & $      -1 $  & $      20 $  & $      11 $  & $      19 $  & $      -1 $  & $      17 $  & $      -1 $  & $      20 $  \\
                 HSC $r$  & $      11 $  & $      15 $  & $       2 $  & $      21 $  & $      16 $  & $      17 $  & $       3 $  & $      16 $  & $       9 $  & $      16 $  & $      -3 $  & $      14 $  & $     -13 $  & $      16 $  \\
              CFHTLS $r$  & $       8 $  & $      21 $  & $     -13 $  & $      23 $  & $      -5 $  & $      19 $  & \multicolumn2c{$\cdots$}  & $       5 $  & $      19 $  & $       1 $  & $      18 $  & \multicolumn2c{$\cdots$}  \\
               SSC $r^+$  & $      20 $  & $      19 $  & $     -34 $  & $      20 $  & $     -21 $  & $      18 $  & $      -8 $  & $      19 $  & $     -14 $  & $      18 $  & $      -1 $  & $      17 $  & $     -18 $  & $      20 $  \\
              SSC  $i^+$  & $       2 $  & $      21 $  & $     -42 $  & $      32 $  & $     -26 $  & $      29 $  & $     -24 $  & $      30 $  & $     -31 $  & $      29 $  & $     -29 $  & $      27 $  & $      -2 $  & $      31 $  \\
              CFHTLS $y$  & $      22 $  & $      26 $  & $     -14 $  & $      25 $  & $      -5 $  & $      25 $  & \multicolumn2c{$\cdots$}  & $      -5 $  & $      25 $  & $      14 $  & $      22 $  & \multicolumn2c{$\cdots$}  \\
               CFHTLS $i$ & $       5 $  & $      29 $  & $     -16 $  & $      27 $  & $     -11 $  & $      26 $  & \multicolumn2c{$\cdots$}  & $      12 $  & $      25 $  & $       8 $  & $      22 $  & \multicolumn2c{$\cdots$}  \\
                 HSC $i$  & $       1 $  & $      23 $  & $       8 $  & $      30 $  & $      -7 $  & $      24 $  & $      -3 $  & $      24 $  & $     -18 $  & $      24 $  & $      14 $  & $      21 $  & $     -16 $  & $      24 $  \\
              CFHTLS $z$  & $     -13 $  & $      59 $  & $     -18 $  & $      55 $  & $      -5 $  & $      55 $  & \multicolumn2c{$\cdots$}  & $      27 $  & $      54 $  & $     -13 $  & $      46 $  & \multicolumn2c{$\cdots$}  \\
                 HSC $z$  & $      17 $  & $      33 $  & $      10 $  & $      40 $  & $     -16 $  & $      35 $  & $     -21 $  & $      35 $  & $      50 $  & $      34 $  & $      43 $  & $      32 $  & $      14 $  & $      34 $  \\
              SSC  $z^+$  & $      69 $  & $      85 $  & $      19 $  & $      77 $  & $      16 $  & $      73 $  & $     -74 $  & $      78 $  & $      54 $  & $      72 $  & $     -75 $  & $      63 $  & $     -41 $  & $      76 $  \\
                 HSC $y$  & $      89 $  & $      80 $  & $      62 $  & $      88 $  & $      55 $  & $      85 $  & $      33 $  & $      87 $  & $      -8 $  & $      85 $  & $      91 $  & $      83 $  & $      14 $  & $      84 $  \\
              UVISTA $Y$  & $      16 $  & $      51 $  & $      -9 $  & $      57 $  & $     -31 $  & $      54 $  & $     101 $  & $      53 $  & $      20 $  & $      53 $  & $     -17 $  & $      45 $  & $     -79 $  & $      53 $  \\
             UVISTA $J$   & $     211 $  & $      53 $  & $     189 $  & $      64 $  & $     192 $  & $      61 $  & $     204 $  & $      59 $  & $     242 $  & $      61 $  & $     158 $  & $      55 $  & $     164 $  & $      58 $  \\
              UVISTA $H$  & $     280 $  & $      66 $  & $     231 $  & $      76 $  & $     241 $  & $      77 $  & $     246 $  & $      78 $  & $     275 $  & $      79 $  & $     297 $  & $      71 $  & $     207 $  & $      76 $  \\
   UVISTA $K_\mathrm{S}$  & $     271 $  & $      82 $  & $     204 $  & $      84 $  & $      85 $  & $      87 $  & $     116 $  & $      87 $  & $     211 $  & $      88 $  & $     201 $  & $      77 $  & $     191 $  & $      83 $  \\
          IRAC $3.6\mu$m  & $     434 $  & $     106 $  & $     235 $  & $     107 $  & $     118 $  & $      81 $  & $     205 $  & $      79 $  & $     153 $  & $      89 $  & $     180 $  & $      54 $  & $     160 $  & $      86 $  \\
          IRAC $4.5\mu$m  & $     598 $  & $     130 $  & $     275 $  & $     102 $  & $     310 $  & $      80 $  & $      92 $  & $     100 $  & $     363 $  & $      78 $  & $     376 $  & $      67 $  & $     186 $  & $      88 $  \\
          IRAC $5.8\mu$m  & $    -643 $  & $    3000 $  & $     488 $  & $    3604 $  & $   -1188 $  & $    2445 $  & $    1378 $  & $    2596 $  & $     368 $  & $    2423 $  & $    -699 $  & $    3257 $  & $    1815 $  & $    2016 $  \\
          IRAC $8.0\mu$m  & $   -3325 $  & $    3803 $  & $    3218 $  & $    5706 $  & $    2049 $  & $    4168 $  & $   -4049 $  & $    2843 $  & $    -207 $  & $    3771 $  & $    2933 $  & $    3972 $  & $    1665 $  & $    3490 $  \\
\enddata
\end{deluxetable*}

\begin{deluxetable*}{lr@{ $\pm$}rr@{ $\pm$}rr@{ $\pm$}rr@{ $\pm$}r}
\tablecaption{Flux density measurements for UVISTA-Y13 to UVISTA-Y16  \label{tab:flx3}}
\tablehead{\colhead{Filter} & \twocolhead{UVISTA-Y13} & \twocolhead{UVISTA-Y14} & \twocolhead{UVISTA-Y15} & \twocolhead{UVISTA-Y16}  \\
\colhead{name} & \twocolhead{[nJy]} & \twocolhead{[nJy]} & \twocolhead{[nJy]} & \twocolhead{[nJy]}
}
\startdata
            CFHTLS $u^*$  & $      -6 $  & $      13 $  & $     -23 $  & $      18 $  & \multicolumn2c{$\cdots$}  & $     -28 $  & $      18 $  \\
                 SSC $B$  & $      -2 $  & $       6 $  & $      -4 $  & $       7 $  & $       8 $  & $       9 $  & $      -6 $  & $      11 $  \\
                 HSC $g$  & $     -10 $  & $      14 $  & $       4 $  & $      17 $  & $      -7 $  & $      19 $  & $      13 $  & $      16 $  \\
              CFHTLS $g$  & $      -2 $  & $      10 $  & $      -1 $  & $      15 $  & \multicolumn2c{$\cdots$}  & $      -4 $  & $      15 $  \\
                 SSC $V$  & $     -11 $  & $      17 $  & $      11 $  & $      20 $  & $      -5 $  & $      21 $  & $      18 $  & $      24 $  \\
                 HSC $r$  & $       5 $  & $      11 $  & $      12 $  & $      13 $  & $     -18 $  & $      17 $  & $       4 $  & $      16 $  \\
              CFHTLS $r$  & $       4 $  & $      15 $  & $       2 $  & $      19 $  & \multicolumn2c{$\cdots$}  & $      -4 $  & $      22 $  \\
               SSC $r^+$  & $       3 $  & $      15 $  & $       4 $  & $      18 $  & $     -11 $  & $      19 $  & $     -18 $  & $      23 $  \\
              SSC  $i^+$  & $       3 $  & $      24 $  & $      -1 $  & $      28 $  & $      22 $  & $      29 $  & $      10 $  & $      35 $  \\
              CFHTLS $y$  & $      14 $  & $      18 $  & $      23 $  & $      26 $  & \multicolumn2c{$\cdots$}  & $     -16 $  & $      27 $  \\
               CFHTLS $i$ & $      -1 $  & $      17 $  & $       8 $  & $      24 $  & \multicolumn2c{$\cdots$}  & $       1 $  & $      31 $  \\
                 HSC $i$  & $       0 $  & $      18 $  & $      20 $  & $      22 $  & $      26 $  & $      26 $  & $      -7 $  & $      24 $  \\
              CFHTLS $z$  & $     -17 $  & $      43 $  & $      54 $  & $      58 $  & \multicolumn2c{$\cdots$}  & $     -11 $  & $      64 $  \\
                 HSC $z$  & $      14 $  & $      26 $  & $      17 $  & $      27 $  & $      47 $  & $      38 $  & $      43 $  & $      35 $  \\
              SSC  $z^+$  & $     -21 $  & $      60 $  & $      29 $  & $      68 $  & $     -41 $  & $      81 $  & $      14 $  & $      75 $  \\
                 HSC $y$  & $      -4 $  & $      69 $  & $      21 $  & $      81 $  & $      24 $  & $      90 $  & $     -25 $  & $      86 $  \\
              UVISTA $Y$  & $      38 $  & $      45 $  & $     -37 $  & $      50 $  & $      69 $  & $      56 $  & $      91 $  & $      56 $  \\
             UVISTA $J$   & $     123 $  & $      47 $  & $     144 $  & $      62 $  & $     189 $  & $      56 $  & $     281 $  & $      64 $  \\
              UVISTA $H$  & $     166 $  & $      65 $  & $     206 $  & $      75 $  & $     215 $  & $      73 $  & $     268 $  & $      81 $  \\
   UVISTA $K_\mathrm{S}$  & $     245 $  & $      66 $  & $      34 $  & $      77 $  & $     106 $  & $      77 $  & $     259 $  & $      93 $  \\
          IRAC $3.6\mu$m  & $     177 $  & $      96 $  & $     191 $  & $      69 $  & $     134 $  & $      70 $  & $     221 $  & $      79 $  \\
          IRAC $4.5\mu$m  & $     362 $  & $      94 $  & $     210 $  & $      93 $  & $      -7 $  & $      83 $  & $     389 $  & $      82 $  \\
          IRAC $5.8\mu$m  & $    -287 $  & $    3140 $  & $   -1202 $  & $    3568 $  & $    1349 $  & $    2446 $  & $   -2535 $  & $    2598 $  \\
          IRAC $8.0\mu$m  & $    7391 $  & $    4246 $  & $    1553 $  & $    4921 $  & $   -1814 $  & $    3403 $  & $    2196 $  & $    3371 $  \\
\enddata
\end{deluxetable*}

\section{LF estimate when UVISTA-Y3 is considered as one single source}
\label{app:lf_bdb}

\begin{figure}
\hspace{-1cm}\includegraphics[width=9.2cm]{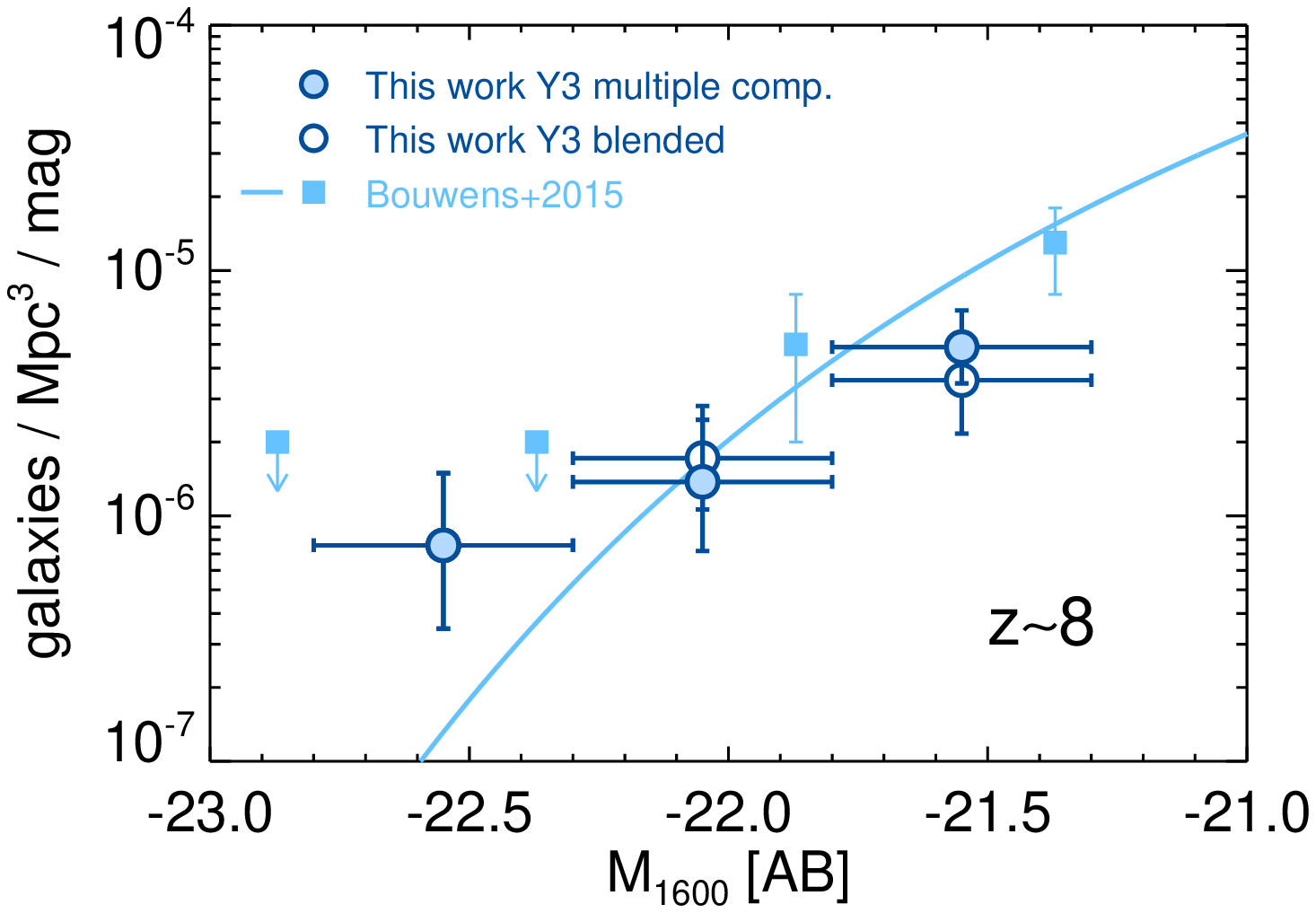}
\caption{Comparison between the LF estimate obtained after deblending UVISTA-Y3 (filled blue circles) and that when considering UVISTA -Y3 as a single source (open symbols). Previous LF determinations at $z\sim8$ from \citet{bouwens2015} are presented for comparison. The systematic differences are within the $1\sigma$ uncertainties. \label{fig:lf_comp_bdb}}
\end{figure}

In this section we present, for completeness,  $V_\mathrm{max}$ measurements of  the $z\sim8$ UV LF when UVISTA-Y3 is considered as a single source. Table \ref{tab:lf_blnd} lists the number densities, while in Figure \ref{fig:lf_comp_bdb} we present these measurements and compare them to the estimates obtained in Sect. \ref{sect:LF} assuming a multiple component nature of UVISTA-Y3. While the impact of removing three sources from the lowest luminosity bin while adding a source to a higher luminosity bin should be obvious, our treatment of this source does not change either of the impacted LF points by more than $1\sigma$

\begin{deluxetable}{cc}
\tablecaption{V$_\mathrm{max}$ determination of the UV LF when UVISTA-Y3 is  treated as a single source. \label{tab:lf_blnd}}
\tablehead{\colhead{$M_\mathrm{UV}$} & \colhead{$\phi$}  \\
\colhead{[mag]} & \colhead{[$ \times 10^{-3} \mathrm{mag}^{-1} \mathrm{Mpc}^{-3} $]} 
}
\startdata
 $ -22.55$ & $0.0008^{+0.0007}_{-0.0004} $ \\
 $ -22.05$ & $0.0017^{+0.0012}_{-0.0007} $ \\
 $ -21.55$ & $0.0036^{+0.0018}_{-0.0012} $ \\
\enddata
\end{deluxetable}

\section{LF estimate after correcting for lensing magnification}
\label{app:lens}
Here we present the $z\sim8$ LF determination after correcting for lensing magnification the luminosities of four objects (Y6, Y8, Y9 and Y13 - see Section \ref{sect:lensing}). None of the sources affected by lensing satisfies the criteria for inclusion in the $z\sim9$ LF. 

The change of solid angle introduced by the lenses affects the volume estimates  and therefore the volume densities. A proper estimate of the volumes that takes into account the lensing effects would require running a simulation with the actual lensing effects from all the sources in the region considered for our search. While this goes beyond the scope of this paper, we can attempt to estimate the average effect based on the statistics in our sample. Considering that $4/18$ objects in our sample have cumulative lensing magnification factors of $\sim1.85, 1.96, 1.39$ and $1.60$ respectively, conserving the surface brightness would imply shrinking by the same amounts the area corresponding to each source (i.e., $\sim1/18$ the area of the UltraVISTA field). Given that the remaining $14/18$ sources are not significantly magnified, the total area available for the search would then be $91\%$ of the area when no lensing is considered. The volumes densities would then be $\sim10\%$ higher (just $\sim0.04$\,dex) than those computed without introducing the lensing magnification. Because of the very marginal effect of this correction compared to the Poissonian and cosmic variance uncertainties, the LF measurements presented in this Section were computed with the same volumes adopted for the LFs presented in Section \ref{sect:LF}, which do not include any correction for lensing effects. In Table \ref{tab:lf_dl} we list the corresponding $V_\mathrm{max}$ determination when the luminosities of those sources are corrected for lensing magnification.
 
\begin{figure}
\hspace{-1cm}\includegraphics[width=9.2cm]{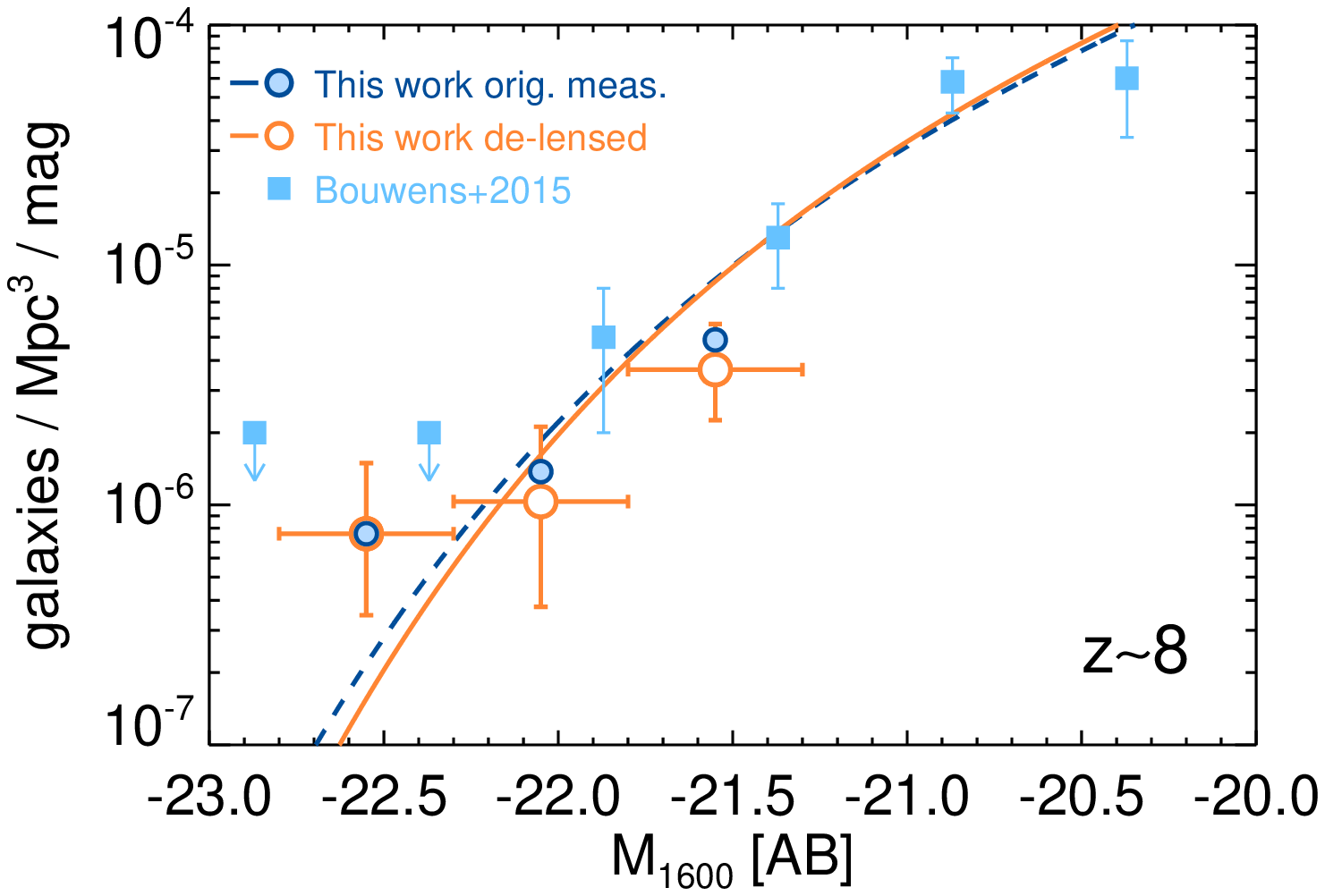}
\caption{Comparison between the LF estimate obtained after correcting sources for lens magnification (open orange circles) to those without any corrections (filled blue circrles). Error bars for the LF without lensing correction were omitted for seek of clarity. Previous LF determinations at $z\sim8$ from \citet{bouwens2015} are presented for comparison. The systematic differences are well within the $1\sigma$ uncertainties. \label{fig:lf_comp_ldl}}
\end{figure}

\begin{figure*}
\hspace{-1cm}\includegraphics[width=18cm]{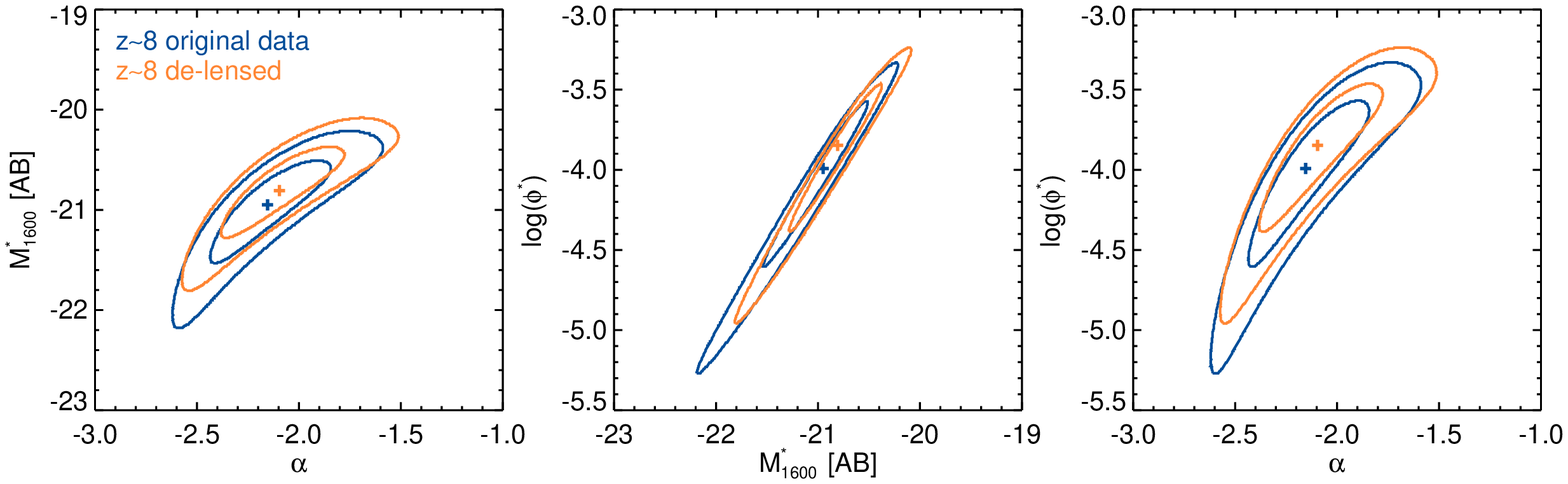}
\caption{68\% and 95\% likelihood contours on the Schechter parameters at $z\sim8$ derived in the present work after correcting the luminosities of the sample for lensing magnification (orange curves) compared to those from the analysis without lensing magnification correction (blue curves). \label{fig:schechter_dl}}
\end{figure*}

In Figure \ref{fig:lf_comp_ldl} we compare the $V_\mathrm{max}$ estimates of the de-lensed sample to those from the original sample and the corresponding Schechter parameterization. A fit to the Schechter form gives $M^*=-20.81^{+0.27}_{-0.30}$\,mag, $\alpha=-2.09^{+0.20}_{-0.19}$ and $\log\Phi^*=-3.84^{+0.26}_{-0.32}$. The $68\%$ and $95\%$ confidence intervals of the Schechter parameters are presented  in Figure \ref{fig:schechter_dl}. The LF determination obtained accounting for lensing magnification is consistent (at $1\sigma$) with that obtained without such a correction.  This suggests that, at least for the small samples that are the subject of the current study, a full accounting for the lensing magnification does not significantly impact on our conclusions.

\begin{deluxetable}{cc}
\tablecaption{V$_\mathrm{max}$ determination of the UV LF when accounting for lensing magnification. \label{tab:lf_dl}}
\tablehead{\colhead{$M_\mathrm{UV}$} & \colhead{$\phi$}  \\
\colhead{[mag]} & \colhead{[$ \times 10^{-3} \mathrm{mag}^{-1} \mathrm{Mpc}^{-3} $]} 
}
\startdata
 $ -22.55$ & $0.0008^{+0.0007}_{-0.0004} $ \\
 $ -22.05$ & $0.0010^{+0.0010}_{-0.0006} $ \\
 $ -21.55$ & $0.0037^{+0.0019}_{-0.0013} $ \\
\enddata
\end{deluxetable}

\bibliographystyle{apj}
\bibliography{mybib}

\end{document}